# The Crystallography of Aluminium and its Alloys

Philip N.H. Nakashima

Department of Materials Science and Engineering, Monash University, Victoria 3800, Australia;
Email:  Philip.Nakashima@monash.edu;  Fax: +61 3 99054940;  Phone: +61 3 99059981

## Abstract

This chapter begins with pure aluminium and a discussion of the form of the crystal structure and different unit cells that can be used to describe the crystal structure.  Measurements of the face-centred cubic lattice parameter and thermal expansion coefficient in pure aluminium are reviewed and parametrisations given that allow the reader to evaluate them across the full range of temperatures where aluminium is a solid.  A new concept called the "vacancy triangle" is introduced and demonstrated as an effective means for determining vacancy concentrations near the melting point of aluminium.  The Debye-Waller factor, quantifying the thermal vibration of aluminium atoms in pure aluminium, is reviewed and parametrised over the full range of temperatures where aluminium is a solid.  The nature of interatomic bonding and the history of its characterisation in pure aluminium is reviewed with the unequivocal conclusion that it is purely tetrahedral in nature.  The crystallography of aluminium alloys is then discussed in terms of all of the concepts covered for pure aluminium, using prominent alloy examples.  The electron density domain theory of solid-state nucleation and precipitate growth is introduced and discussed as a new means of rationalising phase transformations in alloys from a crystallographic point of view.

# Introduction

When it comes to a discussion of the crystallography of aluminium and its alloys, there is vast scope and a semi-infinite number of perspectives that could be adopted. In this chapter, a hierarchic approach is chosen. This means that the focus is initially on pure aluminium before its alloys are considered. The hierarchy breaks the crystallographic discussion down into four aspects: (i) the nature of the crystal structure and efficient ways of describing it; (ii) the magnitude of the lattice parameter at any given temperature where aluminium is a solid (and therefore the linear thermal expansion coefficient); (iii) the amplitude of atomic vibrations in the lattice as a function of temperature, again in the range of temperatures where aluminium is a solid; and (iv) the nature of the bonds between the atoms. The hierarchy outlined, leads to the determination of interatomic bonding, which is the dominant determinant of all materials properties (with the sole exception of radioactivity which is only nuclear). Interatomic bonding is considered the ultimate level of crystallographic characterization of a crystalline material, and the basis that drives all other aspects of structure.

Only after considering the four aspects of the crystallography of pure aluminium given above, can a discussion of aluminium alloys proceed. This chapter examines these four aspects for pure aluminium and then brings them to bear on a discussion of aluminium alloys via a number of significant and illustrative examples.

## The Crystal Structure of Pure Aluminium

Aluminium in its pure form has a face centred cubic crystal structure (fcc), which is a close-packed arrangement (the densest geometric packing of spheres attainable) with a layer sequence of ABCABCA... This is illustrated in figure 1 below.

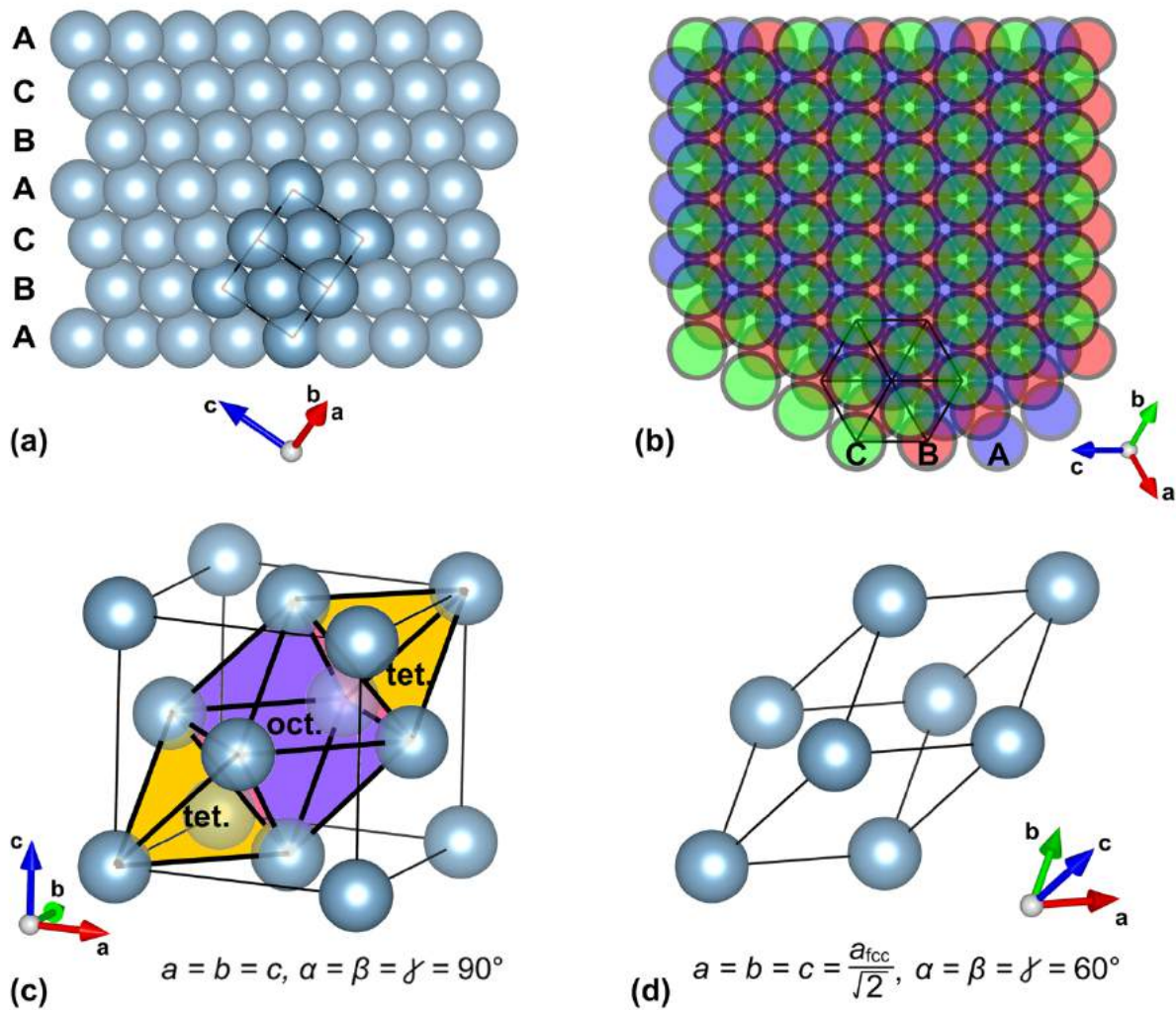

**Figure 1:** The fcc crystal structure of aluminium viewed with [111] pointing up exposes the ABCABCA... stacking of close-packed atoms (a). The atoms in a single unit cell (cell edges shown) have a darker shade. When viewed along the [111] direction (b), the order of packing is revealed by assigning each layer of atoms a different colour (blue for A, red for B and green for C). It is evident from this view that the layers A, B and C do not line up in the [111] stacking direction. The familiar fcc cell is drawn with a smaller atomic radius so that the atoms are not represented by touching spheres (c). Close-packed structures contain twice as many tetrahedral as octahedral interstices, as shown. The illustrated tessellation is a canonical description of the crystal structure as one octahedral interstice sandwiched by two tetrahedral ones form an ensemble that is the primitive rhombohedral cell. This primitive cell, containing just one atom, is, in fact, the most efficient description of the crystal structure (d). The relationships between lattice parameters for the fcc and primitive cells are also given. This figure was drawn with the aid of VESTA (1).

In close-packed structures (both hexagonal close packed, hcp, and fcc), there are always twice as many tetrahedral interstices as there are octahedral ones. This is evident for an elemental fcc structure, such as aluminium, illustrated in figure 1 (c) where the primitive rhombohedral cell, that equivalently describes the structure but contains only a single atom, is drawn within the fcc unit cell. This primitive cell is composed of two tetrahedral interstices sandwiching an octahedral one and because the primitive cell tessellates with periodic repetitions of itself to canonically describe the crystal structure of aluminium, the ratio of tetrahedral to octahedral interstices of two to one applies to the bulk.

"Rules of thumb" for a close-packed elemental structure's ability to accommodate atoms of different elements within its interstices can be established from basic geometric arguments governing each type of interstitial position, as schematically illustrated in figure 2. Considering the atoms of the host structure to be close packed hard spheres allows the size of each interstitial position to be calculated using simple geometry. It turns out that the largest sphere that can be accommodated by an octahedral interstice has a radius of 0.414 times that of the host matrix atom radius in a close-packed structure. For a tetrahedral interstice, the largest sphere that can be accommodated without strain has a radius 0.225 times the radius of the host atoms. If one considers metallic aluminium and its alloys, then there are very few situations in which alloying atoms are located interstitially because they would have to have radii of less than or equal to 0.593Å to be accommodated in the octahedral interstices and less than or equal to 0.322Å to be accommodated, without strain, in the tetrahedral interstices.

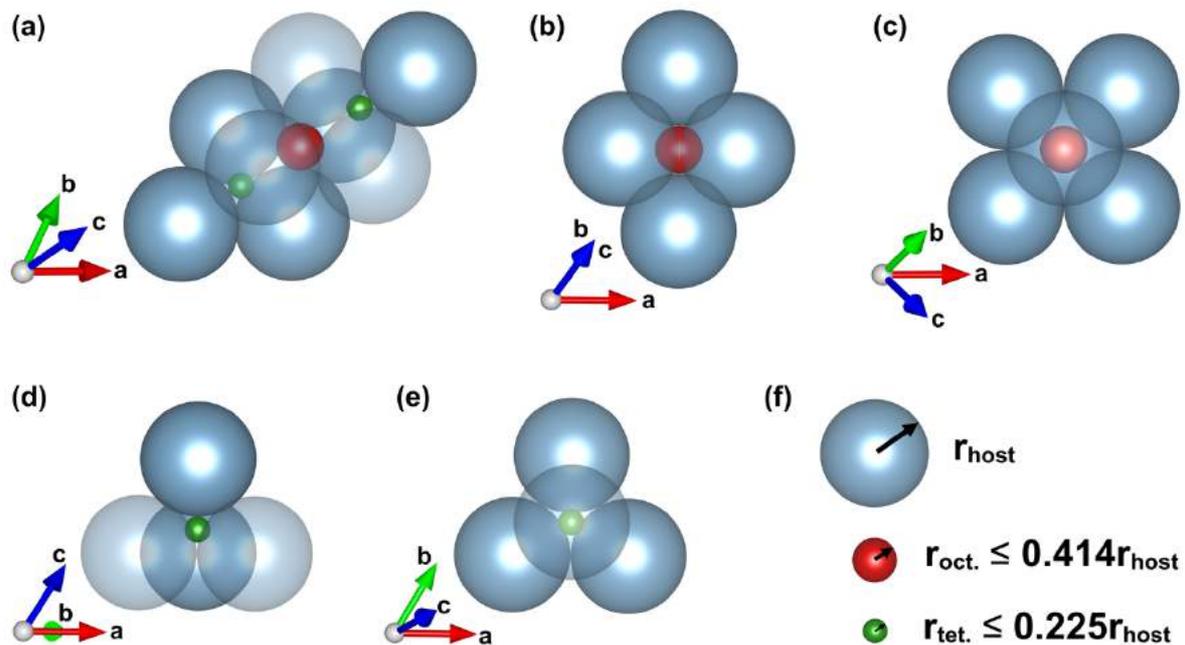

**Figure 2:** Octahedral interstices (a, b and c) can nominally fit interstitial atoms or ions whose radii are no greater than 0.414 times the radius of the host atoms (f) in close-packed structures, like aluminium. Tetrahedral interstices (a, d and e) are significantly smaller and can only accommodate atoms or ions with radii no greater than 0.225 times the radius of the host atoms (f) in close-packed structures. Here, the rhombohedral cell of aluminium is shown (a) with interstitial atoms just fitting in the octahedral interstices (red, a, b and c) and both of the tetrahedral interstices (green, a, d and e). These radius ratios (f) are geometric "rules of thumb" and do not take chemical bonding effects into account. This figure was drawn with the aid of VESTA (1).

Considering only neutral atoms, only hydrogen has a sufficiently small covalent radius (0.37 Å) (2) to be accommodated interstitially in aluminium. Oxygen and fluorine atoms, having covalent radii of 0.66 Å and 0.64 Å respectively (2), can be accommodated in the octahedral interstices with considerable strain, however, these elements will form very strong chemical bonds with aluminium atoms and this will result in a change in crystal structure (e.g. the various phases of aluminium oxide).

Diffusion of hydrogen via the interstices in aluminium can result in the formation of aluminium hydrides, which are very detrimental to the integrity of aluminium and its alloys. Thankfully, in most situations, aluminium and aluminium alloys are protected by a passivating oxide layer that effectively blocks, not only further oxidation of the aluminium, but also the ability of hydrogen to diffuse into the aluminium. One way in which hydrogen can become a problem is if the protective oxide surface of the aluminium or aluminium alloy is being stripped away at a rate that allows hydrogen to enter and diffuse through the exposed aluminium lattice. This can occur, for example, during electropolishing of aluminium and its alloys if the temperature of the electropolishing solution is not maintained at a sufficiently low level (e.g. -20°C for $HNO_3$/methanol solutions).

Some of the most advanced aluminium alloys contain lithium. Only lithium ions can be accommodated interstitially, having ionic radii of 0.60 Å, whilst the neutral atoms have radii of 1.35 Å (2). This becomes important when it comes to characterising aluminium alloys with ionising radiation (as for example, in transmission electron microscopy). Lithium is relatively easily ionised by high-energy beams of radiation and this makes lithium highly mobile in aluminium alloys which contain it, due to migration via interstitial pathways. As a result, lithium is particularly difficult to locate in aluminium alloys using electron microscopy and techniques that probe materials with high-energy radiation.

The size of the interstices in aluminium means that almost all alloy solid solutions in aluminium are substitutional, with different elements having different solid solubilities in aluminium. Upon heat treatment of supersaturated aluminium-based solid solutions, phase transformations resulting in the formation of intermetallic precipitates can occur where the precipitates have entirely different crystal structures from the host fcc aluminium matrix.

Whilst the focus on aluminium alloys is left until the last section of this chapter, it will be useful at this point to consider other ways of describing the atomic structure of pure or elemental aluminium. This is done in figure 3, the purpose of which is to serve as a handy reference for the structural modelling of intermetallic precipitate phases and the surrounding aluminium matrix in aluminium alloys. Many such phases are to a greater or lesser degree, coherent with the aluminium host matrix along interfacial planes that are not necessarily {001} in the fcc cell. For example, the $T_1$ phase ($Al_2CuLi$) in aluminium-copper-lithium alloys is hexagonal and forms fully-coherent platelets with main facets composed of the basal plane, (001), which is coplanar to {111} planes in the fcc cell of the aluminium matrix. To model the ensemble structure of matrix / precipitate / matrix, one could either use the fcc cell of the aluminium matrix and substitute {111} planes with the $T_1$ structure, or one could use the trigonal cell defined in figure 3 (d, e and f) to describe the aluminium matrix and then simply replace {001} planes in this description with the $T_1$ structure, coplanar to the $T_1$ structure's basal plane of (001). The equivalence of these approaches to modelling this alloy structure may make the use of different descriptions of the aluminium matrix structure seem redundant, however, when it comes to applying such structural models to the analysis of experimental data, the defining frames of reference become vital to the task.

An example is the interpretation of electron diffraction patterns or lattice images using the multislice formalism for describing electron scattering from crystals (3). This method requires the material being probed and analysed to be sliced into its constituent planes of atoms whose normals are parallel to the incident electron beam direction. This requires the two-dimensional periodicity of the crystal structure to be defined perpendicularly to the beam direction so that sampling of the structure within slices can be Fourier transformed as part of the multislice algorithm. This is only practical if the beam and slicing directions are defined as [001] throughout the structure being modelled. In the example of $T_1$, this is only possible if the aluminium host matrix is described using a trigonal cell like the one defined in figure 3 (d, e and f).

Whilst figures 1 and 2 show the rhombohedral primitive cell of aluminium in order to illustrate the geometric relationship between the constituent atoms in the two types of interstices that exist in the structure of aluminium, figure 3 presents four additional cells that are defined in relation to the familiar fcc unit cell (and in one case, the rhombohedral primitive cell). These are: a body centred tetragonal (bct) cell where c is the longest cell edge (figure 3 (a, b and c)), a trigonal cell (figure 3 (d, e and f)), an orthorhombic cell (figure 3 (g, h and i)) and a tetragonal cell where c is the shortest cell edge (figure 3 (j, k and l)). For each of these alternative cells, their orientation and position with respect to the fcc description of the structure of aluminium is presented by showing the constituent atoms of the new cells in red. In contrast, the atoms that belong only to the fcc cell and that are not common to the fcc cell and the cell being defined, are shown in blue (figure 3 (a, d, g and j)).

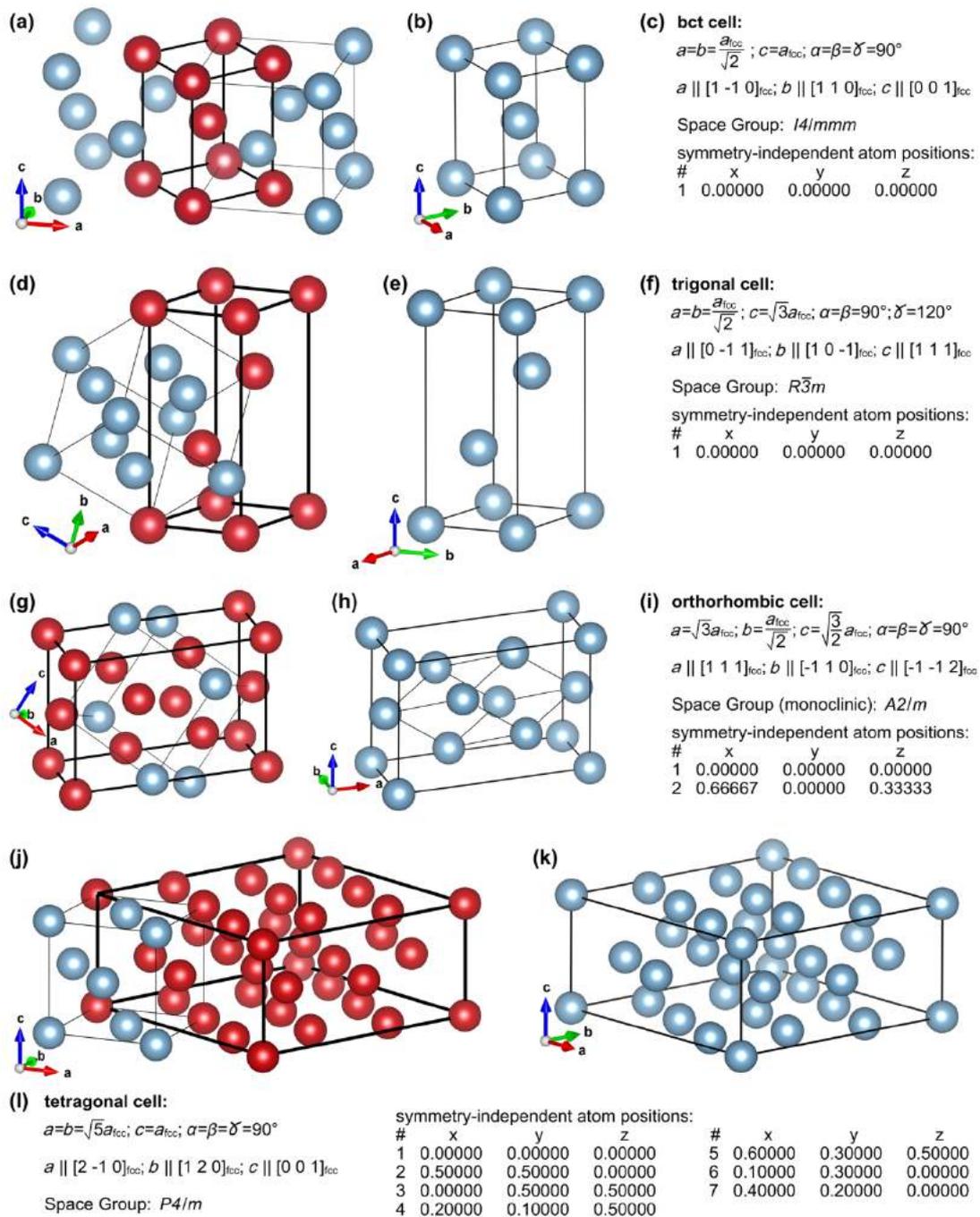

Figure 3: Four alternate unit cells that will fully describe any fcc metal like aluminium. Each cell is shown in relation to the familiar fcc cell (with atoms in the alternate cell in red and those in the fcc cell but outside the alternate cell in blue) (a, d, g and j). Each alternate cell is also shown standing alone (atoms in blue) (b, e, h and k). The alternate cells shown here have axes and cell facets that correspond to the growth axes and planes of many of the precipitate phases encountered in aluminium alloys (see table 4 and the subsection Crystal structures of (some) aluminium alloys). Having the full description of the geometric relationships between these alternate cells and the fcc cell (c, f, i and l) on hand, is useful for modelling and analysing precipitate / matrix interfacial structures in aluminium alloys and for determining the degree of coherence of such interfaces and the amount of strain imparted on the host aluminium matrix by the precipitates. Reference to these geometric relationships is also very useful in the context of simulating electron scattering within aluminium alloys in different directions using the multislice formalism (3). This is becoming an important tool for interpreting transmission electron microscope (TEM) images and diffraction patterns because TEM is one of the primary techniques for characterising alloys at the atomic scale. This figure was drawn with the aid of VESTA (1).

The second frame in the definition of each new cell (figure 3 (b, e, h and k)), shows a single unit cell of the new definition with all atoms coloured blue. In the case of the orthorhombic cell (figure 3 (h)), the two opposing face-centred atoms are also at opposing corners of the primitive rhombohedral cell, so this has been drawn in for further context.

The third frame in each cell definition (figure 3 (c, f, i and l)) explicitly states the geometric relationships between each newly defined cell and the fcc unit cell. The lengths of each of the cell edges are given in terms of $a_{fcc}$, the lattice parameter of the fcc cell, and the angles between cell edges are given explicitly in degrees. The orientation relationships between the edges of the cell being defined and directions in the fcc cell are given next. This is followed by the space group symmetry defining the symmetry-related locations of atoms in each new cell and, in conjunction with the assigned space group, the symmetry-independent atom positions required to locate all of the atoms constituting the new cells.

The unit cells defined in figure 3 cover many of the coherent or semi-coherent systems of matrix/precipitate/matrix encountered in aluminium alloys. The only variable whose value is not specified in the definitions presented in figure 3 is the fcc lattice parameter of aluminium, $a_{fcc}$. The next section examines the lattice parameter of aluminium, and its associated linear thermal expansion coefficient, in sufficient detail to produce an accurate parametrisation of both of these physical characteristics as a function of temperature, allowing the user to obtain their values to high accuracy in pure aluminium at any temperature where aluminium is solid (i.e. 0K – 933K). The value of $a_{fcc}$ at whatever temperature is relevant to a particular experiment, is the value that should be substituted into the geometric relationships listed for each cell definition in figure 3.

# The Lattice Parameter of Pure Aluminium

Fundamental to the crystallography of any crystalline material is knowledge of the lattice parameter of the unit cell. For aluminium, the fcc cell is always taken as the frame of reference (so $a=b=c$ and $α=β=γ=90°$). Here, a summary of the literature is presented with particular focus on experimentally measured lattice parameters for aluminium between 0K and the melting temperature of 933K. From more than 300 measurements spanning this temperature range, a function for the lattice parameter with respect to temperature, $a$(T), has been determined (see equation 1 below). The present parameterisation can be used to give the lattice parameter at any temperature that aluminium is a solid. In addition, the self-normalised derivative of this function should give an accurate function for the thermal expansion coefficient of aluminium, $α$(T). A summary of experimentally measured thermal expansion coefficients for aluminium is also presented in this section.

Measurements of the lattice parameter date back to the 1920s and the early days of X-ray diffraction and crystallography (4 – 6). A large number of measurements of the aluminium lattice parameter have been made since then (7 – 71), using not only powder and single-crystal X-ray diffraction, but also electron diffraction (54, 55). The references provided here may not be exhaustive but are as complete as possible. Appendix A gives a tabular summary of each reference (4 – 71), the temperature at which determinations were made, and the lattice parameter determined at each temperature. Notes about each measurement, where relevant, and sample purity, where available, are supplied in the summary. Prior to the mid-1940s, measurements of lattice parameter were conventionally given in units of kX. The unit, X, was derived from the calcite spacing, thought at the time of its definition to be $1\times10^{-13}$m. Therefore, units of kX (or 1000X) were taken as equivalent to $10^{-10}$m – the unit of length referred to now-a-days as the Ångström. However, it was found that the calcite spacing was in error by approximately 0.2%. In 1947, none other than Bragg and Armstrong Wood made the clarifying statement that 1kX = 1.00202Å (72). Further research settled on the conversion factor of 1kX = 1.00208Å (73). When summarising the measurements of $a$(T) over the last 90+ years, values quoted in kX have been converted to Å using the appropriate conversion factor.

Figure 4 plots all experimental measurements of the aluminium fcc lattice parameter against the temperature of each measurement. The literature includes a number of attempts to extrapolate the low temperature data to a value of $a$(0K) (59, 63, 65) and these are shown as the green points in the graph. The blue line is the best fit to the experimental data (including the T=0K extrapolated values) of the following function:

$$a(\text{T}) = \frac{m}{(n+\text{T})^{14}}\left(\sum_{i=0}^{15} p_i \text{T}^i - \sum_{i=0}^{14} q_i T^i \ln(n+\text{T})\right). \qquad (1)$$

The optimised parameters for equation 1 are listed here to 15 significant figures to prevent rounding errors when evaluating the equation (rounding errors can make this otherwise monotonic function non-monotonic at T<20K):

$m$ = 5.46215569838344x10$^{-4}$, $n$=86.0150000000000,

$p_0$ = 1.80595722249783x10$^{31}$, $p_1$ = 2.96317351342354x10$^{30}$, $p_2$ = 2.25855016769900x10$^{29}$, $p_3$ = 1.06004494945302x10$^{28}$, $p_4$ = 3.42305759706082x10$^{26}$, $p_5$ = 8.04609813573980x10$^{24}$, $p_6$ = 1.41998402393052x10$^{23}$, $p_7$ = 1.91189157411850x10$^{21}$, $p_8$ = 1.97428843905832x10$^{19}$, $p_9$ = 1.55692828415459x10$^{17}$, $p_{10}$ = 9.23827056204128x10$^{14}$, $p_{11}$ = 4.00527886712625x10$^{12}$, $p_{12}$ = 1.20273329860717x10$^{10}$, $p_{13}$ = 2.25714359192389x10$^{7}$, $p_{14}$ = 2.06989439078737x10$^{4}$, $p_{15}$ = 1.00000000000000,

$q_0$ = 2.04335860741967x10$^{30}$, $q_1$ = 3.32581764853519x10$^{29}$, $q_2$ = 2.51326102603950x10$^{28}$, $q_3$ = 1.16875476418741x10$^{27}$, $q_4$ = 3.73664547057534x10$^{25}$, $q_5$ = 8.68835777614448x10$^{23}$, $q_6$ = 1.51514697020481x10$^{22}$, $q_7$ = 2.01313321790028x10$^{20}$, $q_8$ = 2.04788881667471x10$^{18}$, $q_9$ = 1.58723386748412x10$^{16}$, $q_{10}$ = 9.22649460840626x10$^{13}$, $q_{11}$ = 3.90058588445197x10$^{11}$, $q_{12}$ = 1.13369350824041x10$^{9}$, $q_{13}$ = 2.02772058226054x10$^{6}$, $q_{14}$ = 1.68385961108157x10$^{3}$.

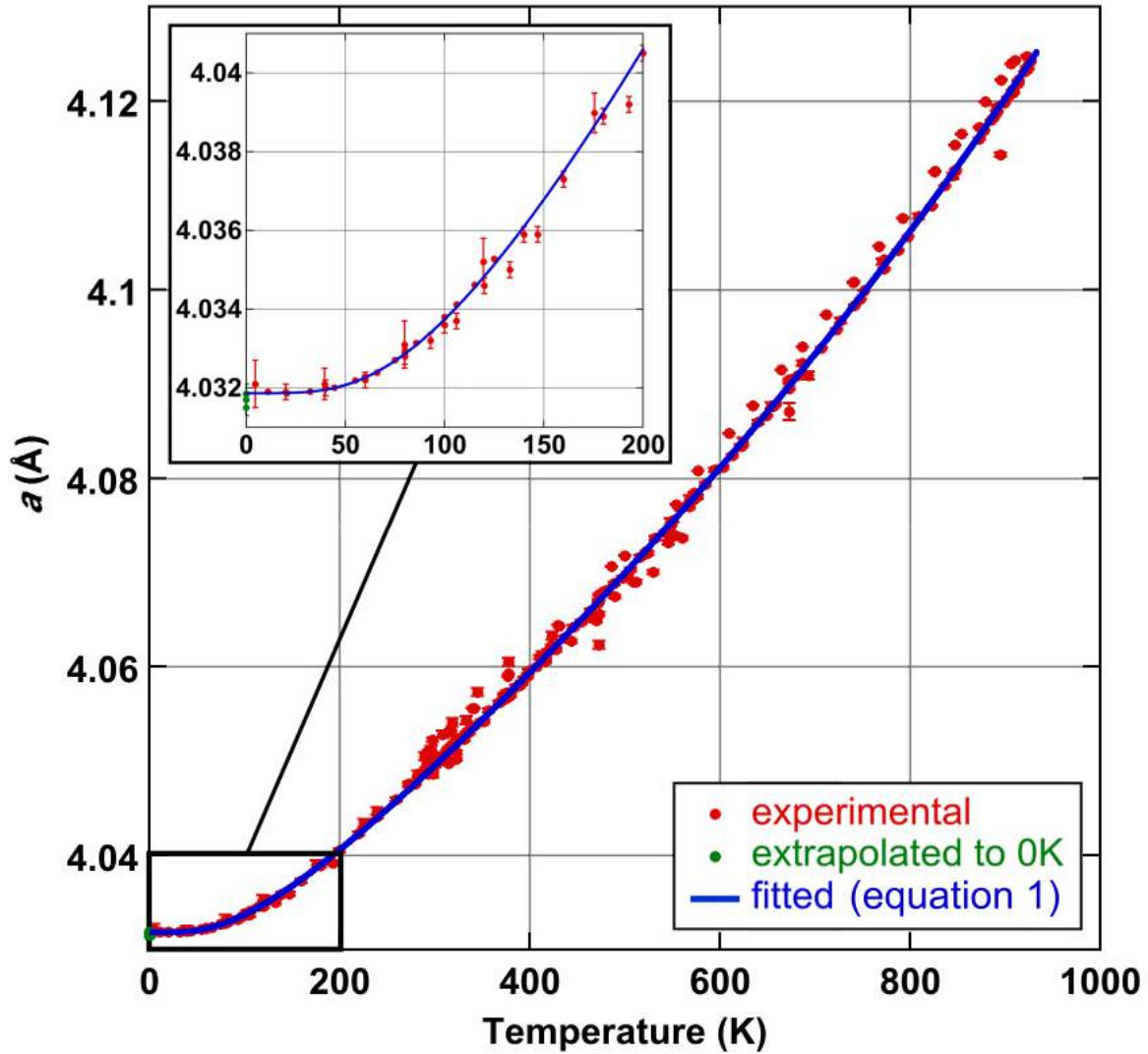

**Figure 4:** A graphical summary of lattice parameter measurements in pure aluminium, from 1925 to the present day (4 – 71). 326 independent experimental measurements are given as red points, ranging from 4.6K to 927K (just below the melting point of 933K). Three points are given at 0K (green) and these were determined by extrapolating low temperature measurements back to absolute zero (59, 63, 65). All points are graphed with error bars; however, some uncertainties are too small to be resolved in the graph. The blue line represents the function fitted to these experimental data (equation 1). The inset expands the graph in the low temperature range to show how equation 1 performs in this region.

Figure 5 compares the present fitted function for $a$(T) with the seminal models of Wang and Reeber for perfect and real crystals (74) as applied to aluminium (75). The distinction between perfect and real is ignoring and accounting for the presence of vacancies respectively. These models deviate significantly from the experimental measurements and thus, from the present fit, at low temperatures. In the middle range of temperature, there is excellent agreement between both models and the present fit, whilst the perfect and real models diverge from each other at higher temperature and bound the present fit. The divergence of the two Wang and Reeber models shows the increasingly significant effect of vacancies on the average lattice parameter in a real crystal, where the concentration of vacancies increases rapidly with temperature as the melting point of aluminium is approached. This becomes even more evident when the linear thermal expansion coefficient, $\alpha$(T), for aluminium is examined.

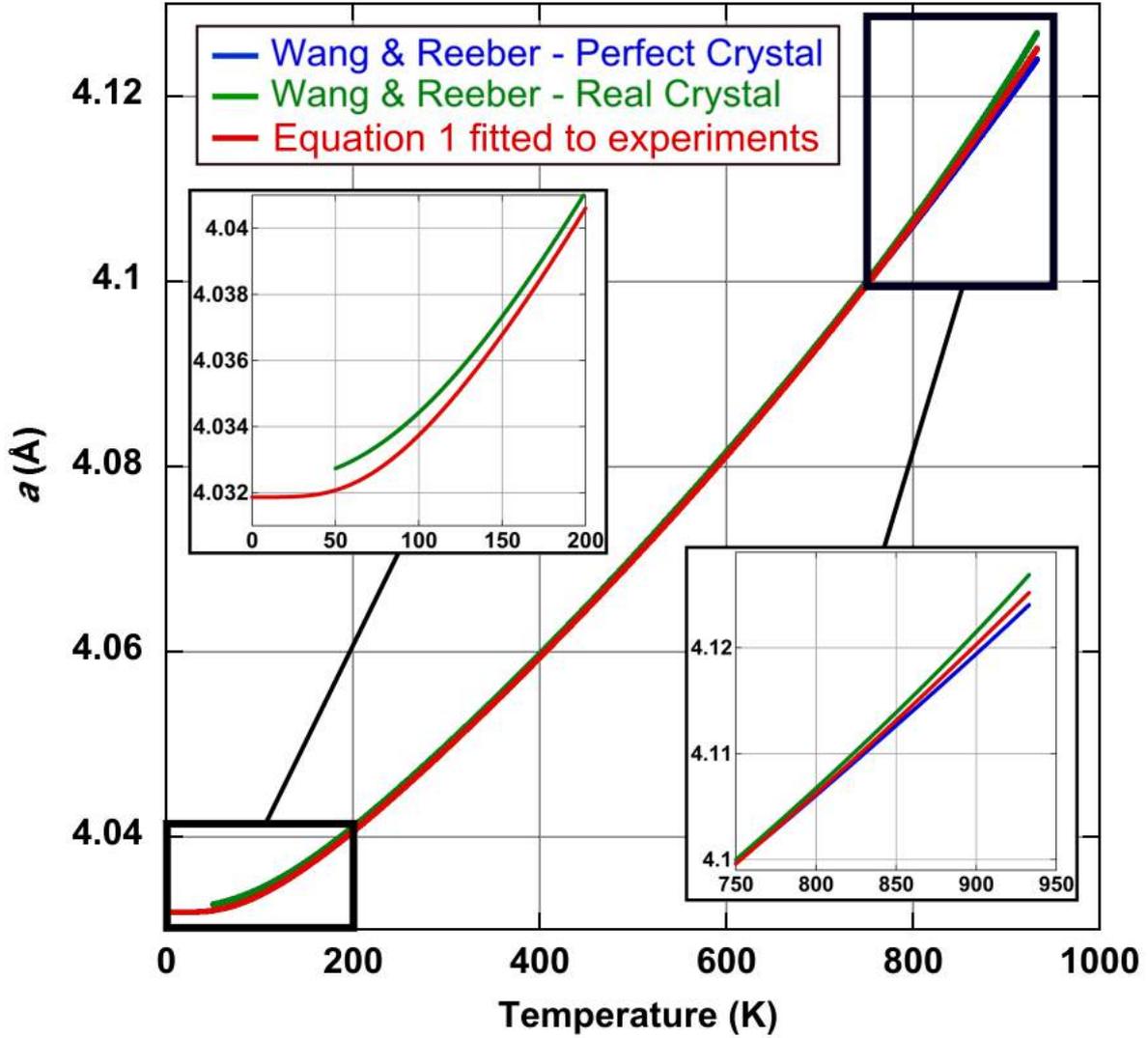

**Figure 5:** A graphical comparison of the present fitted function of lattice parameter versus temperature, $a$(T) (equation 1 and figure 4), with the perfect and real crystal models of aluminium by Wang and Reeber (74, 75). The low and high temperature ranges have been expanded to show the deviations between the present function, $a$(T), and Wang's and Reeber's models in these ranges. At low temperatures the perfect and real crystal models of Wang and Reeber agree exactly due to the low equilibrium concentration of vacancies. Nearer the melting point of aluminium, the vacancy concentration becomes large enough to cause the real crystal model of the lattice parameter to be significantly greater than that of the perfect crystal model.

The functional form and optimised parameters of equation 1 were constrained by the requirement that the self-normalised derivative of equation 1 with respect to temperature (resulting in equation 2) must also be the best fit to experimental determinations of the thermal expansion coefficient, $\alpha$(T), of aluminium. The experimental measurements of $\alpha$(T) span more than 100 years of research (15, 16, 18, 20, 22, 29, 42, 43, 45, 49, 52, 53, 59, 60, 69, 76 – 96) and are plotted in figure 6. Specific values and notes from each reference are given in Appendix B. The experimental data are plotted together with $\alpha$(T) according to the perfect and real crystal models of Wang and Reeber (74, 75) and the function for $\alpha$(T) determined in the present work:

$$\alpha(\text{T}) \;=\; \frac{1}{a(\text{T})}\frac{\mathrm{d}a(\text{T})}{\mathrm{d}\text{T}} \;=\; \frac{\sum_{i=0}^{15} s_i \text{T}^i}{(n+\text{T})\left(\sum_{i=0}^{15} p_i \text{T}^i - \sum_{i=0}^{14} q_i \text{T}^i \ln(n+\text{T})\right)}. \tag{2}$$

The optimised parameters for fitting equation 2 to the experimental data points are:

$s_0 = 8.55692806695981 \times 10^{18}$, $s_1 = 1.09556645819752 \times 10^{24}$, $s_2 = 1.78316852655619 \times 10^{23}$, $s_3 = 2.04804545623587 \times 10^{22}$, $s_4 = 1.60396172331341 \times 10^{21}$, $s_5 = 2.36491757661617 \times 10^{20}$, $s_6 = 2.07873881554774 \times 10^{19}$, $s_7 = 9.19266228575283 \times 10^{17}$, $s_8 = 2.20725652279774 \times 10^{16}$, $s_9 = 2.93361641840158 \times 10^{14}$, $s_{10} = 2.08150841393624 \times 10^{12}$, $s_{11} = 8.47737173962446 \times 10^{9}$, $s_{12} = 5.09073073293741 \times 10^{7}$, $s_{13} = 3.26718741801178 \times 10^{5}$, $s_{14} = -3.93634611081583 \times 10^{2}$, $s_{15} = 1.00000000000000$,

and parameters $n$, $p_i$ and $q_i$ are the same as for equation 1. Again, all parameters have been listed to 15 significant figures to avoid problems associated with rounding errors.

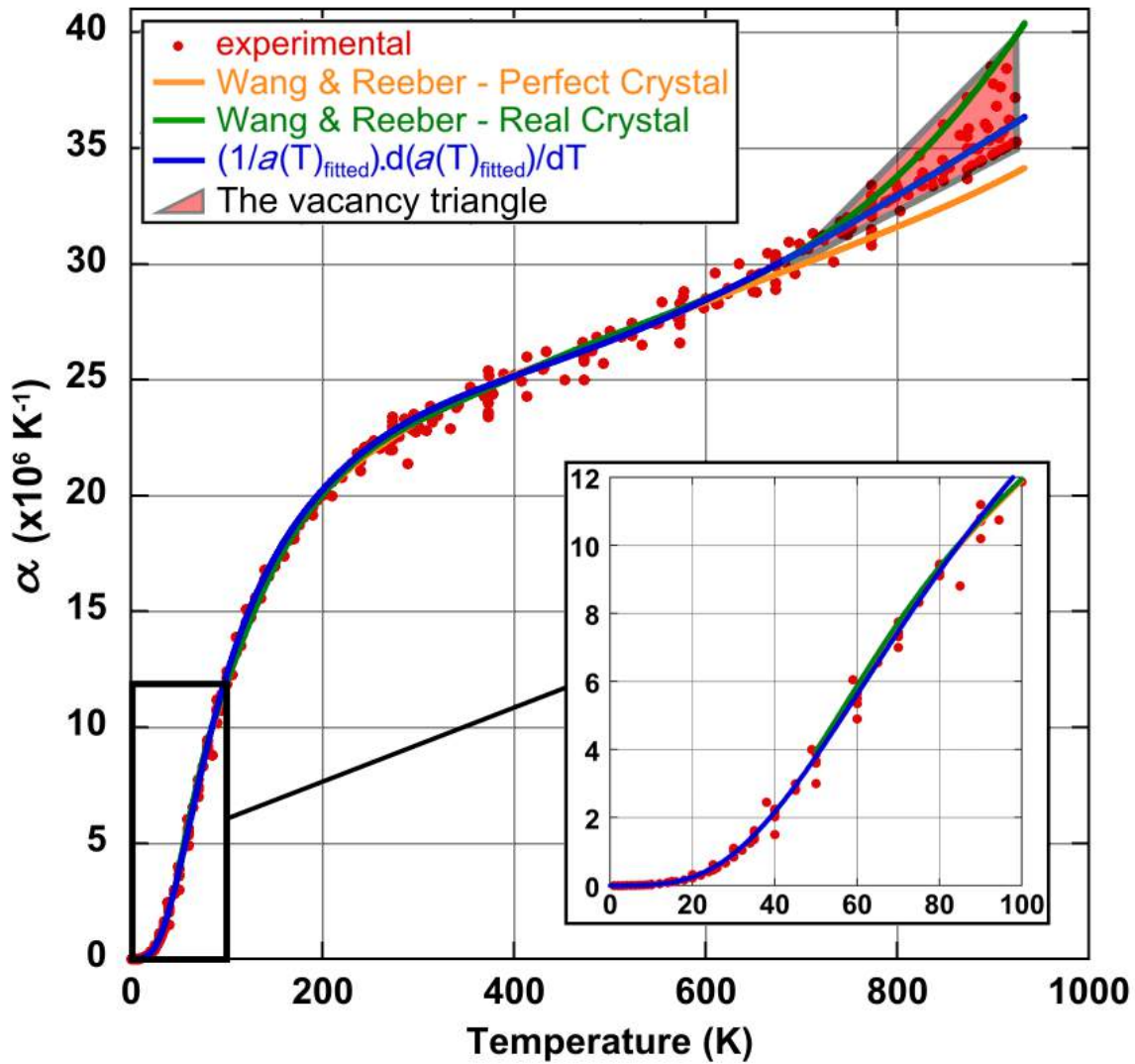

**Figure 6:** A graphical summary of thermal expansion coefficients for aluminium from 0K to 933K (the melting temperature). Experimental measurements date from 1907 to the present day (15, 16, 18, 20, 22, 29, 42, 43, 45, 49, 52, 53, 59, 60, 69, 76 – 96). The thermal expansion coefficient as a function of temperature, $\alpha(T)$, as determined by the present review (equation 2), is plotted and compared with $\alpha(T)$ as determined by the perfect and real crystal models of Wang and Reeber (74, 75). The inset expands the low temperature region of the graph to show the quality of the fit of $\alpha(T)$ determined here with low temperature measurements. A new concept, the "vacancy triangle", is drawn into the graph and spans the divergent area at higher temperatures where $\alpha(T)$ determined from changes in lattice parameter differs from $\alpha(T)$ determined by bulk length dilations, the latter being affected by vacancies.

The constraint that interdependent equations 1 and 2 simultaneously fit the experimental determinations of $a(T)$ and $\alpha(T)$ from 0K to near the melting point of 933K resulted in the large number of terms in both functions to ensure excellent simultaneous fits spanning the whole temperature range where aluminium is a solid. Previous efforts to fit functions to both $a(T)$ and $\alpha(T)$ were based largely on polynomials and were only valid over limited temperature ranges (59, 63, 68, 70, 77, 91, 95, 96). One particular example by Kroeger and Swenson (96), which represents some of the most rigorous work to date for $\alpha(T)$, fits numerous polynomial functions of varying order to four separate temperature ranges that in combination go from 0K to 330K. In total, 24 parameters are used. In the present fits, 32 independent parameters (given $p_{15}$ = 1 for equation 1) are required for $a(T)$ from 0K to 933K, and an additional 15 independent parameters (given $s_{15}$ = 1 for equation 2) are required for $\alpha(T)$ from 0K to 933K.

The present fit functions for $a(T)$ and $\alpha(T)$ lie between the bounds set by the Wang and Reeber perfect and real crystal models at the high-temperature end of the graphs in figures 5 and 6. This can be understood by considering the present fits as averages over the spread in the experimental determinations spanning the literature, whilst the perfect and real crystal models of Wang and Reeber (74, 75) represent the theoretical upper and lower bounds of such measurements respectively.

The experimental thermal expansion data shown in the plot of figure 6 are compiled from measurements of both $\Delta a/a_0$ and $\Delta L/L_0$. The former ratio is the change in lattice parameter from its value at a particular reference temperature, $T_0$, divided by the value of the lattice parameter at that temperature. Measurements of lattice parameter invariably involve a diffraction experiment (usually single crystal (29, 40, 45, 60, 92) or powder X-ray diffraction (15, 16, 20, 22, 29, 42, 43, 49, 52, 53, 59, 69, 82, 84)). The second of the ratios comes from measurements of changes in length of a bulk sample of material. The length dilation, $\Delta L$, with changes in temperature relative to a reference temperature, $T_0$, can be very accurately measured by any number of techniques. These include interferometry (18, 53, 60, 78, 81, 86, 91), capacitance or differential transformer dilatometry (40, 89, 94 – 96), fixed comparative optical microscopy (45, 76, 77, 92) and optical levering (90). The denominator, $L_0$, is simply the absolute length of the bulk sample at the reference temperature. The reference temperature, $T_0$, is a temperature at which the vacancy concentration can be considered to have a negligible effect on length dilation with changes in temperature. In other words, $\Delta a/a_0 \equiv \Delta L/L_0$ at $T_0$.

Measurements of $\Delta a/a_0$ show what is happening in terms of only the unit cell dimension, whilst measurements of $\Delta L/L_0$ reflect what is happening to the bulk material as a result of changes in cell dimension plus dilation of the bulk caused by the presence of vacancies. This means that the difference between thermal expansion coefficients measured crystallographically and by bulk length dilation is a direct measure of vacancy concentration. By definition (40, 42, 45, 53, 60, 90), the vacancy concentration, $C_{vac}$, is given by:

$$C_{\text{vac}} \approx 3\left(\frac{\Delta L}{L_0} - \frac{\Delta a}{a_0}\right). \qquad (3)$$

Using the approximations that $a \approx a_0$ and $T \approx T_0$ (both accurate to within 2% across the entire temperature range that aluminium is a solid) the thermal expansion coefficient, $\alpha(T)$, can be related to $\Delta a/a_0$ and $\Delta L/L_0$ as follows:

$$\alpha_{\text{min}}(T) = \frac{1}{a}\frac{da}{dT} \approx \frac{1}{a_0}\frac{da}{dT}; \quad \alpha_{\text{max}}(T) = \frac{1}{L}\frac{dL}{dT} \approx \frac{1}{L_0}\frac{dL}{dT}. \qquad (4)$$

Here, $\alpha_{\text{min}}(T)$ is essentially the linear thermal expansion coefficient for a perfect crystal whilst $\alpha_{\text{max}}(T)$ is that for a crystal containing the equilibrium concentration of vacancies at temperature T. It then follows that:

$$\int_{T_0}^{T} \alpha_{\text{min}}(T)dT \approx \frac{1}{a_0}\int_{a_0}^{a} da = \frac{\Delta a}{a_0}; \int_{T_0}^{T} \alpha_{\text{max}}(T)dT \approx \frac{1}{L_0}\int_{L_0}^{L} dL = \frac{\Delta L}{L_0}, \qquad (5)$$

and therefore:

$$\left(\frac{\Delta L}{L_0} - \frac{\Delta a}{a_0}\right) \approx \int_{T_0}^{T} \alpha_{max}(T)dT - \int_{T_0}^{T} \alpha_{min}(T)dT. \tag{6}$$

Because each integral is just the area under $\alpha_{max}(T)$ and $\alpha_{min}(T)$ respectively, equation 6 is equivalent to the area spanned by the spread in experimental determinations of $\alpha(T)$ by measurements of both $\Delta a/a_0$ and $\Delta L/L_0$ as the temperature gets high enough for vacancies to have an appreciable effect on length dilation. The spread in experimental determinations of $\alpha(T)$ is bounded by $\alpha_{max}(T)$ and $\alpha_{min}(T)$ and the area is approximately represented in the graph of $\alpha(T)$ (figure 6) as the shaded triangle. According to equations 3 and 6, the area of this triangle should be equivalent to one third of the vacancy concentration at the temperature at which the vertical edge of the triangle is located (927K in the present summary). The area of the triangle as drawn in figure 6 is $\approx 4.46 \times 10^{-4}$ and therefore $C_{vac} \approx 1.3 \times 10^{-3}$. The present triangle encompasses almost all of the experimental points and therefore represents an upper bound. Depending on where the left vertex of the triangle is positioned, the area of the triangle can change somewhat and a lower bound on the area has also been measured. It is $2.7 \times 10^{-4}$, which equates to $C_{vac} \approx 8.1 \times 10^{-4}$. Thus, the vacancy concentration reported using the new "vacancy triangle" estimation method presented here for the first time is $C_{vac} \approx (1.1 \pm 0.3) \times 10^{-3}$. This is comparable with determinations published previously by individual studies (40, 42, 45, 53, 60, 92) as shown in table 1. The $C_{vac}$ determined from the "vacancy triangle" in figure 6, agrees, within the given margins of error, with all previous determinations of $C_{vac}$ near the melting point of aluminium, with the exception of the determination of (40). The result reported by (40) is significantly different to all other determinations from the literature, summarised here.

**Table 1:** A comparison of previously published vacancy concentration determinations for aluminium near its melting point with the determination using the "vacancy triangle" method derived in the present work. Previous determinations used measurements of $\Delta L/L_0$ and $\Delta a/a_0$. The "vacancy triangle" approach is described here in the present review for the first time and is a quantitatively useful way of interpreting the spread in the experimental graph of $\alpha(T)$ Vs T in figure 6 in terms of the vacancy concentration at the temperature bounding the vertical side of the vacancy triangle (927K in the present experimental summary). Note that the "vacancy triangle" method gives a vacancy concentration that agrees, within the margins of error, with all previous determinations summarised below, with the exception of the result from (40). The measurement of (40) appears to significantly underestimate $C_{vac}$ near the melting point and is also at odds with all other determinations presented in the table.

| $C_{vac}$ | T (K) | Ref. |
|---|---|---|
| $(1.1 \pm 0.3) \times 10^{-3}$ | 927 | present |
| $3 \times 10^{-4}$ | 933 | (40) |
| $(1.1 \pm 0.2) \times 10^{-3}$ | 933 | (42) |
| $9.4 \times 10^{-4}$ | 933 | (45) |
| $(9 \pm 1) \times 10^{-4}$ | 933 | (53) |
| $9.8 \times 10^{-4}$ | 933 | (60) |
| $8.5 \times 10^{-4}$ | 928 | (92) |

The present determination is on the higher side of the range of estimations previously published and can be explained by the fact that the triangle only approximates the region bounded by $\alpha_{max}(T)$ and $\alpha_{min}(T)$. A more accurate shape would be a triangle with the top edge replaced by a concave-up arc, which would reduce the enclosed area and thus, the estimated vacancy concentration. In addition, the triangle is drawn to contain almost all of the experimental measurements of $\alpha(T)$, whilst the true spread may not be as large.

Vacancies are fundamental crystal defects that have a very strong influence on the diffusion of solutes in alloys. This and the fact that they can act as nucleation sites for precipitate phases and phase transformations, means that vacancies play a pivotal role in the evolution of alloy microstructure along any heat treatment route. Vacancies alone are known, under conducive conditions, to aggregate into nano-scale voids (97 – 104), which can be deleterious to mechanical and electrical properties (105 – 110).  On the other hand, voids have great potential in the fabrication of novel electronic, catalytic and plasmonic materials (111 – 113).  Voids have received significant attention over the years (97 – 123) and a very recent paper revisits them in aluminium with new findings about how they evolve by the migration of vacancies through different crystallographic facets (123).

The primary focus of this section is, however, on the lattice parameter of aluminium, its dependence on temperature and the properties that can be derived from it.  This section closes with a comparison of the aluminium lattice parameter and linear thermal expansion coefficient obtained from equations 1 and 2 respectively (using the coefficients listed above), with average experimental values determined from published measurements (4 – 71, 76 – 96) at six key temperatures.  These are 0K, 50K (the lower limit of Wang's and Reeber's models (74, 75)), 273K (freezing point of water), 298K (room temperature), 400K (near the Debye temperature of aluminium (86, 124)) and 923K (10K below the melting point of aluminium as measurements at the melting point are not feasible).  This is done in table 2.

Table 2:  A comparison of the lattice parameter, $a$(T), and linear thermal expansion coefficient, $\alpha$(T), at six key temperatures (0K, 50K, 273K, 400K, 298K and 923K).  The values of $a$(T) and $\alpha$(T) obtained from equations 1 and 2 respectively, using the given parameters, are compared to the average of the experimentally measured values for $a$(T) and $\alpha$(T) from the literature (4 – 71, 76 – 96) reviewed here and Wang's and Reeber's perfect (WRP) and real (WRR) crystal models (74, 75) (also plotted in figures 5 and 6).

| T(K) | $a$(T) (Å) Expt. | $a$(T) (Å) WRP | $a$(T) (Å) WRR | $a$(T) (Å) (Eq. 1) |
|---|---|---|---|---|
| 0 | 4.0317±0.0002 | ----- | ----- | 4.0319 |
| 50 | 4.0321±0.0001 | 4.0327 | 4.0327 | 4.0321 |
| 273 | 4.0475±0.0002 | 4.0475 | 4.0474 | 4.0470 |
| 298 | 4.0497±0.0006 | 4.0498 | 4.0498 | 4.0493 |
| 400 | 4.059±0.001 | 4.0598 | 4.0598 | 4.0594 |
| 923 | 4.1237±0.0006 | 4.1226 | 4.1252 | 4.1237 |
| | $\alpha$(T)x$10^6$ Expt. | $\alpha$(T)x$10^6$ WRP | $\alpha$(T)x$10^6$ WRR | $\alpha$(T)x$10^6$ (Eq. 2) |
| 0 | 0.000 | ----- | ----- | 0.000 |
| 50 | 3.7±0.4 | 3.969 | 3.962 | 3.795 |
| 273 | 22.7±0.4 | 22.53 | 22.63 | 22.80 |
| 298 | 23.2±0.2 | 23.19 | 23.20 | 23.40 |
| 400 | 25.1±0.2 | 25.23 | 25.16 | 25.17 |
| 923 | 36±1 | 33.93 | 39.80 | 36.10 |

From the comparisons in table 2, there is little surprise that the present functions for $a$(T) and $\alpha$(T) agree with the experiments as they were fitted to them.  As discussed previously, the agreement of the present fits with the models of Wang and Reeber (74, 75) is best in the middle temperature range and not as good at low and high temperatures.  The perfect crystal model of Wang and Reeber (WRP in the table) ignores the vacancies that are present in significant concentrations at temperatures approaching the melting point, whilst their real crystal model (WRR) does not.  The present functions for $a$(T) and $\alpha$(T) are flanked by WRP and WRR at high temperatures in both cases because WRP gives a theoretical lower bound for both $a$(T) and $\alpha$(T) whilst WRR gives an upper bound and, in principle, the present functions (equations 1 and 2) should give the average of the two models.  At any rate, the functions for $a$(T) and $\alpha$(T) presented here in equations 1 and 2 respectively, together with the listed optimum fit parameters, are intended to provide a useful resource to researchers requiring information about the lattice parameter or linear thermal expansion coefficient of aluminium at any temperature where aluminium is a solid.

## The Debye-Waller Factor of Elemental Aluminium

In the pursuit of a fundamental understanding of materials properties, all routes lead back to the bonding electron density that is the dominant determinant of almost all materials properties (except radioactivity). Aluminium is no exception. Whilst the bonding electron density of aluminium is the subject of the next section, its determination depends on three fundamental pieces of *á priori* information. As can be gleaned from the layout of this chapter, the first is the type of unit cell and the location of the atoms within it, whilst the second is the dimension of the unit cell (or the lattice parameter). The third is the thermal vibration amplitude of the aluminium atoms. This is quantified by the Debye-Waller factor (DWF) (124, 125).

Figure 7 communicates the ambiguity of whether electron distribution is due to bonding or due to thermal motion of the atoms. When viewed in connection with equations 7 and 8, the origin of this ambiguity becomes obvious. The starting point is the fact that any periodic object can be described by a Fourier sum. In the case of a crystal, this sum is:

$$\rho(\mathbf{r}) = \frac{\sum_{\mathbf{g}} F_{\mathbf{g}} e^{-2\pi i \mathbf{g} \cdot \mathbf{r}}}{V_{\text{cell}}}, \tag{7}$$

where $\rho(\mathbf{r})$ is the electron density at a position in the unit cell with real space vector **r**. The sum is over all reciprocal lattice vectors, **g**, and $F_\mathbf{g}$ are the Fourier coefficients of the crystal electron density, known as structure factors. $V_{\text{cell}}$ is the volume of the unit cell. The structure factors, $F_\mathbf{g}$, are determined according to the following equation:

$$F_{\mathbf{g}} = \sum_{j} \underbrace{f_{j}(s) e^{-2\pi i \mathbf{g} \cdot \mathbf{r}_{j}}}_{\text{static electron distribution}} \underbrace{e^{-B_{j} s^{2}}}_{\text{temp. factor}}, \tag{8}$$

where the sum is over all atoms in the unit cell and $f_j(s)$ is the atomic form factor for the j$^{\text{th}}$ atom at position $\mathbf{r}_j$ in the unit cell, and $s = (\sin\theta_B)/\lambda$ for the set of atomic planes with reciprocal lattice vector **g**, having Miller indices *hkl*. $\theta_B$ is the Bragg angle for this set of planes, corresponding to the wavelength, $\lambda$, of the radiation being diffracted from the planes with structure factor $F_\mathbf{g}$. The first component of the equation (shaded in blue) essentially quantifies the T=0K electron distribution contribution to the structure factor. The second exponential in equation 8 (shaded in red) is referred to as the temperature factor, where $B_j$ is the Debye-Waller factor of the j$^{\text{th}}$ atom at a particular temperature.

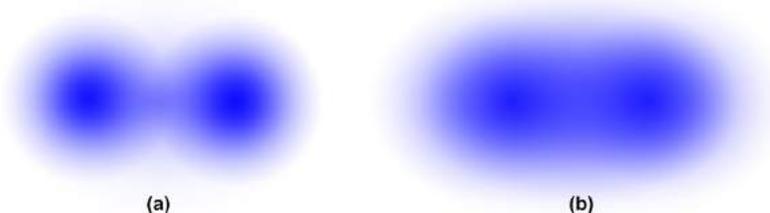

**Figure 7:** Two pairs of bonded atoms of the same elements are compared (a and b) with clearly more electron density between the atoms in the second case (b). Is the larger electron density between the atoms in case b due to stronger bonding or larger thermal motion than in case a? Without specific knowledge of the Debye-Waller factors for each case, this cannot be answered. See equation 8.

The temperature factor "smears" the electron distribution from its static form to give its time-averaged dynamic form due to the thermal vibration of the constituent atoms. Therefore, if the Debye-Waller factor is not accurately known, then one has no way of knowing the relative contributions of the static and dynamic components to the structure factor, and therefore, it is impossible to accurately determine the true bonding electron distribution. This is illustrated by figure 7.

In theory, the Debye-Waller factor deals only with the thermal vibrations of atoms. In practice, crystal defects can serve to "smear" measurements of structure factors in a similar way that the thermal vibrations do. As a result, the International Union of Crystallography (IUCr) has introduced the term "atomic displacement parameters" (ADPs), which acknowledges that static as well as dynamic (thermal) displacements of atoms from their modelled sites result in a reduction of the scattering power of planes of atoms in a particular direction, quantified by the associated structure factor.

Given that many crystalline materials can be annealed to remove most displacive crystal defects (especially true for metals), the remaining obstacle to the accurate experimental determination of bonding electron densities is accurate knowledge of the Debye-Waller factors of the constituent atoms at the temperature of the experiments. In the present section, a summary of $B(T)$ (the Debye-Waller factor as a function of temperature) is presented for elemental aluminium, as determined from previous literature (126 – 155). The aim, again, is to provide a resource for conveniently obtaining $B(T)$ for aluminium over the range of temperatures where aluminium is a solid.

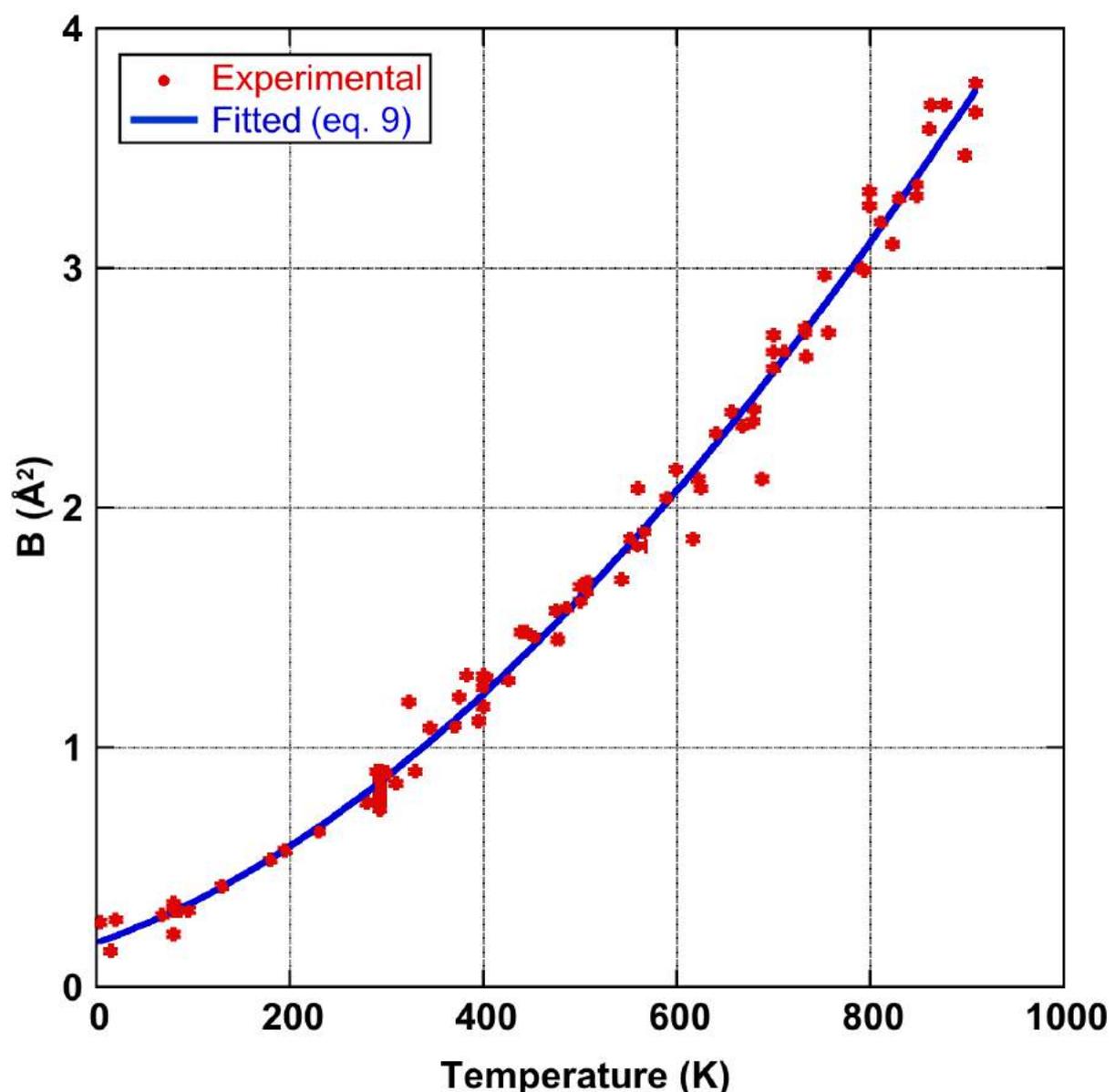

**Figure 8:** A graphical summary of published experimental measurements of the Debye-Waller factor for aluminium spanning the range of temperatures where aluminium is a solid (126 – 155). The blue line is the best fit of equation 9 to the experimental data points (red dots with error bars), yielding the optimised parameters listed in the text. This should be used as a reference for $B(T)$ in analyses which require the Debye-Waller factor as input without measuring it in an associated experiment.

Figure 8 plots experimental measurements of the Debye-Waller factor against temperature where the data points were obtained from an extensive survey of the literature (126 – 155). For specific values, their sources and some notes associated with each source, see Appendix C. Analytical derivations of the Debye-Waller factor for the general case where anharmonic vibrations are also considered, have concluded that the temperature dependence of $B$(T) in its most general form for fcc structures, should be a cubic function of temperature (141, 147, 151, 156, 157), i.e:

$$B(\text{T}) = m + n\text{T} + p\text{T}^2 + q\text{T}^3. \qquad (9)$$

This function is fitted to the experimental data points in figure 8, giving the blue line in the graph and the optimised parameters:

$m$ = 0.18488, $n$ = 1.3926x10$^{-3}$, $p$ = 3.1939x10$^{-6}$ and $q$ = -4.6582x10$^{-10}$.

The fit of equation 9 to the experimental data with these coefficient values cannot be improved by adding higher orders as Peng et al. have done in their parametrisation of $B$(T) for many elements (158 – 160), which involve a quartic function of temperature.

Equation 9 with the optimised parameters listed above (plotted in figure 8), predicts a significant zero-point energy with $B$(0K) = 0.18±0.04 Å$^2$. Sternemann et al. (155) report $B$(15K) = 0.15 Å$^2$, which is lower than other measurements at 4K and 20K (131), but more in alignment with the low temperature powder X-ray diffraction work of Rantavuori and Tanninen (149), which was aimed at measuring very accurate structure factors in aluminium as part of a bonding electron density study. Their study gave $B$(80K) = 0.22Å$^2$. This value is substantially lower than other measurements at T=80K which have a very narrow spread about a central value of 0.33Å$^2$ (126, 131, 137 – 139).

It might well be argued that the alignment between the precision measurements of Sternemann et al. (155) and Rantavuori and Tanninen (149) is suggestive of their accuracy. On the other hand, it could equally be argued that the weight of measurements suggests $B$(80K) is nearer to 0.33Å than 0.22Å, calling into question the lower zero-point energy that Sternemann et al. (155) and Rantavuori and Tanninen (149) would suggest. It is also possible that higher concentrations of crystal imperfections may have affected the measurements of $B$(4K), $B$(20K) and $B$(80K) in many of the cases leading to elevated values. Regardless, the present fit effectively averages out these differences and as a result, has a significant level of uncertainty associated with its prediction of $B$(0K).

Theoretical work in the derivation of $B$(T) from interatomic forces, calculated from a variety of ab initio potentials and electron gas screening functions, has yielded some varied results for aluminium (148, 156, 157, 161 – 166). These are plotted, together with the parameterisation of $B$(T) by Gao and Peng (159, 160) and the fit of equation 9 to the experimental results (126 – 155), in figure 9. Only Killean's nearest neighbour central force pair interaction model (148) is a good fit to $B$(T) determined by experiments.

The other models plotted in figure 9 are derived from: (i) the Ashcroft pseudopotential with a Vashishta – Singwi screening function (AVS) (157, 161); (ii) the Ashcroft pseudopotential with a Hubbard screening function (AH) (157, 163, 164); (iii) a Harrison modified point ion pseudopotential with a Hubbard-Sham screening function (HHS) (157, 163, 164); and (iv) a Morse potential (Morse) (157, 162). Three separate models (numbered in the graph) are derived from each of these potentials and they are: (1) quasi-harmonic; (2) quasi-harmonic plus low order perturbation terms as per Maradudin and Flinn (156, 157); and (3) fully inclusive of all anharmonic contributions via a Green's function approach (157, 165, 166).

Gao and Peng's parametrisation (159, 160) was based on the phonon density of states determinations of Gilat and Nicklow (138), which relied on the measurements of Stedman and Nilsson (137). This parametrisation has been a useful reference where input of the Debye-Waller factor has been required in analyses of experimental data and where it has not been measured in situ. At T=293K, the Gao and Peng parametrisation (159, 160) gives $B(293K) = 0.82$Å whilst the present function and optimised parameters give $B(293K) = 0.86\pm0.04$Å$^2$. Gao and Peng's parametrisation is in best agreement with the present function for $B(T)$ in the range 100K – 300K. Therefore, analyses of near-room-temperature data will not be strongly affected by the choice of $B(T)$ function, whether it be that of Gao and Peng (159, 160) or the present function. However, in regions of higher or lower temperatures, Gao and Peng's parametrisation deviates significantly from the present function for $B(T)$ and therefore also from the experimental measurements of the Debye-Waller factor. The deviation is especially large at higher temperatures, where use of the present function for $B(T)$ is strongly advised.

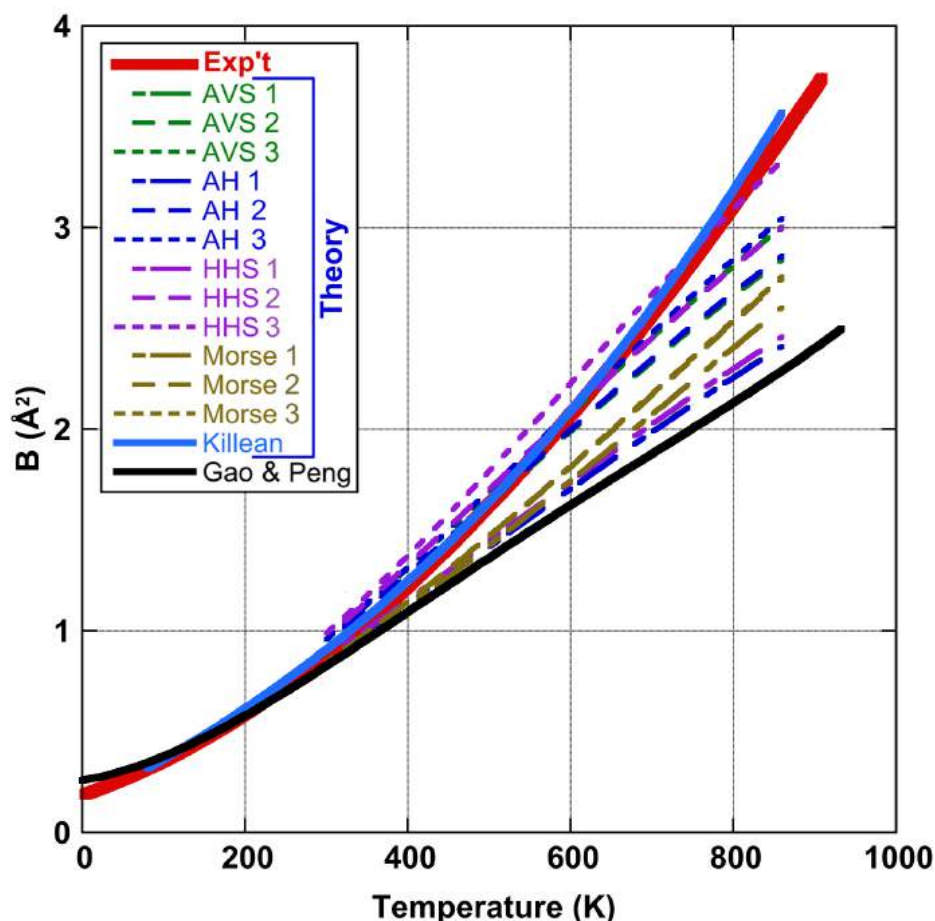

**Figure 9:** A comparison of the present function for $B(T)$ (see equation 9 with the optimised parameters listed in the text) with different theoretical models for $B(T)$ (148, 156, 157, 161 – 166) and the parametrisation by Gao and Peng (159, 160). The present function for $B(T)$ is the best fit to the experimental data surveyed in this work (126 – 155) and is the thickest solid line in the graph (red). The nearest neighbour central force pair interaction model of Killean (148) (solid bright blue line) is in closest agreement with experiments across the range of temperatures where aluminium is a solid. On the other hand, the parametrisation by Gao and Peng (159, 160) does not fit the experimental measurements well at all (solid black line). All of the other theoretical determinations of $B(T)$ are plotted with dashed lines and the abbreviations with which they are tagged, name the potentials and screening functions used to derive the corresponding $B(T)$ curves (described in the text).

Equation 9 with its optimised parameters, given above, represents the summary of all of the published experimental measurements of $B(T)$ (126 – 155) and is useful as a simple reference tool when $B(T)$ is required as input and is not measured in situ. The next section deals with the determination of the bonding electron distribution in aluminium and, as illustrated by equation 8 and figure 7, a reliable knowledge of $B(T)$ at the temperature at which structure factors are measured, is vital to the accurate determination of bonding. This is true for any material.

# The Bonding Electron Distribution in Aluminium

Aluminium has been the focus of considerable charge density research for almost a century (126, 127, 129, 130, 133, 140, 144, 149, 152, 154, 167 – 186). What makes this metal so interesting is that it closely approximates an ideal Drude metal (187) which is commonly described as a lattice of "cations immersed in a sea of delocalised electrons". This is in fact the crude description applied to metals in general at high school and undergraduate levels in order to describe metallic bonding and the general properties of malleability, ductility and thermal, electrical and acoustic conduction that it gives rise to (188). Whilst this description is a gross approximation, aluminium fits it remarkably well and better than most other metals. Aluminium is one of nature's best approximants to a "free-electron gas" or "jellium" and, as a consequence, is an excellent thermal conductor, is highly malleable, is one of the best known electrical conductors and is the most efficient reflector of visible radiation (a direct result of the oscillation of its nearly free valence electrons, i.e. plasmons).

The bonding electron distribution in materials is key to all of their properties (with the exception of radioactivity, which is entirely nuclear). Therefore, to gain a fundamental understanding of the properties of aluminium, one must closely examine the nature of metallic bonding between the atoms in it.

Experimental measurements of the electron density in aluminium were largely confined to X-ray diffraction experiments (126, 127, 129, 130, 133, 140, 144, 149, 167 – 170, 172) from powders or single crystals. Some higher precision results from electron diffraction were obtained by the critical voltage (CV) method (152, 173, 174) and the highest precision measurements to date were presented in a recent study using quantitative convergent-beam electron diffraction (QCBED) (175).

Conventional single-crystal and powder X-ray diffraction techniques variably suffer from errors caused by extinction, which originates from the single scattering approximation (or kinematic approximation) made in the analysis of the diffracted intensities (189). A number of approaches (190 – 193) have been developed and applied to correct for the multiple scattering (or dynamical diffraction) that inevitably occurs in crystals with small unit cells and relatively high degrees of crystal perfection, however, these approaches all involve significantly limiting approximations. Experiments seeking dynamical diffraction data and applying a full dynamical scattering analysis eliminate the concept of extinction and in turn, should result in more accurate measurements. Pendellösung experiments with X-rays and single crystals have been attempted with aluminium and are included in this review (172). The problem with this method is that it is difficult to obtain perfect single crystals of the sizes needed for such X-ray experiments, especially when it comes to metals because metallic bonding supports crystal defects very readily, which is associated with the property of ductility and malleability – the defining characteristics of metals. Crystal imperfections cannot be avoided in the volumes of metal needed to perform these dynamical X-ray diffraction experiments and this in turn, leads to error in the measurement of structure factors by these techniques.

Whilst X-rays are scattered by the total electron density in a crystal, electrons, being charged, are scattered by the crystal potential. Potential and electron density are related (via a simple electrostatic relationship called the Mott formula (194)) in such a way that makes electron diffraction more sensitive to bonding than the rest of the electron density. In addition, electrons, due to their charge, interact with matter about 1,000 times more strongly than X-rays, and can be focused into sub-nanometre probes with electromagnetic lenses. This combination means that convergent-beam electron diffraction (CBED) is able to probe volumes of material with $<10^5$ atoms. This is about $10^{10}$ times smaller than is possible with conventional X- ray diffraction techniques, thereby allowing defect-free regions of crystal to be probed selectively by CBED in electron microscopes.

CBED gave rise to the critical voltage method for measuring bonding-sensitive structure factors and several studies using this technique were carried out for aluminium (152, 173, 174). The problem with this approach is that the range of structure factors that can be measured is extremely limited. The smaller the structure factor magnitude, the higher the electron energy required to reach the point at which there is a change in contrast in the CBED pattern that acts as the indicator in the method. The electron energy at which this occurs is used to determine the magnitude of the relevant structure factor(s). The electron energy is a direct product of the accelerating voltage in an electron microscope and higher accelerating voltages, and thus higher electron energies, require larger and larger electron guns to accelerate the electrons to the required energy. This means that the critical voltage method has a limited range of applicability because it is impractical to make electron guns huge enough to measure more than just the strongest 2 or 3 structure factors in a material. In aluminium, the practical limit is just the two strongest structure factors, $F_{111}$ and $F_{200}$. This was insufficient to unequivocally determine the bonding electron density in aluminium as it was long thought that bonding information also resides in the next structure factor, $F_{220}$, that is inaccessible by the critical voltage method.

Quantitative convergent-beam electron diffraction (QCBED) has emerged in the last 2 decades as a very accurate technique for measuring bonding-sensitive structure factors (175, 195 – 208). It involves calculating convergent-beam electron diffraction (CBED) patterns and fitting them iteratively to experimental ones by adjusting the parameters to which the patterns are sensitive (these include the bonding-sensitive structure factors). The precision AND accuracy come from the fact that a full dynamical scattering calculation of intensity as a function of scattering angle is being fitted to an experimental intensity distribution, i.e. a CBED pattern, as opposed to the integrated intensities of reflections in point diffraction patterns, as is the case in X-ray diffraction. QCBED results in massive over-determination of the refined structure factors as of order 10 parameters are outnumbered by $\sim 10^4$ data points in the matched intensity distribution of a CBED pattern. QCBED is more computationally intensive than X-ray diffraction analysis, however, with fast computers and highly linear electron-sensitive area detectors on modern electron microscopes, QCBED is emerging as the technique of choice when it comes to requiring very high precision and accuracy in bonding–sensitive structure factor measurements.

In aluminium, the need for precision and accuracy is possibly even more crucial than in bonding electron density studies of other materials. This is because aluminium's maximum bonding electron density is tiny (~0.047 e$^-$Å$^{-3}$ based on the structure factors of (175)) compared with the experimental benchmark in X-ray diffraction of diamond, for example, where the maximum bonding density is more than an order of magnitude greater (~0.66 e$^-$Å$^{-3}$ using the structure factors of (209)). In other words, to measure bonding in aluminium, the experiments have to be at least an order of magnitude more sensitive than in a material like diamond. This is where QCBED comes into its own as an experimental technique. It is QCBED that gave rise to the experimental measurements presented in (175), which can be considered the modern benchmark for bonding in aluminium, due to the precision and accuracy obtained from the technique.

The same requirements of precision and accuracy can be imposed on theoretical, ab initio, modelling and calculation of the bonding electron density in aluminium. Since the advent of density functional theory (DFT) (210), a significant number of publications on the theoretical calculation of the bonding electron distribution in aluminium have appeared using different approximations within the framework of DFT (175 – 186). The results are varied and depend on the approximations made. The two historical approaches that are closest to the experimental benchmark set by QCBED in (175), are the augmented plane wave (APW) calculations of Perrot (180) and the "atom in jellium" model of Rantala (186). The former made fewer approximations by extending beyond the non-muffin tin constraints prevalent in contemporary calculations. It is in fact very similar in nature to the DFT calculation presented in (175), which used the full potential linearly augmented plane wave approach (FP-LAPW) and the generalised gradient approximation (GGA) with local orbital (lo) and local screening (ls) pseudopotentials. The proximity of the benchmark QCBED measurements in (175) to the "atom in jellium" model of Rantala (186) is in itself a testament to aluminium being an excellent approximation to the Drude model of ideal metals being a lattice of cations in a sea of delocalised electrons.

The importance of precision and accuracy in bonding studies is highlighted by the following review of all bonding electron density studies, both experimental and theoretical, published to date for aluminium. In order to present all of these historical results in a coherent fashion, key locations within the crystal structure of aluminium are considered. Figure 10 presents the fcc unit cell of aluminium and revisits the tetrahedral and octahedral interstices discussed in the first section of this chapter. These are marked in the figure together with the bridge centre, which is defined as the midpoint between nearest neighbour atoms. Each of the positions is marked with a cross in the unit cell and the dotted lines represent the coordination of each of these positions with their nearest atoms. The bonding electron density at each of these positions can be calculated for any set of published structure factors, thus allowing each published study to be graphed as a point in a 3-dimensional plot with axes corresponding to the bonding electron densities at the tetrahedral, octahedral and bridge centres.

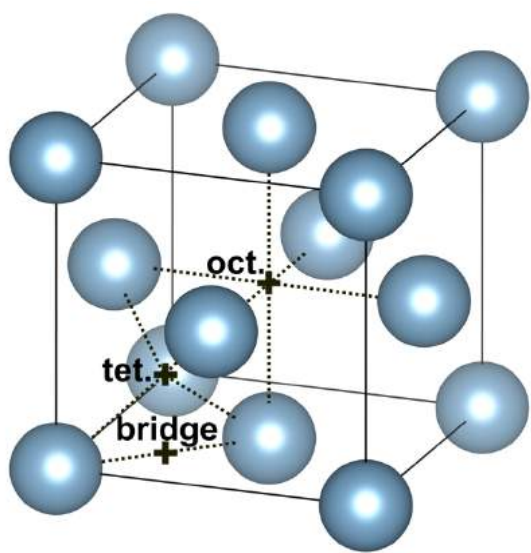

**Figure 10:** A diagram of the fcc unit cell of aluminium showing the key bonding locations and their coordination to nearest neighbour atoms. The tetrahedral and octahedral interstitial positions (marked "tet." and "oct." respectively), were previously identified in figure 1 (c). A third position of importance in bonding studies is the bridge centre (as marked) which is at the midpoint between nearest neighbour atoms. As shown by the dotted lines, the bridge centre has a coordination number of 2 atoms, whilst the tetrahedral and octahedral centres have coordination numbers of 4 and 6 atoms respectively. It is interesting to note that of the three positions, only the tetrahedral centre does not lie on a line between any of its coordinated atoms. The tetrahedral, octahedral and bridge centres have coordinates of 0.25 0.25 0.25, 0.5 0.5 0.5 and 0.25 0.25 0.00 respectively. This figure was drawn with the aid of VESTA (1).

Here, as in (175), the bonding electron density is calculated for each of the published sets of structure factors by subtracting the independent atom model (IAM) structure factors based on Doyle's and Turner's landmark relativistic Hartree-Fock calculations of electron density for non-interacting, isolated neutral atoms (unbonded) (211). This proceeds according to the following equations, which follow on from equation 7:

$$\Delta\rho(\mathbf{r}) = \rho(\mathbf{r})^{\text{actual}} - \rho(\mathbf{r})^{\text{IAM}} = \frac{\sum_g F_g^{\text{actual}} e^{-2\pi i \mathbf{g} \cdot \mathbf{r}}}{V_{\text{cell}}} - \frac{\sum_g F_g^{\text{IAM}} e^{-2\pi i \mathbf{g} \cdot \mathbf{r}}}{V_{\text{cell}}}, \quad (10)$$

$$\therefore \quad \Delta\rho(\mathbf{r}) = \frac{\sum_g (F_g^{\text{actual}} - F_g^{\text{IAM}}) e^{-2\pi i \mathbf{g} \cdot \mathbf{r}}}{V_{\text{cell}}}. \quad (11)$$

Here, $\Delta\rho(\mathbf{r})$ is generally referred to as the deformation electron density. It is a measure of the deviation of the measured or calculated total electron density, $\rho(\mathbf{r})^{actual}$, from the total electron density given by the IAM, $\rho(\mathbf{r})^{IAM}$. As equation 11 shows, it is simply the Fourier sum of the differences between the structure factors measured in an experiment or calculated by a theory that models a bonded crystal, $F_{\mathbf{g}}^{actual}$, and those calculated for a procrystal of unbonded atoms, $F_{\mathbf{g}}^{IAM}$. Division is always by the cell volume, $V_{cell}$, which results in $\Delta\rho(\mathbf{r})$ having units of $e^-\text{Å}^{-3}$. Note that if $\Delta\rho(\mathbf{r})$ is positive, then this is known as the bonding electron density, whilst if it is negative, it is known as the anti-bonding electron density.

Equation 11 and the IAM structure factors for aluminium from Doyle and Turner (211), have been applied to all published sets of structure factors (126, 127, 129, 130, 133, 140, 144, 149, 152, 154, 167 – 186) for aluminium, which nominally constitute $F_{\mathbf{g}}^{actual}$, details of which, are given in Appendix D. The definitions, $\Delta\rho_{tet} = \Delta\rho(0.25, 0.25, 0.25)$ (the tetrahedral centre in figure 10), $\Delta\rho_{oct} = \Delta\rho(0.5, 0.5, 0.5)$ (the octahedral centre in figure 10) and $\Delta\rho_{bridge} = \Delta\rho(0.25, 0.25, 0)$ (the bridge centre in figure 10), establish the three axes in figure 11.

Figure 11 gives a 3-dimensional plot of $\Delta\rho_{tet}$, $\Delta\rho_{oct}$ and $\Delta\rho_{bridge}$ with each published set of structure factors determined experimentally, prior to (175), constituting a red cube, each published set of theoretically calculated structure factors (prior to (175)) constituting a blue sphere and the most recent experimental and theoretical work of (175) constituting the green cube and purple sphere respectively. The 3-dimensional plot of all of these points occupies the centre of the figure and the left, right and bottom sides show projections of this plot onto separate 2-dimensional plots perpendicular to the $\Delta\rho_{tet}$, $\Delta\rho_{oct}$ and $\Delta\rho_{bridge}$ axes respectively. The error bars, shown only in the 2-dimensional plots in order to reduce crowding in the 3-dimensional plot, are very large for all experimental determinations prior to the work of (175).

The QCBED determination of (175) has much smaller error bars than previous experiments and defines a much narrower range of uncertainty in the distribution of the deformation electron density. The green point, which represents these latest measurements by QCBED, is very close to and encloses within its margins of error, the DFT calculation of (175), shown here as the purple point. The calculation used the full potential linearly augmented plane wave approach (FP-LAPW) and the generalised gradient approximation (GGA) with local orbital (lo) and local screening (ls) pseudopotentials.

The DFT calculation of (175) (green point in the plots of figure 11) is not far from the "atom in jellium" model of Rantala (186) and the APW calculation of Perrot (180), discussed previously. Agreement with the model of Perrot (180) (marked by the black "P" in figure 11) is understandable because both calculations are very similar in nature as previously explained. Agreement with the QCBED experiment (175) and the "atom in jellium" model of Rantala (186) (marked by the black "R" in figure 11) suggests that the theoretical treatment of aluminium as very closely approximating an ideal Drude metal is remarkably close to the truth as well as being well-modelled by the theoretical approach taken in (175). The points in the graph resulting from Perrot's (180) and Rantala's (186) calculations are also within the range of uncertainty of the QCBED measurements.

The only experimental point that lies within the range of uncertainty of the QCBED measurements of (175) comes from the X-ray diffraction study of Inkinen et al. (144) (marked with a black "I" in figure 11. They used powder samples that were pressed into slabs with pressures just below 50MPa. These pressures were found low enough to cause no orientational texture within the pressed samples yet were sufficiently high to eliminate significant effects in the integrated diffracted intensities caused by surface roughness or specimen porosity. The advantage of powders with small grain sizes in X-ray diffraction is that extinction effects caused by multiple scattering are minimised by a short path length through any given grain. This is the likely reason for the agreement between this X-ray study (144) and the QCBED study of (175), within the error associated with the latter study.

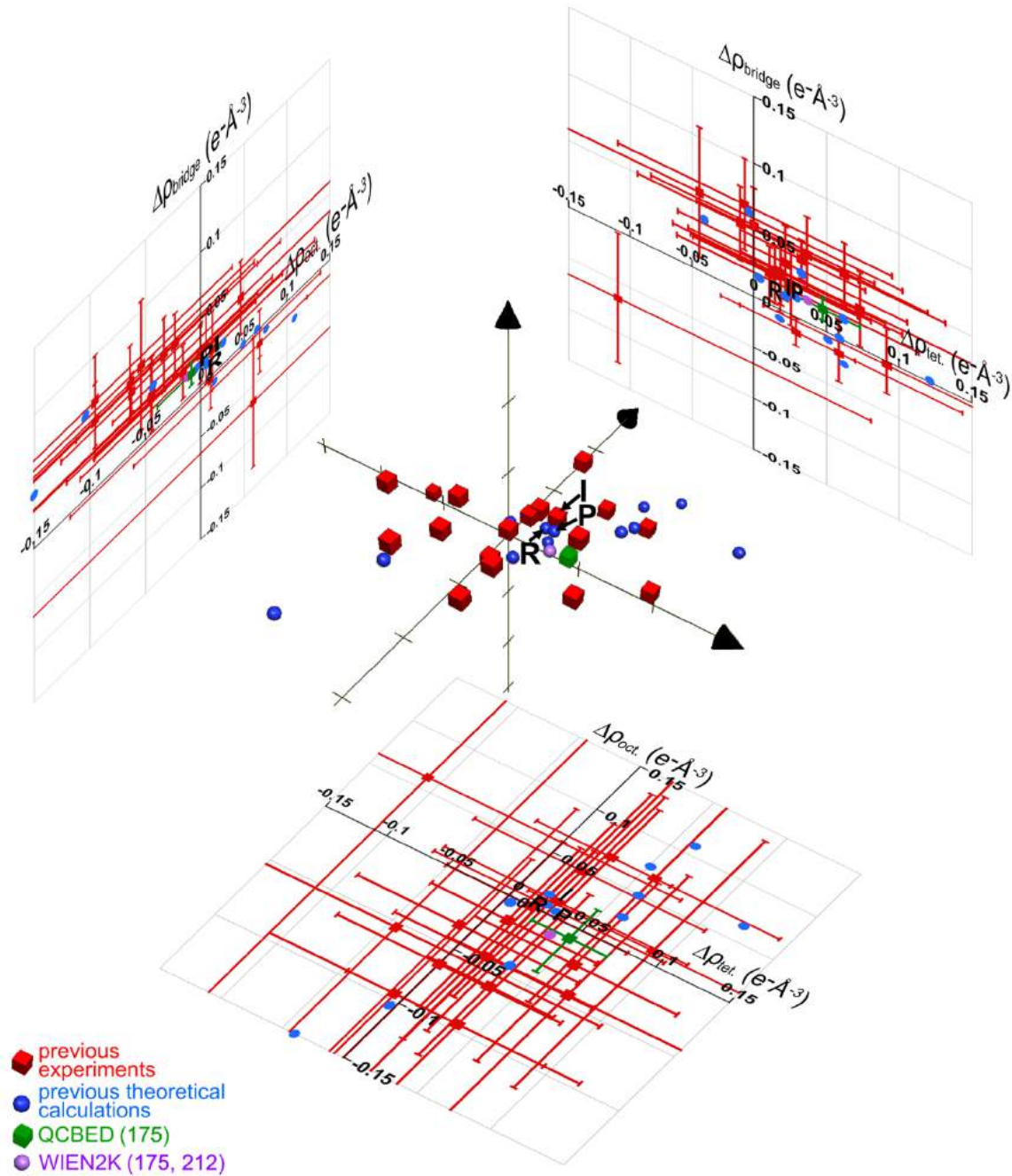

**Figure 11:** Plots of all experimental (126, 127, 129, 130, 133, 140, 144, 149, 152, 154, 167 – 175) and theoretical (175 – 186) determinations of deformation electron density, $\Delta\rho$, in aluminium to date. Note, the deformation electron density is computed by subtracting the independent atom model (IAM) structure factors according to Doyle and Turner (211) from each set of published structure factors. This allows the computation of the Fourier sum that gives the deformation of the electron density from neutral, unbonded atoms. The 3-dimensional axes (labelled only in their 2 dimensional projections in order to reduce clutter) are the deformation electron densities at the tetrahedral ($\Delta\rho_{tet.}$), octahedral ($\Delta\rho_{oct.}$) and bridge ($\Delta\rho_{bridge}$) centres in the x, y and z axes of the plot respectively. The red points show the distribution of all experimentally determined $\Delta\rho_{tet.}$, $\Delta\rho_{oct.}$ and $\Delta\rho_{bridge}$, prior to the most recent experiments of Nakashima et al. using QCBED (175) (green point). The blue points in the plot comprise all of the theoretical determinations prior to the latest calculations of (175) in which, the FP-LAPW approach and GGA (+lo +ls) were applied using the WIEN2K package (212) (purple point). To give a better impression of the spread of all of the determinations, the 3-dimensional plot is projected along each of its axes to form the 2-dimensional plots shown. Error bars are intentionally omitted from the experimental points in the 3-dimensional plot in order to minimise the obscuration of points by error bars from nearby points. It is noteworthy that the errors associated with the latest QCBED measurements of bonding in aluminium are much smaller than previous experiments. The points marked "P", "R" and "I" refer to the separate theoretical calculations of Perrot (180) and Rantala (186) and the experimental measurements of Inkinen et al. (144) respectively. These are the only points from structure factor determinations prior to the work of (175) that fall within the bounds of error associated with the benchmark QCBED measurement (green point) presented in (175).

Evident from figure 11 is the large spread in the experimental and theoretical determinations of the deformation electron density in aluminium. This is what necessitated a more accurate and precise study, furnished by QCBED (175), to resolve the ambiguities of all the preceding bonding studies in this nearly free electron gas. The mean and uncertainty of all experimental and theoretical structure factor determinations in aluminium prior to the work of (175) are now considered and compared with the QCBED measurements and the DFT calculation presented in (175), in table 3.

Table 3: Summary of the four lowest order (lowest $(\sin\theta)/\lambda$) structure factors of aluminium as determined by all experimental and theoretical work prior to (175). The values and uncertainties reflect the mean values of all published structure factors and their standard deviation from their respective means. A comparison is made with the experimental and theoretical results from (175), which is taken as the most accurate reference for bonding in aluminium to date. The final column in the table shows the independent atom model (IAM) values as determined by the relativistic Hartree-Fock calculations of Doyle and Turner (211), which, at present, is taken as the standard reference for electron distribution around neutral, independent (unbonded and non-interacting) atoms.

| h k l | $F_{hkl}$ (e$^-$/atom) experiments prior to (175) | $F_{hkl}$ (e$^-$/atom) theory prior to (175) | $F_{hkl}$ (e$^-$/atom) QCBED (175) | $F_{hkl}$ (e$^-$/atom) WIEN2K GGA/FP-LAPW +lo +ls (175, 212) | $F_{hkl}$ (e$^-$/atom) IAM (211) |
|---|---|---|---|---|---|
| 1 1 1 | 8.8±0.2 | 8.86±0.07 | 8.87±0.01 | 8.87 | 8.95 |
| 0 0 2 | 8.4±0.2 | 8.40±0.08 | 8.37±0.01 | 8.38 | 8.50 |
| 0 2 2 | 7.3±0.1 | 7.31±0.07 | 7.31±0.03 | 7.30 | 7.31 |
| 1 1 3 | 6.6±0.1 | 6.65±0.05 | 6.64±0.06 | 6.64 | 6.65 |

Table 3 summarises the four lowest order (i.e. lowest scattering angle) structure factors of aluminium for all experiments prior to (175), all theoretical calculations prior to (175), the QCBED measurement of (175), and the DFT calculation of (175). The table gives the mean value of each structure factor and the associated uncertainty, which is determined from the spread in the determinations plus the error bars associated with each determination in the case of the experimental studies. The DFT calculation of (175) used the well-known WIEN2K software package developed by Blaha et al. (212). The final column of the table presents the IAM calculated structure factors. Differences between structure factors listed in the other columns and those of the IAM are associated with the deformation of the electron density from that of spherical non-interacting neutral atoms to the bonded real structure determined from the averages of the different approaches. It can immediately be seen in table 3, that the uncertainties associated with the experimental structure factors (column 2) are much larger than the differences between these structure factors and those of the IAM (column 6). It can therefore be said that the averages of the experimental structure factors measured prior to (175) are unable to determine bonding in aluminium with any certainty whatsoever.

Considering the averages of the structure factors from all theoretical determinations prior to (175), the uncertainties due to the spread of these determinations is somewhat smaller than the experimental uncertainties in column 2 of table 3. The uncertainties are in fact smaller than the differences between the averages of the theoretically calculated structure factors and the corresponding IAM values. However, the spread in these determinations is still large enough to leave significant doubt as to where the truth lies.

The QCBED measurements of (175) have uncertainties associated with each structure factor that are an order of magnitude smaller than the preceding results in the table. This increases the confidence in these results being able to say something definitive about the distribution of bonding electron density within aluminium. Furthermore, the agreement with the most recent DFT calculation, using the most up-to-date formalisms, approximations and software (WIEN2K (212)), presented in (175), is well within the margins of error of the QCBED measurements.

In figures 12 to 15 inclusive, the average structure factors and their uncertainties in each of the columns in table 3 are explored in greater detail with respect to the bonding electron distribution determined from these sets of structure factors. The first of these figures, figure 12, plots the deformation electron density determined from each set of structure factors in table 3 by first subtracting the IAM values from them and applying equation 11. These determinations constitute the solid points plotted in the three-dimensional plot whose axes are the same as those in figure 11, namely $\Delta\rho_{tet}$, $\Delta\rho_{oct}$ and $\Delta\rho_{bridge}$. In addition to these central points, the uncertainties associated with the corresponding mean structure factors in table 3 are used to calculate the range of possible $\Delta\rho_{tet}$, $\Delta\rho_{oct}$ and $\Delta\rho_{bridge}$ values associated with each set of structure factors. These ranges are shaded: red for all experimental determinations prior to the work of (175), blue for all theoretical calculations prior to (175) and green for the QCBED measurements of (175) and their associated uncertainties. The focus in figure 12 is on the experiments prior to (175) and the spread (red region) is significantly larger than the spread associated with the historical theoretical determinations (blue region), and much larger than the spread in the QCBED determination of $\Delta\rho(\mathbf{r})$ (green region).

Specific plots of the bonding electron density (positive $\Delta\rho(\mathbf{r})$) in the fcc cell of aluminium are presented in figure 12 for specific positions in the range of $\Delta\rho_{tet}$, $\Delta\rho_{oct}$ and $\Delta\rho_{bridge}$ spanned by the historical experimental measurements prior to (175). Position A is at the point of minimum $\Delta\rho_{tet}$ in the region, whilst position B is at the point of maximum $\Delta\rho_{bridge}$ in the region. Position C is at the point of minimum $\Delta\rho_{oct}$, D is at the point of minimum $\Delta\rho_{bridge}$ and maximum $\Delta\rho_{oct}$ and E is at the point of maximum $\Delta\rho_{tet}$. The bonding electron density plot labelled F is from the point at the centre of the region of uncertainty, marked by the red cube, and corresponds to the mean structure factors in column two of table 3.

All bonding electron density iso-surface plots in the present review are drawn with an iso-surface level at 50% of the maximum $\Delta\rho(\mathbf{r})$ in a cell for the set of structure factors being used.

In the present case, considering the historical experimental determinations of $\Delta\rho(\mathbf{r})$ (prior to (175)), the variation in types of bonding presented in each cell from each position in the region of uncertainty, is large. In cell A, the iso-surface at 50% of the maximum $\Delta\rho(\mathbf{r})$ in the cell encloses bonding volumes centred within the octahedral interstices with holes at the centres. Cell B shows strong transverse bridge bonding (i.e. where the bridge bonds are elongated perpendicularly to the line between the bridged atoms) and holes at the tetrahedral centres. Cell C shows very strong linear bridge bonding (i.e. the bridge bonds are elongated along the line between the bridged atoms) and cell D shows highly concentrated octahedrally-centred bonding. Cell E shows strongly concentrated tetrahedrally-centred bonds and F shows a hole at the octahedral centres with elevated bonding density at the tetrahedral centres and in transverse bridge bonds.

Figure 13 has the experimental spread stripped away to reveal the spread in theoretical determinations prior to (175) more vividly. The form of the figure is the same as figure 12. The blue shaded region is more constricted in comparison to the red region which embodied the historical experimental measurements prior to (175). This is to be expected as the uncertainties in table 3 are much smaller for the theoretical determinations than for the historical experimental determinations. Cells A to F show the bonding electron density iso-surface (at 50% of the maximum $\Delta\rho(\mathbf{r})$ in each cell) for the same points in the blue region of uncertainty as points A to F in figure 12 for the red region. Cell A shows a bonding network with holes at the tetrahedral and octahedral centres, with the main concentration occurring in the bridges. Cell B shows strong transverse bridge bonding with holes at the octahedral centres, whilst C shows more linear bridge bonding and very significant holes at the octahedral centres. Cell D, as in the case of the experimental spread, shows strong octahedrally-centred bonds and E shows small octahedrally-centered bonding density concentrations and much stronger concentrations at the tetrahedral centres. Cell F, showing the average of all of the theoretical determinations prior to (175), is indicative of tetrahedrally-centred bonding.

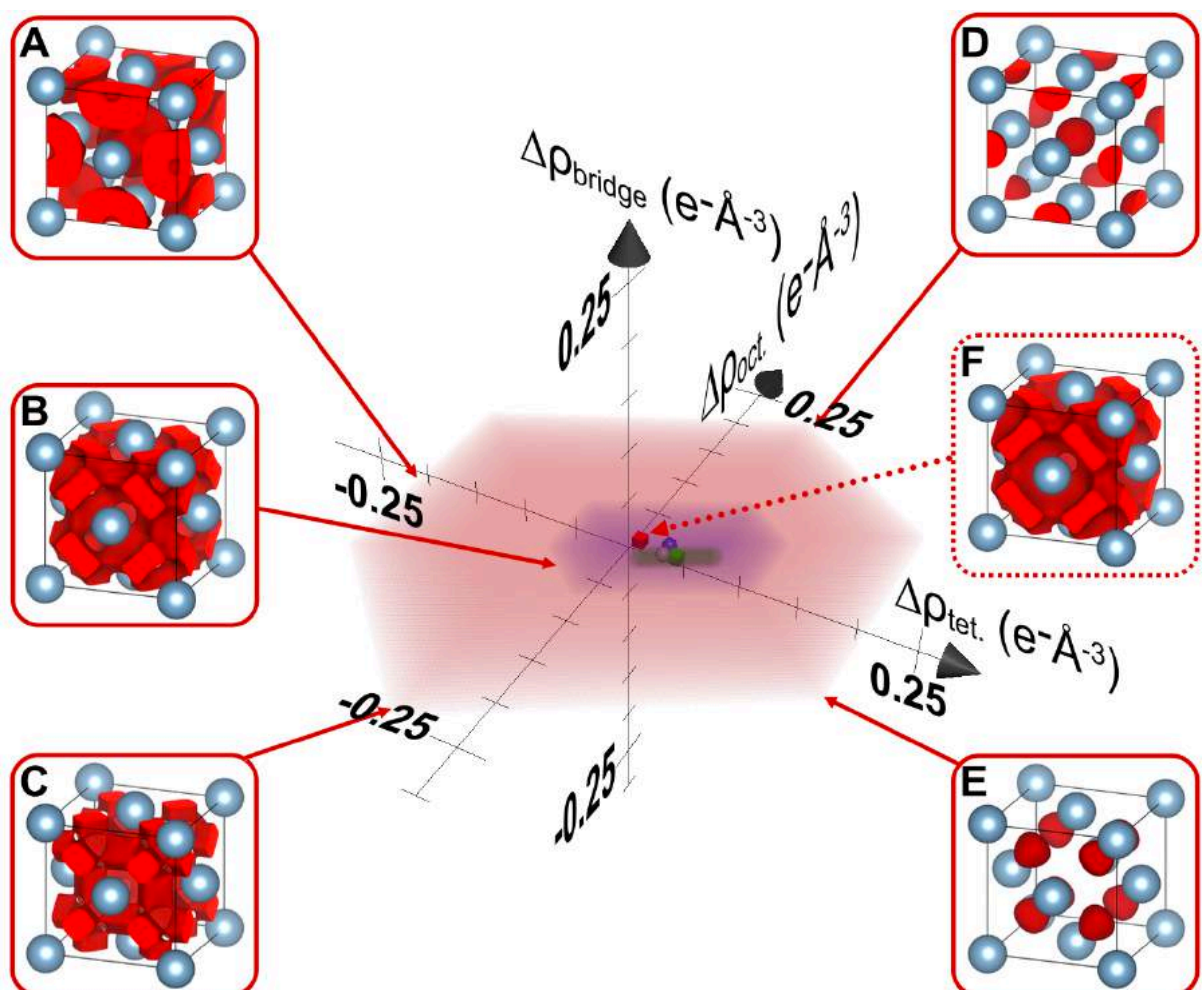

Figure 12: Plots of the spread in deformation electron density predicted by the mean and margin of error of the all structure factors in table 3. The red shaded region in the three-dimensional plot shows the extent of all possible outcomes from the mean structure factors and their uncertainties determined from all experimental studies prior to that of (175), listed in column 2 of table 3. The actual distributions of the bonding electron density at various points in the region are shown for single unit cells with arrows pointing to the corresponding region of the plot. The locations have the following significance in bounding the region: (A) the point of minimum $\Delta\rho_{tet.}$, (B) the point of maximum $\Delta\rho_{bridge}$, (C) the point of minimum $\Delta\rho_{oct.}$, (D) the point of minimum $\Delta\rho_{bridge}$, which is also the point of maximum $\Delta\rho_{oct.}$, (E) the point of maximum $\Delta\rho_{tet.}$, and (F) the point at the centre of the region. The iso-surface plots are at a level of 50% of the maximum bonding density in each cell. Also shown in the graph are the regions of deformation electron density spanned by the mean structure factors and their uncertainties determined from all theoretical calculations preceding (175) (column 3 of table 3) – shaded in blue, and from the QCBED measurements of (175) (column 4 of table 3) – shaded in green. The points corresponding to the mean deformation density distribution for each of these sets of structure factors are shown with red, blue and green points respectively. In this figure, the relevant point, for which the mean bonding density is shown in the dotted cell (F), is the red cube at the centre of the red shaded region. An additional point – a purple sphere, shows the location in the plot of the WIEN2K (212) calculated results from (175) (column 5 of table 3). All deformation densities plotted were determined by subtraction of the IAM (211) structure factors (listed in column 6 of table 3) from each of the other sets of structure factors listed in table 3. The cells in this figure were plotted with VESTA (1).

So far, in view of the different bonding distributions possible within the spread of experimental and theoretical studies prior to (175), a definitive conclusion is difficult to draw. This is where an experimental technique with very high precision and accuracy, such as QCBED, can make a definitive assessment.

Figure 14 strips away the blue shaded region of theoretical uncertainty examined in figure 13, to reveal the highly constricted green region that represents the uncertainty associated with the QCBED measurements of (175). Again, cells A to F correspond to the same significant points in the green (QCBED) region of uncertainty as in previous graphs for the red (previous experiments) and blue (previous theoretical determinations) uncertainties. In all cases, the bonding is dominantly tetrahedrally-centred, with no indication of any octahedrally-centred bonding at all. Cells A to C show some significant concentrations of bonding electron density at the bridge centres and in every case, the form of the bonding is transverse in the bridges. Cells D to F all show nearly identical bonding concentrations, which are exclusively tetrahedrally-centred. It is important to note that morphologically, tetrahedrally-centred bonds that have a larger spatial extent will lead to significant bonding electron density at the bridge centres. Furthermore, what is evident from the consistency of the iso-surfaces near the bridge centres in cells A to C, is that the shapes of the associated tetrahedrally-centred bonds are all the same in cells A to C. Close inspection of the iso-surfaces in cells D to F reveals that if the surfaces were expanded about the tetrahedral centres, the intersection with the cell faces at the bridge positions would result in transverse forms similar to those observed in cells A to C.

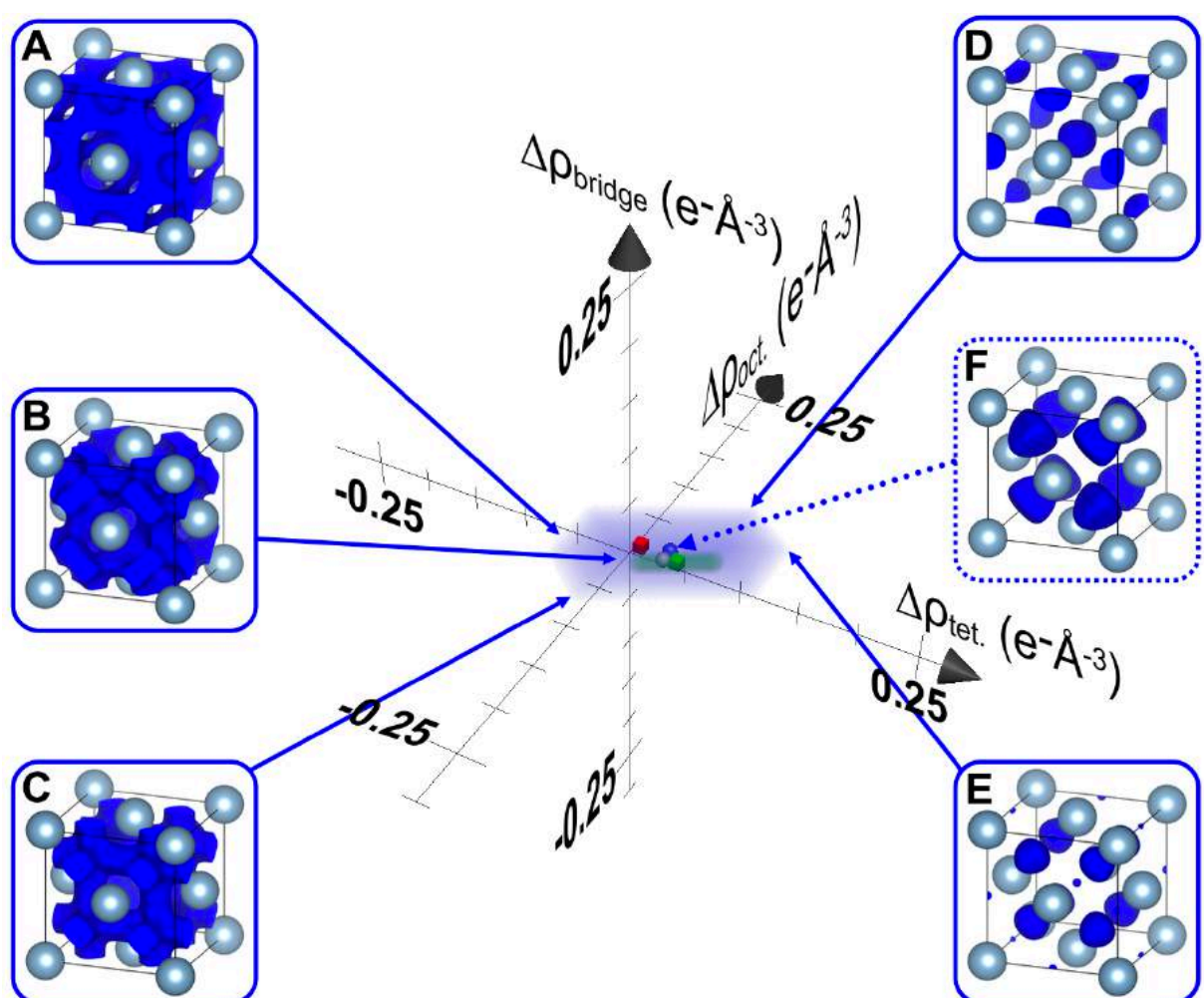

Figure 13: As per figure 12, except that the red region and corresponding bonding density plots are hidden, with the focus here being on all theoretical calculations preceding (175) (column 3 of table 3) – shaded in blue. Again, the letter of each cell corresponds to the same point for the blue region as described for the red region in figure 12, and the dotted cell (F) indicates the mean bonding electron density determined from all theoretical calculations preceding (175), which is the centre of the blue region indicated by the blue sphere. All iso-surfaces are drawn at a level of 50% of the maximum bonding density in a cell for each region. The cells in this figure were plotted with VESTA (1).

The very high precision of the QCBED measurements of (175) affords the first definitive experimental conclusion that bonding in aluminium is almost purely tetrahedral in nature. As discussed in (175), even relatively recent theoretical studies by Kioussis et al. (213) and Ogata et al. (214), published almost simultaneously but without structure factors, are at odds about the bonding in aluminium. The former study (213) was correct in asserting that the bonds are tetrahedrally-centred, whilst the latter study

(214) asserted that they are octahedrally-centred. This means that the mechanical properties of aluminium derived in (214) from that assessment of bonding was based on an incorrect bonding electron distribution.

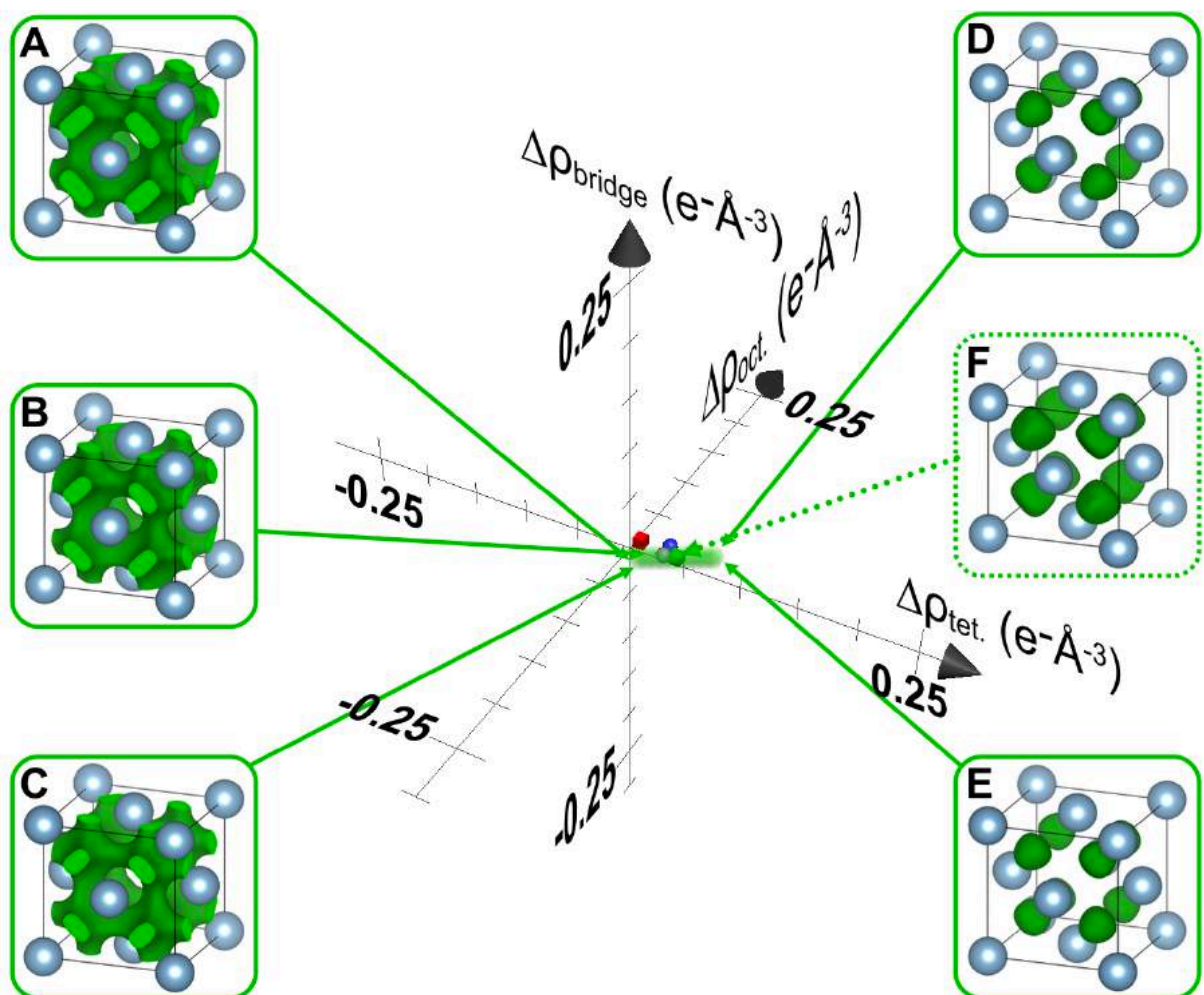

**Figure 14:** As per figure 12, except here the focus is on the most recent experimental measurement of bonding in the literature (175), coloured in green and corresponding to column 4 in table 3. Again, each cell corresponds to the same point for the green region, examined here, as described for the red and blue regions of the preceding figures (figures 12 and 13). The dotted cell (F) indicates the mean bonding electron density determined by the QCBED study of (175), which is at the centre of the green region indicated by the green cube. All iso-surfaces are drawn at a level of 50% of the maximum bonding density in each cell. The cells in this figure were plotted with VESTA (1).

Finally, it remains to make one final comparison based on the summary of structure factors in table 3. This is done in figure 15, where the mean values in columns 2 to 5 of the table are plotted in terms of $\Delta\rho_{tet}$, $\Delta\rho_{oct}$ and $\Delta\rho_{bridge}$. It is worth noting in this comparison, that all of the points (representative of the centres of the spreads illustrated in figures 12 to 14) in the graph, are very close together. It is therefore not surprising that they yield very similar plots of the bonding electron density iso-surface. In all cases, the morphologies indicate tetrahedrally-centred bonding. All cases with the exception of cell A, show iso-surfaces (at 50% of the maximum bonding density) that fully enclose only the tetrahedral centre. In the case of cell A, which represents the mean of all experimental studies prior to (175), the iso-surface encloses a much greater expanse, centred on the tetrahedral interstices and including the bridge centres. The iso-surface intersects the cell walls in a manner very similar to the extremes of the QCBED measurements shown in cells A to C in figure 14, namely in a form that is transverse to the bridge between nearest neighbour atoms. Cell D in figure 15 shows the WIEN2K (212) DFT calculation of the bonding electron density, carried out in (175) using the full potential linearly augmented plane wave approach (FP-LAPW) and the generalised gradient approximation (GGA) with local orbital (lo) and local screening (ls) pseudopotentials. It is in very close agreement with the QCBED result as well as the average of all previous theoretical studies.

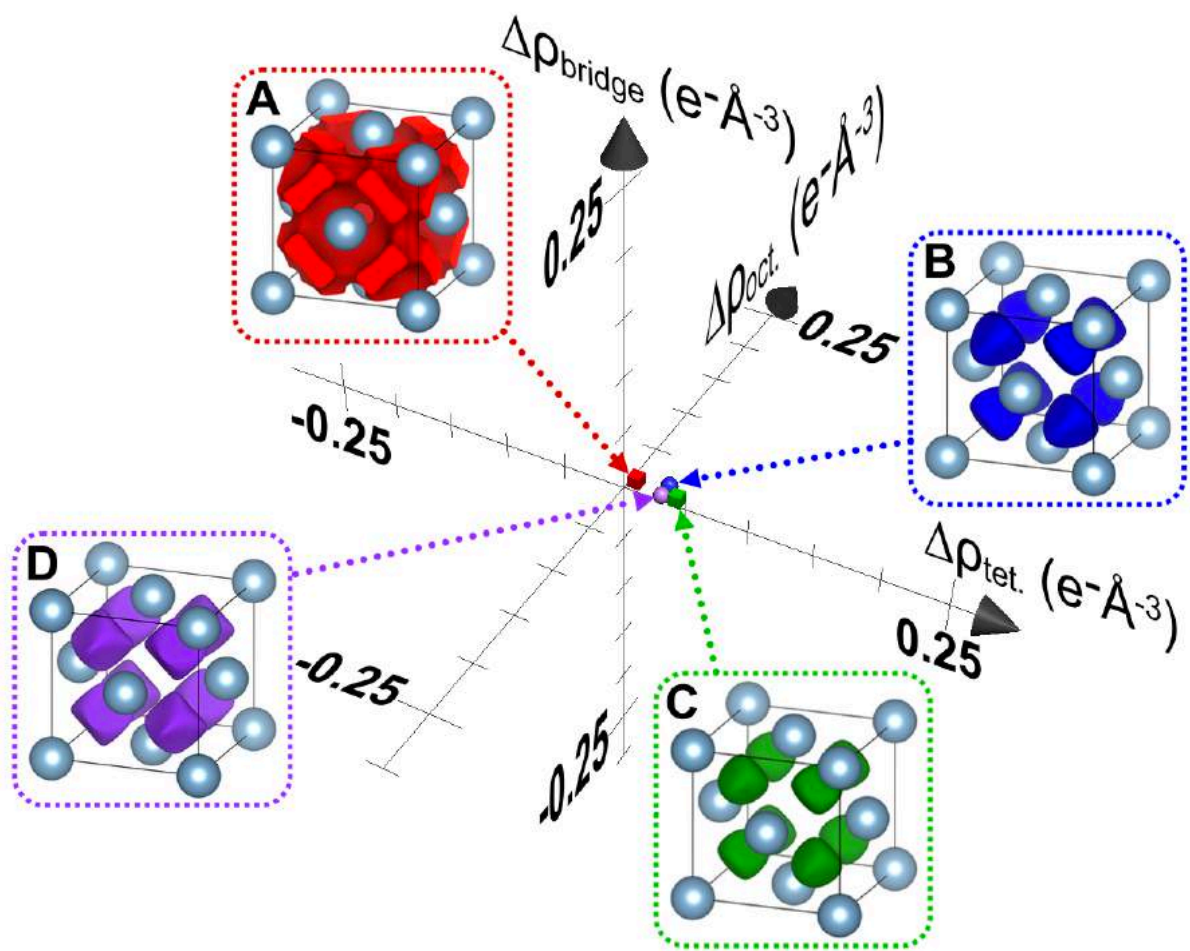

Figure 15: Comparison of the bonding electron densities determined from the structure factors in each column of table 3. (A) The mean of all experimental measurements prior to the QCBED study of (175) constitutes the red cube in the deformation density graph and the red iso-surface plot of bonding electron density in the red dotted cell (as in figure 12 (F)). (B) The mean of all theoretical determinations prior to the WIEN2K calculation of (175) constitutes the blue sphere in the deformation density graph and the blue iso-surface plot of bonding electron density in the blue dotted cell (as in figure 13 (F)). (C) The mean bonding electron density determined by the QCBED study of (175) is given here by the green cube in the graph and the green iso-surface plot of bonding electron density in the green dotted cell (as in figure 14 (F)). (D) The purple sphere and purple iso-surface plot of bonding electron density correspond to the WIEN2K (212) calculation performed in (175) (column 5 in table 3). All iso-surfaces are drawn at a level of 50% of the maximum bonding density in each cell. The cells in this figure were plotted with VESTA (1).

A final observation, based on figure 15, is that in the absence of the QCBED result of (175), the graph would contain two closely related, theoretically derived points, giving rise to almost equivalent bonding electron density distributions as plotted in cells B and D. With experimental measurements being the only true window onto what exists in nature, the morphological differences between A and B and D would only serve to question the accuracy and validity of the latest solid state theory (DFT), even in such a very simple system as aluminium. By providing an accurate and highly precise experimental benchmark, the QCBED results of (175) serve to validate some forms of DFT over others, with some approximations being shown to be better than others. Using the average of past experiments as a benchmark for validating different approaches to DFT may have led to the wrong conclusions, even in such a simple material. This would not bode well for theoretical studies of more complex systems such as aluminium alloys, for example.

# The Crystallography of (some) Aluminium Alloys

In the first four sections of this chapter, the fundamental aspects of the crystallography of pure aluminium were discussed at length. These were: (I) the crystal structure, (II) the lattice parameter, (III) the thermal vibration amplitude (Debye-Waller factor) of elemental aluminium, and (IV) the bonding electron distribution in aluminium. In this section, all of these aspects are brought to bear on the crystallography of aluminium alloys in the form of a general discussion, using a number of well-known examples, and a review of key literature on the crystallography of some significant aluminium alloys.

For the sake of brevity, it is impossible to discuss the crystallography of all aluminium alloys in this section. It is also impossible to detail and review all of the techniques used to characterise aluminium alloys and their constituent phases. The present section therefore provides a perspective for considering the crystallography of aluminium alloys and illustrates the ideas presented with a small number of examples which have gained significant attention in the scientific literature.

Before commencing, however, it is worth giving a very brief review of the vast literature discussing the characterisation and application of aluminium alloy crystallography. The following, very simple and brief summary is given in order to provide a few illustrative examples of pioneering, developmental and state-of-the-art research into the atomic structure of aluminium alloys.

Aluminium alloys fall into the following classes or series (references are a selection of works involving crystallographic characterization of alloys that fall within each class and that may not otherwise have been cited in other parts of the present review. Note that where a code has not been designated to the alloy being studied, the alloy has been assigned a class based on the main alloying element):

1XXX - commercially pure aluminium (having a minimum purity of 99%) (215 – 219).
2XXX - copper (Cu) is the main alloying element (219 – 283).
3XXX - manganese (Mn) is the main alloying element (284 – 292).
4XXX - silicon (Si) is the main alloying element (219, 293 – 303).
5XXX - magnesium (Mg) is the main alloying element (275, 304 – 312).
6XXX - magnesium (Mg) and silicon (Si) are the main alloying elements (230, 234, 242, 243, 267, 278, 287, 313 – 359).
7XXX - zinc (Zn) is the main alloying element (230, 234, 267, 278, 287, 360 – 379).
8XXX - other elements including rare earths are used as the main alloy elements (216, 245, 286, 287, 380 – 408).
High entropy alloys (HEAs) (409).

An extensive generic review in terms of structure and properties is given by Mondolfo (410), giving almost complete coverage of all aluminium alloys with the exception of those that have been developed since the 1980s.

The characterization, analysis, prediction and modelling of aluminium alloy crystallography, structure and composition, have involved the following techniques (references from the list above are associated with each of the techniques involved in each of the works).

Experimental Methods:

*X-ray diffraction:*
(218, 220 – 225, 237, 264, 270, 277, 285, 286, 289, 293, 294, 376, 386, 387, 394, 397, 402 – 404, 409).

*Selected area electron diffraction (SAED):*

(217, 226 – 228, 230, 232, 235, 236, 238, 239, 241, 243, 245, 247, 248, 250 – 253, 257, 264, 265, 268, 270, 273, 281 – 287, 289 – 292, 295, 298, 300, 301, 305, 310, 312 – 314, 316, 318 – 323, 325, 327, 329, 331, 333 – 335, 337, 342, 348, 360 – 365, 367, 368, 370 – 372, 374, 381 – 385, 389, 390, 394, 395, 401, 402).

*Quantitative SAED atomic structure solution via Multislice Least Squares (MSLS):*
(315, 316, 331, 332, 339).

*Convergent-beam electron diffraction (CBED):*
(226, 230, 253, 270, 282, 283, 287, 297, 298, 303 – 306, 311, 320, 335, 337, 374, 381 – 383, 385, 403, 405).

*Position-averaged convergent-beam electron diffraction (PACBED):*
(267, 309).

*Quantitative CBED (QCBED) bonding measurements in monolithic intermetallic phases:*
(382).

*Neutron diffraction:*
(268).

*Precession electron diffraction (PED):*
(287, 337, 369, 371).

*Small angle X-ray scattering (SAXS):*
(259, 268, 367, 370).

*Extended X-ray absorption fine structure (EXAFS):*
(302).

*X-ray absorption near-edge structure (XANES):*
(302).

*X-ray texture analysis:*
(215).

*Positron annihilation lifetime spectroscopy (PALS) / Coincidence Doppler broadening (CDB):*
(232, 312).

*Differential scanning calorimetry (DSC):*
(248, 251, 317, 322, 323, 325, 379, 394).

*Scanning electron microscopy (SEM):*
(215, 218, 236, 273, 288, 296, 297, 366, 384, 386, 394, 402, 404).

*Electron Backscatter Diffraction (EBSD):*
(249, 288, 296, 297, 337, 403).

*Energy dispersive (X-ray) spectroscopy (EDS) / Spectrum imaging:*
(226 – 228, 236, 247, 253, 266, 273, 274, 278, 280, 281, 296, 297, 301, 303, 305, 312, 333, 337, 348, 366, 372, 374, 378, 379, 381, 388, 400, 403, 404).

*Optical Microscopy:*
(215, 272, 394).

*Nuclear magnetic resonance (NMR) spectroscopy:*
(240, 246, 264, 376).

*Muon spin relaxation (MSR):*
(355).

*Confocal laser scanning microscopy (CLSM):*
(272).

*High-resolution transmission electron microscopy (HRTEM):*
(217, 227, 229 – 231, 233, 238, 239, 241, 243, 245, 247, 250, 252 – 254, 256, 265, 269, 273, 283, 287, 290 – 292, 300, 301, 303, 305, 311 – 321, 324, 326, 327, 329, 331 – 336, 339, 340, 342, 344, 345, 354, 363 – 365, 371, 380, 385, 391, 393, 394, 396, 399 – 401).

*Energy-filtered transmission electron microscopy (EFTEM) / Electron energy-loss spectroscopy (EELS) / Spectrum imaging:*
(259, 273, 274, 353, 377, 379, 388).

*Through-focal series reconstruction HRTEM:*
(260, 328).

*3D Atom Probe / Field Ion Microscopy (3D APFIM also known as atom probe tomography (APT)):*
(219, 230, 239, 245, 250, 255 – 257, 265, 273, 277, 279, 301, 302, 308, 317 – 319, 332, 336, 341, 348, 362, 364, 368, 370, 373 – 375, 385, 393, 399, 407, 411).

*High angle annular dark field (HAADF) / annular dark field (ADF) scanning transmission electron microscopy (STEM):*
(262, 269, 292, 301, 307, 326, 333, 334, 344, 348, 400, 408).

*Annular bright field (ABF) STEM:*
(334).

*Aberration-corrected HAADF / ADF-STEM:*
(257 – 262, 266 – 268, 271 – 274, 278 – 282, 307, 309, 338 – 340, 345, 347, 349 – 354, 356 – 359, 377 – 379, 405, 407, 408).

*Aberration-corrected bright-field STEM:*
(257, 357).

*Quantitative HAADF-STEM:*
(259, 260, 309, 351, 405).

*Electron tomography:*
(269, 292, 407).

Theoretical Modelling / Data Analysis Techniques:

*Density functional theory (DFT):*
(234, 242, 260, 261, 271, 276, 307, 328, 330, 339, 343, 346, 350, 351, 358, 378, 408).

*Lattice matching:*
(299, 392, 398, 406).

*Multislice simulations of lattice images:*
(227, 238, 247, 250, 260 – 262, 265, 274, 309, 351, 380, 405).

*Multislice calculations of electron diffraction patterns:*
(250, 315, 316, 331, 332, 339).

*Bloch-wave simulations of CBED patterns:*
(337, 382).

*Lattice rectification (solid solutions):*
(308).

*Cluster identification techniques in 3D APFIM:*
(341).

*Phase-field modelling:*
(263, 276).

*Monte Carlo / molecular dynamics simulations of structural evolution:*
(275).

*Semi-empirical modelling based on data mining:*
(293).

These lists are only small subsets of all of the existing literature on the experimental and theoretical investigation of the crystallography, composition and structure of aluminium alloys. They are biased towards more recent work but also include some examples of landmark pioneering research. They do not cover all systems that have been explored or that are being explored. It is emphasised again that the references given here serve to provide just a taste for the vast research into the crystallography of aluminium alloys.

At this point, it is worth returning to the key concepts developed and explored for pure aluminium, in order to apply them to aluminium alloys.

## Crystal structures of (some) aluminium alloys

Alloys can take the form of a solid solution, which has a homogeneous crystal structure of the same form (single phase) everywhere (allowing for changes in orientation from grain to grain in a polycrystalline microstructure and also allowing for crystal defects within each grain). Alternatively, alloys can consist of a dominant matrix phase, or parent crystal structure, containing a dispersion of intermetallic precipitates that have different crystal structures and compositions to the matrix and make up a small volume fraction of the alloy. Eutectic mixtures consist of interleaved metallic phases of differing compositions which dominate the microstructure (412).

The majority of commercial and research interest in aluminium alloys is in age hardenable alloys (or precipitation hardening alloys), which begin as supersaturated solid solutions (SSSS). These are subjected to ageing heat treatments which result in the migration of solute atoms to points of nucleation of intermetallic phases. Often, a crystallographic phase transformation ensues in the formation of a critical nucleus for a precipitating intermetallic phase, which grows until the local concentration of solute atoms decreases below a critical concentration or crystallographic barriers to precipitate growth stop the process.

In cases where the intermetallic precipitates act as strong barriers to the movement of dislocations, strengthening occurs. Where they promote or mediate the movement of dislocations, the alloy is embrittled or weakened. In a rather hand-waving manner of describing precipitation hardened alloys, the metaphor is that the precipitates act as a strengthening scaffold in an otherwise relatively weak matrix.

Pure aluminium (99.999+% purity) has an ultimate tensile strength (UTS) of 40 to 50 MPa (410), whilst some of the strongest precipitation-hardened alloys have UTS's exceeding 700 MPa (e.g. Weldalite 049 which is an Al-Cu-Li-Mg-Ag-Zr alloy) (413). This massive increase in strength means that these types of alloys have a very high strength-to-weight ratio that makes them competitive with high-strength steels. The alloying elements are present at concentrations of only a few atomic percent and therefore, changes in density compared to pure aluminium are negligible in the face of the 10- to 20-fold gains in strength. These gains in strength are largely due to the scaffolding effect of high-aspect-ratio strengthening precipitates that inhibit dislocation movement within the alloys.

This section makes examples of the most prominent strengthening precipitate phases that have been the focus of significant research. These include the θ" ($Al_3Cu$), θ' ($Al_2Cu$) and θ ($Al_2Cu$) precipitate phases in Al-Cu based alloys (104, 220, 230, 239, 245, 261 – 263, 268, 271), the $T_1$ phase ($Al_2CuLi$) in Al-Cu-Li based alloys (225, 241, 245, 259, 260, 268, 272, 279 – 281), the Ω phase ($Al_2Cu$) in Al-Cu-Mg-Ag alloys (226, 230, 231, 235, 239, 245, 274) and the S phase ($Al_2CuMg$) in Al-Cu-Mg alloys (229, 230, 234, 239, 245, 251, 252, 256, 258). All of these alloy systems fall within the 2XXX and 8XXX series of aluminium alloys but the crystallographic concepts explored here are equally applicable in the other series and, consequently, to other intermetallic precipitate phases and solid solutions.

Figure 16 begins in part a with an illustration of a binary Al-Cu solid solution with copper atoms present at an atomic concentration of approximately 4%. Given that the maximum solid solubility of copper in aluminium is 2.4 atomic percent (413), figure 16 (a) represents a supersaturated solid solution (SSSS).

Copper atoms are too large to fit into any interstitial sites within the fcc lattice of aluminium, and they are thus substitutional. Note that distortions of the lattice due to the smaller copper atoms are not drawn into this depiction of the Al-Cu solid solution. Heat treatment (artificial ageing) of this SSSS results in the possible formation of numerous phases including Guinier-Preston (GP) zones (414, 415) (which are single continuous planes of copper atoms substituted on aluminium lattice sites), θ" precipitates (consisting of multiple GP zones separated by three {002} planes of aluminium (see table 4)), θ' precipitates (shown in figure 16 (b) and table 4) and eventually, the stable θ phase (see table 4).

GP zones, θ" and θ' are all metastable phases, whilst the θ phase is the end product of the solid-state reaction: SSSS -> GP + θ" + matrix -> θ' + matrix -> θ + matrix. It is the θ' phase that is the most effective strengthening precipitate and processing routes for Al-Cu based alloys are frequently tailored to promote a high number density of θ' precipitates. Figure 16 (b) shows a block of Al-Cu alloy containing a θ' precipitate which is sandwiched on both major {001} facets by the aluminium matrix. Parts c and d of the figure show the unit cells of θ' and the aluminium matrix in the orientation relationship they have within the alloy. The tetragonal cell of θ' has very similar *a* and *b* lattice parameters to the corresponding lattice parameters in the $Al_{fcc}$ cell shown in part d, with only a 0.2% linear misfit (see also table 4). The *c* lattice parameter of θ', $c_{θ'}$, is significantly different and can impart a large amount of strain in the direction normal to the θ' precipitate. This strain is minimised for certain multiples of $c_{θ'}$, which span thicknesses that are very close to integer multiples of the aluminium matrix *c* lattice parameter, $c_{Al, fcc}$. One such case is drawn in both parts b and e of figure 16 where $3.5c_{θ'}$ is approximately equal to $5c_{Al,fcc}$ (i.e. to within 0.3%). This is known as a "magic thickness" (262). Other magic thicknesses for θ' precipitates are given in table 4 where the magic thicknesses of $2c_{θ'}$ and $3.5c_{θ'}$ are highlighted in red because these are the most commonly observed thicknesses of θ' in Al-Cu alloys (262). Figure 16 (e) shows a section through the modelled θ' precipitate and the surrounding matrix with all atoms drawn in a non space-filling manner and with a depth of only one unit cell so that the orientation relationship between the θ' unit cell and the Al matrix fcc unit cell can be clearly seen.

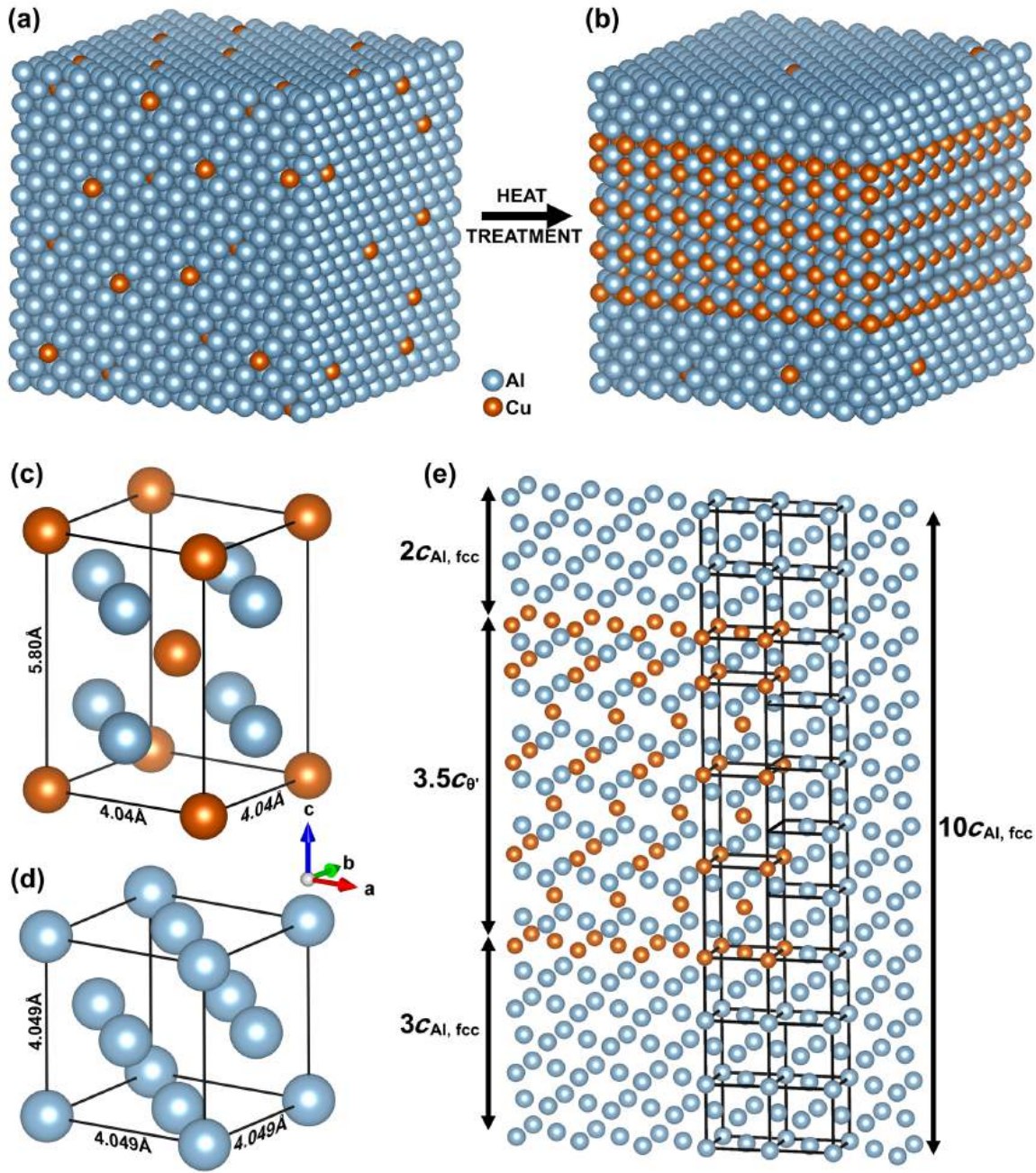

**Figure 16:** An illustration of the θ' phase ($Al_2Cu$) embedded in the aluminium alloy matrix. The precursor supersaturated solid solution is illustrated with a 10x10x10 fcc unit cell block containing randomly placed copper atoms at a concentration of approximately 4 at. % (a). Aluminium alloy solid solutions are usually always substitutional, as shown in the present case. After heat treatment, precipitates can form, such as the segment of θ' phase shown in part b. The unit cell of θ' is shown in part c and the fcc unit cell of aluminium is shown for comparison beneath it (d). In part e, a slice one unit cell thick is taken through the precipitate structure and the surrounding matrix to show the relationship between the precipitate and matrix structures. A situation of minimal volumetric strain is illustrated which involves a precipitate thickness of 3.5 unit cells of θ' in the c-axis of θ', which corresponds to nearly 5 fcc unit cells of aluminium. This θ' thickness of $3.5c_{θ'}$ is known as a "magic thickness" (262). This figure illustrates a binary alloy (2 elements) and was drawn with the aid of VESTA (1).

Whilst the θ' unit cell has no face-centred copper atoms in the basal plane of the cell, the interfaces of the main facets of θ' precipitates with the aluminium matrix are observed experimentally to have extra face-centred copper atoms (261). Density functional theory calculations show that this configuration has a lower energy than one in which the θ' precipitates terminate without the extra face-centred copper atoms (261). This is an excellent example of where new capabilities in electron microscopy, due to aberration correction, have allowed such crystallographic detail to be resolved.

The crystallographic relationships (orientation relationships) between GP zones, θ", θ', θ and the aluminium matrix can all be simply depicted using the fcc cell to describe the aluminium matrix crystal structure (see table 4). This makes electron scattering from these systems relatively easy to simulate using the multislice formalism (3) explained earlier in the first section of this chapter. Figure 17 examines the $T_1$ phase in Al-Cu-Li alloys where this is no longer the case.

The $T_1$ phase (nominally $Al_2CuLi$ in composition) has been the subject of considerable uncertainty and debate with regard to its crystal structure (225, 241, 245, 259, 260, 416 – 423). The most recent structure assessment combined TEM through-focal-series reconstruction with quantitative aberration-corrected HAADF-STEM, and DFT to present the most reliable structure for $T_1$ to date (260) (see figure 17 (d)). The $T_1$ structure is hexagonal with *P6/mmm* space group symmetry according to (260). These results are closest to the previous X-ray diffraction study by Van Smaalen et al. (421), but not in absolute agreement. Precipitates of the $T_1$ phase form very high aspect ratio platelets with major facets coplanar to the basal plane of the hexagonal unit cell that describes the atomic structure of $T_1$. These facets are coplanar to {111} in the fcc cell that describes the aluminium matrix. A section through a $T_1$ precipitate surrounded by the supporting aluminium matrix is modelled in figure 17 (b).

In order to understand the orientation relationship of $T_1$ with the aluminium matrix and the magnitude of any strain imparted on the matrix by misfit, it is simpler to deal with the aluminium matrix described in terms of the trigonal cell shown in figure 17 (c). In this representation, the small trigonal cell derived earlier in figure 3 (d, e and f) is also shown in relation to the larger trigonal cell derived here. The present cell (dark blue atoms to make it distinct from the previous trigonal cell of figure 3 (d, e and f)) is described in relation to the fcc unit cell of aluminium in the specifications beneath the cell diagram, together with the space group of this larger trigonal cell and all symmetry-independent atom positions.

The orientation relationship of the $T_1$ structure with this new trigonal cell description of the matrix is such that the *a*, *b* and *c* axes of one cell are parallel with their counterparts in the other cell. Such a description of the matrix structure facilitates analysis by the multislice approach along the common *c* axes, which are normal to the $T_1$ precipitate / Al matrix interfaces. This description and representation of the orientation relationships of the precipitate and matrix structures also allows immediate assessment of the degree of lattice misfit, which is very small (see table 4).

Most $T_1$ precipitates are just a single unit cell thick as they strongly resist growth along the c axis in the hexagonal structure (i.e. $<111>_{Al, fcc}$ or $<001>_{Al, trig.}$). The fit of the $T_1$ unit cell into the matrix structure shown in figure 17 (e) illustrates the equivalence of the a and b lattice parameters for the matrix and $T_1$ cells and the relationship $c_{T1} \sim 2c_{Al, trig.}$ with a misfit of 0.9% (see table 4).

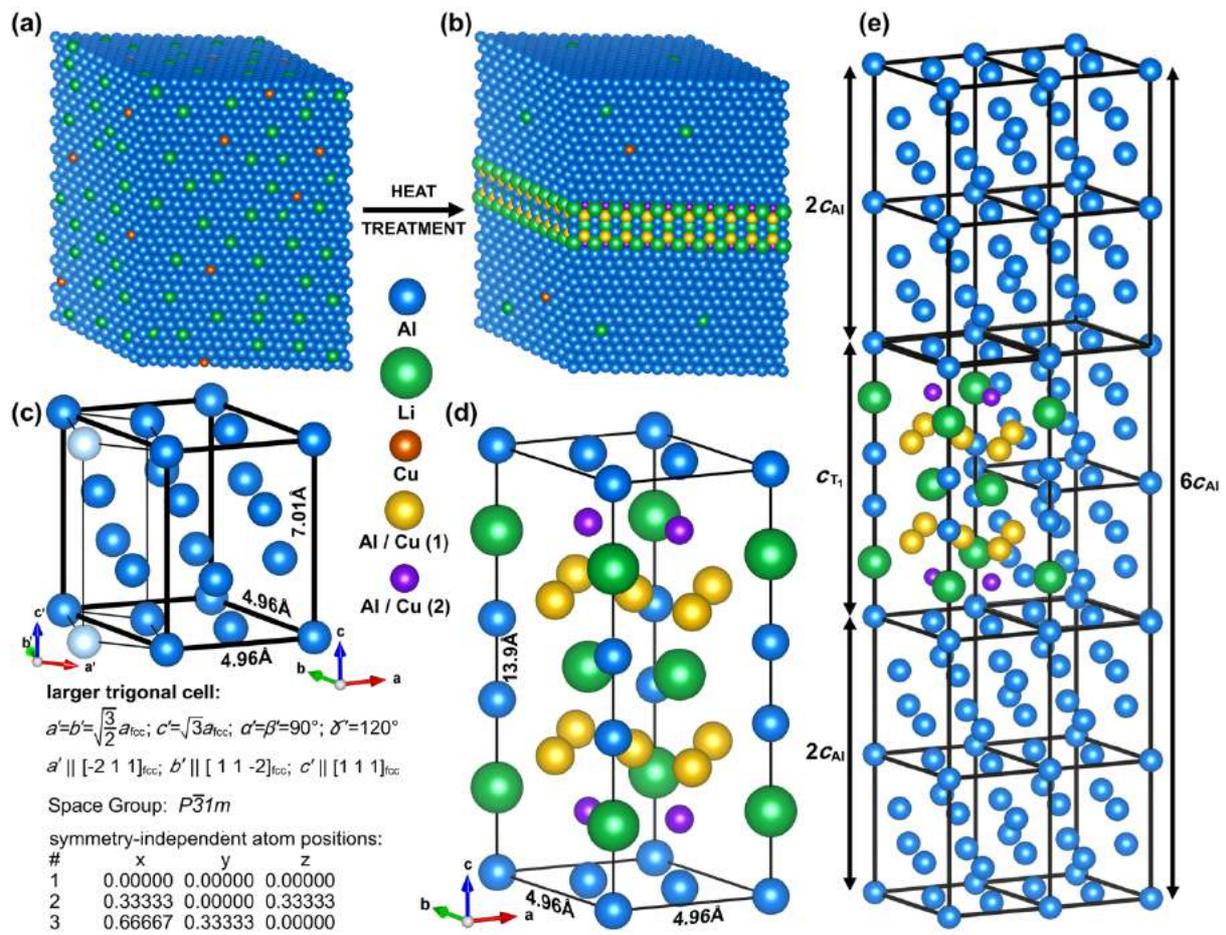

Figure 17: This figure is similar to figure 16, except that in this case, a more complex alloy is illustrated. The supersaturated solid solution (a) shows copper and lithium atoms dissolved in the aluminium matrix (a ternary alloy). Copper and lithium are present at concentrations of approximately 2% and 6% respectively (i.e. an 8XXX alloy). The $T_1$ phase is illustrated in part b, after the heat treatment of the solid solution, and it is surrounded by the aluminium matrix which still contains solute atoms but at a much lower concentration. In parts c and d, the structures of the aluminium matrix and the $T_1$ phase (260) are compared via the matrix unit cell that is most conducive to such a comparison. $T_1$ has a hexagonal structure (shown in part d), with a basal plane that is coplanar with {1 1 1} in the fcc cell of aluminium. It is therefore much easier to describe the aluminium matrix with a trigonal cell (c). The trigonal cell shown in figure 3 (d, e and f) gives rise to the present description, as drawn and described fully here in part c. The cell of figure 3 (d, e and f) is outlined with lighter lines and lighter atoms where they are not shared with the larger trigonal cell described here. The relationships between the matrix and $T_1$ cells (c and d) in the actual alloy are shown in part e, where it is evident that the lattice mismatch in all axes is very small and therefore imparts almost no volumetric strain within the alloy. This figure was drawn with the aid of VESTA (1).

It is worth noting that the $T_1$ phase in Al-Cu-Li alloys is one of the most efficient strengthening phases in all aluminium alloys.

Other efficient strengthening precipitate phases include the Ω phase, found in Al-Cu-Mg-Ag alloys (226, 230, 231, 235, 239, 245, 274), and the S phase found in Al-Cu-Mg systems (229, 230, 234, 239, 245, 251, 252, 256, 258). Table 4 presents these structures and their orientation relationships and lattice misfits with the most convenient unit cell for describing the aluminium matrix.

In the case of the Ω phase (determined unequivocally by CBED to be orthorhombic (226)), the most convenient matrix structure description is via the orthorhombic cell derived in figure 3 (g, h and i). The relationship between this orthorhombic matrix cell and the Ω unit cell, described in table 4, is $3b_{Al, ortho} = b_\Omega$, and $c_{Al, ortho.} = a_\Omega$.

| Alloy System | Phase | Space Group | Lattice Parameters (Å, °) a b c α β γ | Unit Cell Structure Element x y z OCC | Precipitate/Matrix O.R. | Misfit | References |
|---|---|---|---|---|---|---|---|
| Al-Cu based | θ" (Al$_3$Cu) | P4/mmm | 4.040 4.040 7.680 90.0 90.0 90.0 | Cu 0.00000 0.00000 0.00000 1.000<br>Cu 0.50000 0.50000 0.00000 1.000<br>Al 0.50000 0.00000 0.23698 1.000<br>Al 0.00000 0.00000 0.50000 1.000<br>Al 0.50000 0.50000 0.50000 1.000 | 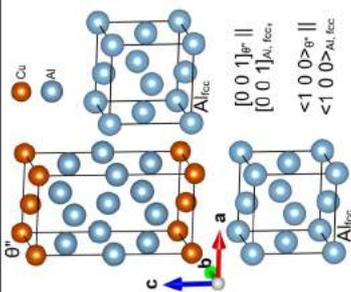<br>[0 0 1]$_{θ"}$ ∥ [0 0 1]$_{Al, fcc}$,<br><1 0 0>$_{θ"}$ ∥ <1 0 0>$_{Al, fcc}$ | [1 0 0]$_{θ"}$ 0.2%<br>[0 1 0]$_{θ"}$ 0.2%<br>[0 0 1]$_{θ"}$ 5.2% | (234, 245, 413, 424 – 428) |
| Al-Cu based | θ' (Al$_2$Cu) | I-4m2 | 4.040 4.040 5.800 90.0 90.0 90.0 | Cu 0.00000 0.00000 0.00000 1.000<br>Al 0.00000 0.50000 0.25000 1.000<br>Al 0.50000 0.00000 0.25000 1.000<br>NB: There is an additional copper atom in the terminal ab-planes located at 0.5 0.5 0.0 relative to the copper atoms in these planes. See (261). | 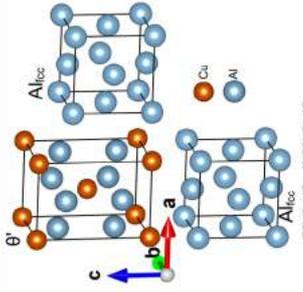<br>[0 0 1]$_{θ'}$ ∥ [0 0 1]$_{Al, fcc}$,<br><1 0 0>$_{θ'}$ ∥ <1 0 0>$_{Al, fcc}$ | [1 0 0]$_{θ'}$ 0.2%<br>[0 1 0]$_{θ'}$ 0.2%<br>[0 0 1]$_{θ'}$ 43% for 1$c_θ$, 4.5% for 2$c_θ$, 0.3% for 3.5$c_θ$, 4.5% for 4$c_θ$, 2.3% for 5$c_θ$, 1.5% for 5.5$c_θ$ | (220, 245, 261, 262, 413) |
| Al-Cu based | θ (Al$_2$Cu) | I4/mcm | 6.067 6.067 4.877 90.0 90.0 90.0 | Cu 0.00000 0.00000 0.00000 1.000<br>Al 0.15810 0.65810 0.07500 1.000 | 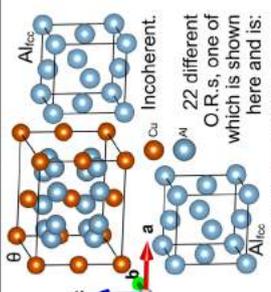<br>Incoherent. 22 different O.R.s, one of which is shown here and is:<br>[0 0 1]$_θ$ ∥ [0 0 1]$_{Al, fcc}$,<br><1 0 0>$_θ$ ∥ <1 0 0>$_{Al, fcc}$ | For the O.R. shown:<br>[1 0 0]$_θ$ 49.8% for 1$a_θ$, 0.11% for 2$a_θ$<br>[0 1 0]$_θ$ 49.8% for 1$b_θ$, 0.11% for 2$b_θ$<br>[0 0 1]$_θ$ 20.4% for 1$c_θ$, 0.37% for 5$c_θ$ | (245, 413, 429 – 432) |

Table 4: A summary of selected common precipitate phases occurring in aluminium alloys, their unit cell structures, orientation relationships and misfit with the aluminium matrix. The different unit cells that can be used to describe the structure of the aluminium matrix, explored in figure 3, are used here to simplify the description of orientation relationships and lattice mismatch. All unit cells and axes were drawn with VESTA (1).

| Alloy System | Phase | Space Group | Lattice Parameters (Å, °) | | | | | | Unit Cell Structure | | | | | Precipitate/Matrix O.R. | Misfit | References |
|---|---|---|---|---|---|---|---|---|---|---|---|---|---|---|---|---|
| | | | a | b | c | α | β | γ | Element | x | y | z | OCC | | | |
| Al-Cu-Li based | $T_1$ ($Al_2CuLi$) | P6/mmm | 4.960 | 4.960 | 13.900 | 90.0 | 90.0 | 120.0 | Al | 0.00000 | 0.00000 | 0.00000 | 1.000 | $[1\,0\,0]_{T_1} \parallel [1\,0\,0]_{Al,\,trig.}$, $[0\,1\,0]_{T_1} \parallel [0\,1\,0]_{Al,\,trig.}$, $[0\,0\,1]_{T_1} \parallel [0\,0\,1]_{Al,\,trig.}$ | $[1\,0\,0]_{T_1}$ 0.00% $[0\,1\,0]_{T_1}$ 0.00% $[0\,0\,1]_{T_1}$ 0.9% | (259, 260, 280, 423) |
| | | | | | | | | | Al | 0.66667 | 0.33333 | 0.00000 | 1.000 | | | |
| | | | | | | | | | Al | 0.00000 | 0.00000 | 0.40500 | 1.000 | | | |
| | | | | | | | | | Al | 0.66667 | 0.33333 | 0.16200 | 1 − m | | | |
| | | | | | | | | | Cu | 0.66667 | 0.33333 | 0.16200 | m | | | |
| | | | | | | | | | Al | 0.50000 | 0.00000 | 0.31800 | n | | | |
| | | | | | | | | | Cu | 0.50000 | 0.00000 | 0.31800 | 1 − n | | | |
| | | | | | | | | | Li | 0.00000 | 0.00000 | 0.19930 | 1.000 | | | |
| | | | | | | | | | Li | 0.33333 | 0.66667 | 0.50000 | 1.000 | | | |
| Al-Cu-Mg-Ag | Ω ($Al_2Cu$) | Fmmm | 4.960 | 8.590 | 8.480 | 90.0 | 90.0 | 90.0 | Al | 0.75000 | 0.08333 | 0.75000 | 1.000 | $[0\,0\,1]_\Omega \parallel [1\,0\,0]_{Al,\,ortho}$, $[1\,0\,0]_\Omega \parallel [0\,0\,\bar{1}]_{Al,\,ortho}$, $[0\,1\,0]_\Omega \parallel [0\,1\,0]_{Al,\,ortho}$ | $[1\,0\,0]_\Omega$ 0.00% $[0\,1\,0]_\Omega$ 0.00% $[0\,0\,1]_\Omega$ 20.9% for $1c_\Omega$, 0.76% for $5c_\Omega$ | (226, 245, 433, 434) |
| | | | | | | | | | Al | 0.75000 | 0.75000 | 0.41667 | 1.000 | | | |
| | | | | | | | | | Cu | 0.00000 | 0.00000 | 0.00000 | 1.000 | | | |

Table 4 (continued): A summary of selected common precipitate phases occurring in aluminium alloys, their unit cell structures, orientation relationships and misfit with the aluminium matrix. The different unit cells that can be used to describe the structure of the aluminium matrix, explored in figure 3, are used here to simplify the description of orientation relationships and lattice mismatch. All unit cells and axes were drawn with VESTA (1).

| Alloy System | Phase | Space Group | Lattice Parameters (Å, °) | | | | | | Unit Cell Structure | | | | | Precipitate/Matrix O.R. | Misfit | References |
|---|---|---|---|---|---|---|---|---|---|---|---|---|---|---|---|---|
| | | | a | b | c | α | β | γ | Element | x | y | z | OCC | | | |
| Al-Cu-Mg based | S (Al$_2$CuMg) | Cmcm | 4.000 | 9.056 | 7.245 | 90.0 | 90.0 | 90.0 | Al | 0.00000 | 0.35600 | 0.05600 | 1.000 | 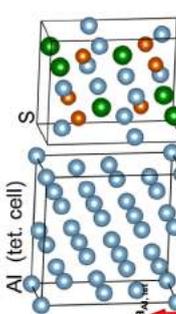 [1 0 0]$_S$ ∥ [0 0 1]$_{Al, tet.}$, [0 1 0]$_S$ ∥ [1 0 0]$_{Al, tet.}$, [0 0 1]$_S$ ∥ [0 1 0]$_{Al, tet.}$ | [1 0 0]$_S$ 1.2% [0 1 0]$_S$ 0.02% [0 0 1]$_S$ 20.0% for 1$c_S$, 0.03% for 5$c_S$ | (66, 229, 234, 245, 252, 435 - 437) |
| | | | | | | | | | Mg | 0.00000 | 0.07200 | 0.25000 | 1.000 | | | |
| | | | | | | | | | Cu | 0.00000 | 0.77800 | 0.25000 | 1.000 | | | |

Table 4 (continued): A summary of selected common precipitate phases occurring in aluminium alloys, their unit cell structures, orientation relationships and misfit with the aluminium matrix. The different unit cells that can be used to describe the structure of the aluminium matrix, explored in figure 3, are used here to simplify the description of orientation relationships and lattice mismatch. All unit cells and axes were drawn with VESTA (1).

For the S phase, the most convenient matrix structure description is by the large tetragonal cell defined in figure 3 (j, k and l). The S phase imparts a small amount of misfit strain in the $[100]_S$ direction, almost none in $[010]_S$ and large strain for multiples of $c_S$ that are not multiples of five.

In the analysis of precipitate crystallography, quantitative electron diffraction shows great potential. Examples of pattern-matching-based quantitative electron diffraction structure determination are sparse and limited to the multislice least squares (MSLS) approach of Zandbergen, Andersen and Jansen (315, 316), also applied in the work of Vissers et al. (331), Hasting et al. (332) and Holmestad et al. (339). Whilst MSLS is a parallel-beam diffraction analysis tool, the same multislice-based approach could be readily applied to CBED patterns collected from individual precipitates, such as the example in figure 18.

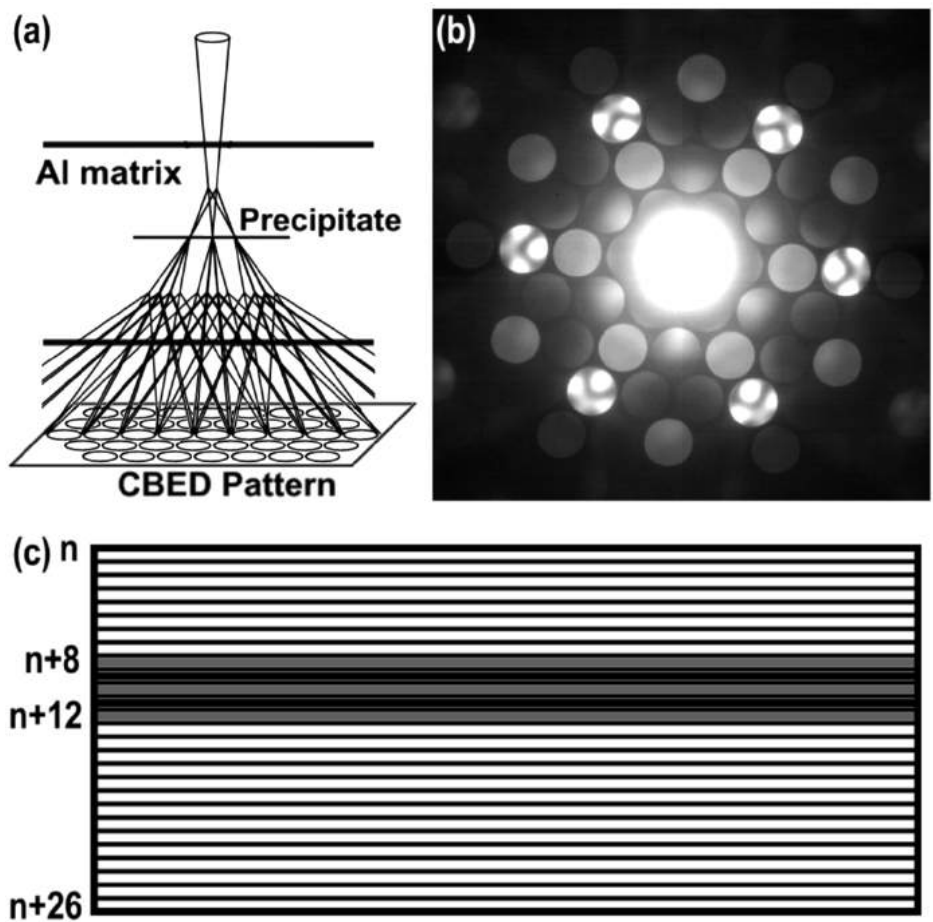

Figure 18: Using convergent-beam electron diffraction (CBED) to probe and analyse the structures of precipitate phases in alloys. Part a shows a schematic diagram of electron scattering when the crystal structure of the matrix (aluminium) is interrupted by a precipitate having a different structure. In the present example, a $T_1$ precipitate is intercepted by the incident and matrix-scattered electron beams. Due to the crystallography of the $T_1$ phase (see figure 17 and table 4), these beams are scattered at three times as many angles by $T_1$ as the aluminium matrix (shown only in 1 dimension in part a to reduce clutter). The resulting experimental CBED pattern (b), collected with 200keV electrons along $<1\ 1\ 1>_{Al,\ fcc}$ (or $<0\ 0\ 1>_{Al,\ trig.}$), shows bright reflections at positions three times the shortest scattering vectors in the pattern from the central beam (which floods the centre of the pattern). These are the reflections that originate from initial scattering by the matrix before the beam electrons enter the precipitate. All other reflections originate from the $T_1$ precipitate structure and subsequent rescattering by the matrix below the precipitate. Part c is a schematic representation of the specimen in the volume surrounding the $T_1$ precipitate and shows it being sliced into individual atomic planes. This is required by the multislice theory of electron scattering (3) for a quantitative analysis of the CBED pattern in part b by multislice simulation and pattern-matching. The effects of each plane of atoms on the scattering of the electrons is summed up over the entire probed thickness of the specimen in the direction of the specimen normal (coincident with the beam direction and precipitate normal). The precipitate constitutes only a few layers, shown as slices n+8 to n+12 inclusive (c), as it is about 1 nm thick. In contrast, there may be hundreds of slices through the matrix, which can be 50 to 200 times thicker (only slices n to n+26 are shown here).

CBED uses a focused electron probe and is therefore more capable than parallel beam techniques of position-selective probing of nanostructures such as intermetallic precipitates in alloys. CBED also has the advantage that the reflections possess an intensity distribution over a range of incident angles that reveal the symmetry of the crystal structure along the incident beam direction in a highly visual and easily interpretable fashion. This is not the case with parallel-beam diffraction. Not only does CBED allow an immediate in situ determination of structural symmetry from the intensity distribution in the pattern (226, 299), it has the capability of constraining pattern-matching refinements of crystal structure and bonding-sensitive structure factors to a level that parallel beam diffraction is incapable of.

The CBED pattern in figure 18 (b) was taken through a single $T_1$ precipitate embedded in an aluminium matrix in an Al-Cu-Li based alloy. The 6mm symmetry of the pattern is entirely commensurate with the determination of *P6/mmm* space group symmetry and the structure of Dwyer et al. (260). The incident electron beam is parallel to the <001> direction illustrated in figure 17 and table 4.

Part a of figure 18 shows the nature of electron scattering in the present scenario schematically. The incident electron beam diffracts from the simpler crystal structure of the aluminium matrix before all scattered beams are intercepted and re-scattered by the $T_1$ structure, which has a three-fold larger periodicity in projection in real space than the aluminium matrix. The scattered beams exiting the $T_1$ precipitate are then further scattered by the matrix below the precipitate. This scenario is illustrated in only one dimension in part a of the figure to reduce clutter and simplify the diagram, however, this entanglement process of scattering by the different crystal structures being probed, occurs in two dimensions within the slice approximation used by the multislice formalism (3).

The slicing of the probed region along the zone axis and incident beam direction (coincident with the interface normal of the matrix/precipitate/matrix system) is illustrated schematically in figure 18 (c). In practice, the structural model for multislice simulations would consist of many layers of aluminium atoms spanning the total specimen thickness, with a few of these layers replaced by monolayers describing the $T_1$ atomic structure. The intensities in the CBED pattern are very sensitive to the total thickness as well as the depth of the $T_1$ precipitate in the specimen. This sensitivity comes from the entangled nature of the scattered and re-scattered beams from each region of the specimen. The perturbation caused by even the thinnest of precipitates, like $T_1$, will be compounded by the subsequent re-scattering by the matrix below the precipitate.

To gain a better understanding of the intensities in the experimental CBED pattern of figure 18 (b), figure 19 is presented. Part a gives a schematic representation of the pattern, which has been fully indexed, based on the commensurate matrix and $T_1$ unit cells illustrated in figure 17 and table 4. This in itself shows the importance of describing the structure of the matrix in a manner that is commensurate with the cell of the precipitate phase because cells that do not have common *a* and *b* lattice parameters would require a diffraction pattern to have two sets of indices (one for the matrix cell and one for the precipitate phase cell). In figure 19 (a), the colourations of the symmetry equivalent reflections in the pattern match the colours of the corresponding loci of the lattice planes that give rise to the reflections, plotted in the *c*-axis cell projection in figure 19 (b). Part c shows both the $T_1$ and matrix cells so that the positions of different groups of atoms can be seen in terms of where they lie along the *c* axes in both cells. Figure 19 (d-g) show each set of atoms in the $T_1$ structure and their positions in the *a* and *b* axes, projected along the *c* axis. Below the structure diagrams, the simulated diffracted intensities from each group of atoms is given. A strong correlation between atomic positions and the strongest intensities contributed to particular reflections is evident. Part h of the figure shows how the much shorter periodicity of the aluminium matrix results in scattering by only the {300} scattering vectors.

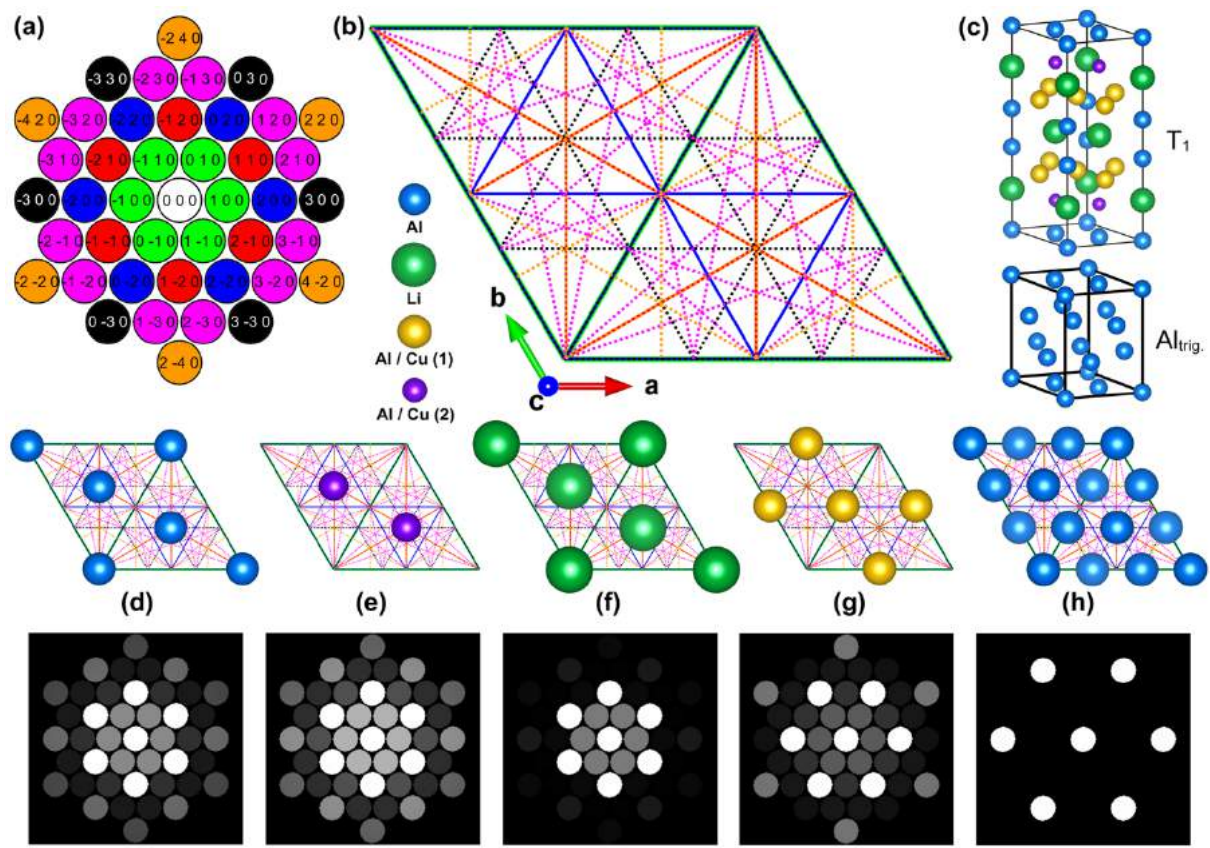

Figure 19: The CBED pattern from figure 18 (b) is considered in more detail from the point of view of the atomic structure in this figure sequence. Part a shows a schematic representation of the pattern (fully indexed according to the unit cell definitions of figure 17 and table 4), with each set of symmetry-equivalent reflections being assigned a colour. These colours are repeated for the corresponding lines in part b which show the positions of the corresponding crystal planes in projection along the c axis of the unit cell of $T_1$ (c). For parts d – h, the different atomic species and their sites in projection along the c axis are plotted in superposition on the plane locus diagram of part b. Note that in the case of part h, the set of atoms plotted is the projected structure of the aluminium matrix, given in the second trigonal cell shown in part c. Different sets of atoms lie at the intersections of different sets of atomic planes and give rise to different intensity contributions to the reflections in the CBED pattern. The lower halves of parts d – h show simulations of the relative contributions of each set of atoms to a CBED pattern for just a single unit cell in the cases of both $T_1$ and the aluminium matrix. The intensities correlate with the locations of atoms in specific planes in the structures and allow a CBED pattern (such as the one in figure 18) to be inspected and interpreted structurally, in an approximate fashion. The unit cells and their projections were drawn in VESTA (1).

Returning to figure 18, the scattering that gives rise to the CBED pattern (b), can be described in the following sequence: prior to intercepting the $T_1$ precipitate, the electron beam is only scattered into directions that are linear combinations of the {300} scattering vectors (figure 19 (h)). Upon entering the precipitate, each atomic layer scatters every diffracted beam entering the precipitate according to the scattering vectors and relative intensities plotted in figure 19 (d-g). The scattering within the precipitate is any linear combination of the {100} scattering vectors. This triples the number of reflections in all directions. Travelling through the matrix below the precipitate, all of the resultant beams are re-scattered by all linear combinations of only the {300} scattering vectors.

The variations in intensity within the reflection discs in a pattern such as figure 18 (b), not only allow the total thickness of the specimen to be measured from the {300} reflections, but will also allow the depth of the precipitate to be determined quite accurately from all of the reflections in the pattern. In a more approximate fashion, the relative intensities of the reflections in between the {300} reflections, are an indication of how strongly the atoms at particular positions are scattering the electron beams. In other words, the average intensities in the reflections are a rough map of atomic mass in the precipitate structure. For example, the pattern in figure 18 (b) shows a higher average intensity in the

{200} reflections (yellow atomic positions, figure 19 (g)) compared to the other reflections (excluding the {300} reflections). This suggests a higher projected atomic mass at the yellow atom positions in the precipitate structure. This may be indicative of a higher copper occupancy at the yellow sites than at the purple sites.

Work is currently underway on QCBED pattern matching of figure 18 (b) via the multislice formalism with results yet to be published.

One final tantalising point to be made about CBED patterns through precipitates having crystallographically coherent interfaces with the surrounding matrix, is that whilst the frequency of the intensity distributions within reflections is indicative of thickness, the shape of the intensity distribution is strongly influenced by the structure factors of the scattering atomic planes. This presents the open question of whether it may be possible to measure bonding effects across the interfaces as well as within the precipitates themselves in a matrix/precipitate/matrix type CBED geometry such as the one presented in figure 18. Work is also currently underway in Al-Cu alloys where QCBED is being applied to CBED patterns taken through θ" and θ' precipitates within the aluminium matrix. Further discussion of this theme is left for the segment on bonding.

### Lattice parameters

Diffraction experiments with both X-rays and electrons have yielded lattice parameters for precipitate phases in aluminium alloys that can be roughly verified by HRTEM imaging or aberration-corrected HAADF-STEM. The work of Dwyer et al. (260) on the $T_1$ phase also compares the experimentally measured lattice parameters with the relaxed lattice constants calculated by DFT.

In the case of lattice parameters in alloy structures, the more difficult assessment is associated with the continuous changes in lattice parameter from one position to another in a solid solution. The standard approach for relating lattice parameters to compositional variation within a single-phase material (such as aluminium-based solid solutions in the present context), is via Vegard's Law (438 – 440).

Examples from the literature of experimental lattice parameter measurements in aluminium-based solid solutions include very early work on Al-Cu alloys (441), investigation of: the Ti-Al system (442, 443), Al-Mn solid solutions (444), Al-Zn solid solutions (63) and, more recently, the Al-Si-Cu-Mg system with particular attention paid to Al-Mg solid solutions and the aluminium matrix lattice parameter as a function of heat treatment time (293). An excellent summary of size-factors (lattice parameters) as a function of composition in hundreds of different binary alloy solid solutions, is given by King (445).

Data mining and semi-empirical methods of modelling lattice parameters in alloy solid solutions are becoming modal via the computer CALculation of PHAse Diagrams (CALPHAD) methodology. The semi-empirical models are, at their foundation, based on Vegard's Law. Some of the binary solid solutions in aluminium that have been investigated by CALPHAD include Al-Ni (446), Al-Li (447), Al-Mg (447) and Al-Si (447, 448).

### Debye-Waller factors

The third section of this chapter covered the Debye-Waller factor of aluminium atoms within pure aluminium. When aluminium alloys are considered, the environment surrounding each aluminium atom can vary. The thermal vibration of atoms is constrained by the interatomic forces exerted on them in bonding with their neighbours. Thus, an aluminium atom that has a copper atom as a neighbour in a 2XXX series alloy, for example, will have a very different Debye-Waller factor than another aluminium atom in another region of the same alloy where it is surrounded by only aluminium atoms. Because most aluminium alloys are very dilute (in terms of the concentration of alloying elements), most aluminium atoms in aluminium alloys are surrounded by other aluminium atoms as nearest neighbours. As a result, the function for $B$(T) presented in the present work for pure aluminium is valid for these atoms, whilst it is invalid for those neighbouring solute atoms or contained within intermetallic precipitates. The work of determining Debye-Waller factors for aluminium and solute atoms in alloys is limited compared to work in the pure metal, however it is not insignificant. Examples include Ti-Al (449 – 452), Al-Cu (453, 454), Al-Si (453), Al-Ge (453), Ni-Al (455, 456) and Al-Co-Ni-Fe-Mn-Rh (457) systems.

The correlation between bond strengths between atoms, and thus mechanical strength, and Debye-Waller factors is highly intuitive (i.e. the stronger the interatomic bonding, the greater the mechanical strength and the smaller the thermal vibration amplitude or Debye-Waller factor). A strong correlation between Young's moduli and Debye-Waller factors has in fact been shown to exist in the limited literature on the subject, which is mainly restricted to elements and simple compounds (458 – 460). This is an area that has not gained much research attention but is well worth exploring in future for more complex systems such as alloy solid solutions and intermetallics.

## Bonding

As stated in the section on bonding in pure aluminium, the electronic structure associated with chemical bonding in all materials is the dominant determinant of all materials properties (except radioactivity). Until now, high-resolution bonding studies have been confined to highly pure, homogeneous and highly ordered crystal structures in relatively simple materials (small unit cells and uniform, stoichiometric compositions throughout the probed material).

The possibility of performing bonding measurements by QCBED in matrix/precipitate/matrix scattering geometries, using the multislice formalism for electron scattering (3), has been alluded to already in the discussion of figure 18 earlier in this section.

At this juncture it can be argued that the ultimate resolution to be attained in the crystallography of any crystalline material is the measurement of interatomic bonds. Therefore, the ultimate level of characterisation that is to be achieved in aluminium and indeed any of its alloys is to be able to map bonding structure as a function of position.

Another challenge in canonically describing the structural evolution of an alloy is to accurately and fully understand the processes of nucleation and growth of precipitate phases from solid solutions and all the physico-chemical driving forces associated with these processes from the most fundamental denominator, namely interatomic bonding.

This chapter concludes with the following conjecture and proposed means of experimentally testing it, that combines the two challenges just described. An electron density domain theory (EDDT) for nucleation and growth of precipitate phases from solid solutions is proposed. This theory is described with the aid of figure 20.

It begins with a supersaturated solid solution where the solute atoms of element A are uniformly dispersed in a matrix of element M. For the sake of simplicity, a binary alloy is considered. Atoms of element A in pure form, have a crystal structure with a bonding electron density distribution which will be correspondingly labelled as type a. Pure element M that forms the host matrix would have a bonding electron distribution of a different form to type a, which shall be called type m in correspondence with matrix M.

In figure 20, the solute atoms, A, are shown as red dots dispersed in the blue matrix atoms M. Bonding electron density of type a is given a red shade, whilst that of type m is shaded blue. When atoms A and M form a precipitate with a particular stoichiometry and structural order, a third type of bonding electron density distribution exists within the precipitate phase.  This shall be called type p and this is given a yellow colour.

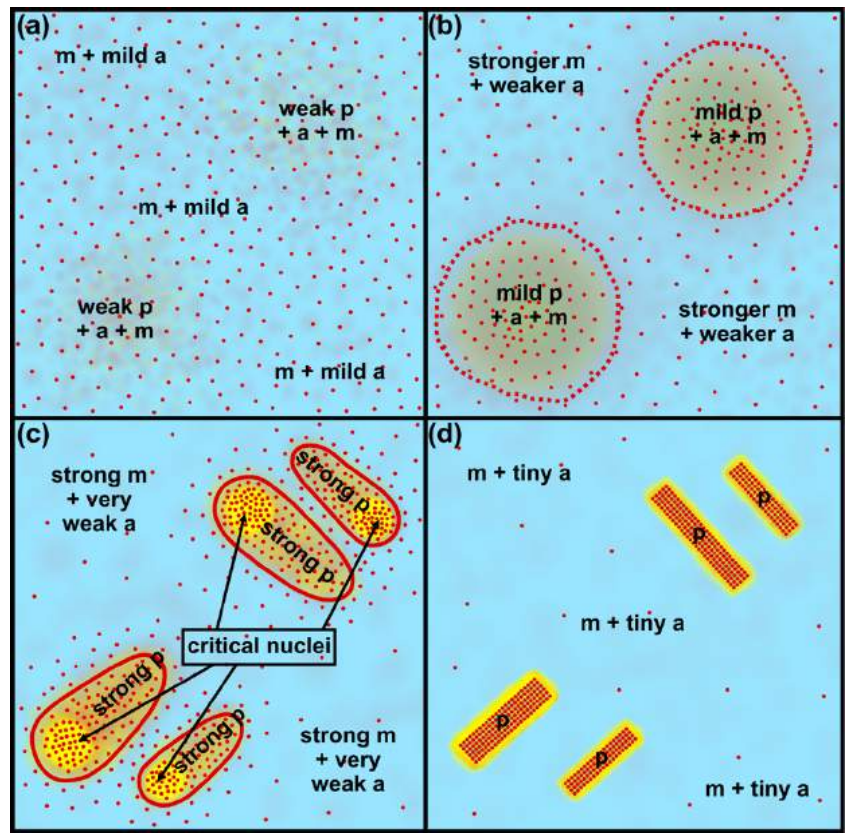

**Figure 20:** A schematic illustration of the electron density domain theory (EDDT) of phase transformation by nucleation and precipitate growth in a binary alloy. (a) A nearly homogeneous solid solution of A atoms (red dots) in a matrix of M atoms (blue). In their pure forms, A and M have bonding electron densities of types "a" (red shading) and "m" (blue shading) respectively. In the solid solution, the superposition of electronic configurations can give rise to other metastable electronic configurations. This figure considers only one such configuration for simplicity, type "p" (shaded in yellow). Slight inhomogeneities present in even the most well-mixed solid solutions, cause very weak segregations between different electronic configurations. (b) Heating causes atoms to vibrate with greater amplitude and this agitates the "electron sea" as well as causing the diffusion of solute atoms, A (red), towards preferred electronic environments. This in turn results in increased segregation into weak domains of different electronic environments (outlined by the dotted lines). (c) Heating continues, causing the domains of type p and m to strengthen via the migration of atoms into their preferred bonding environments. Critical nuclei of a secondary phase stabilised by the type p bonding environment are formed within the strengthening type p domains. (d) The nuclei grow until almost all of the surrounding solute atoms, A (red), are incorporated into the mature precipitates. During this process, domain segregation approaches completion.

In the initial solid solution (figure 20 (a)), bonding type m dominates with some mild type a bonding having a minor influence due to the generally low concentration of atoms A in solution. There will inevitably be regions of slightly higher and slightly lower concentration of atoms A in the solid solution and in those regions of slightly higher concentration, there may be a very weak influence or component of type p bonding electron distribution. Increased mobility due to elevated thermal vibration as the temperature is increased during an ageing heat treatment can cause the bonding electron distribution to fluctuate. This may be enhanced in the presence of dislocations or lattice defects (not included in the figure), which would cause sharp localised bonding perturbations which may help promote bonding electron distribution type p.

In figure 20 (b), the solute atoms A (red) are starting to cluster and be driven towards a structural arrangement with atoms M that further promotes bonding type p, causing weak domains (surrounded by dashed red lines) of type p to form. The influence of this mild type p bonding within these domains is to lower the energy barrier for atoms A and M to form an ordered intermetallic compound. Clustering then accelerates with the increased mobility of solute atoms at the elevated heat treatment temperature. The increasingly strong type p bonding within these domains further accelerates the solid-state chemical reaction that is occurring and provides the driving force underlying the precipitation process. At some point (figure 20 (c)), critical nuclei within the ever-strengthening domains of bonding type p (outlined with solid red lines and having a strong yellow hue) form the basis of the precipitates with bonding type p and stoichiometry of $M_XA_Y$. Within these nuclei the atomic structure that is commensurate with bonding type p and is actually driven to form by bonding type p, is established and stabilized as a secondary phase in the alloy. The reaction is driven to completion by the self-sustaining process that strengthens bonding type p, which provides the driving force for atoms to arrange themselves in the new crystal structure. The reaction stops when the number of solute atoms remaining in the surrounding matrix is too small to sustain the growth of the new phase and, thereby, the extension of the range of influence of bonding type p (figure 20 (d)).

How could such a theory ever be validated? By "scanning QCBED". The basis of such a technique is already commonplace. Every time a STEM image is obtained, a focused electron probe or, in other words a convergent electron beam, is rastered across the area of interest of the specimen and CBED patterns are formed for every probe position spanning that area. STEM detectors integrate over large angular ranges in these CBED patterns to produce a single number, which is the intensity assigned to the corresponding pixel in the STEM image. There is a large push in the electron microscopy community for faster detectors with greater electron detection efficiency and faster readout electronics so that the entire CBED pattern can be collected and stored for every point in a STEM image. This of course requires many terabytes of storage for a single data set! Some, like the groups of Tsuda (461) and Zuo (462 – 465), have managed to collect scanning CBED data sets for areas with small scan dimensions (few pixels in the scan) and have used these data sets to obtain position-sensitive information about subtle changes in crystal structure across domain boundaries. The main focus by the groups of Zuo and Tsuda has been the investigation of polarization domains in ferroelectric materials (461 – 465), but scanning CBED combined with QCBED to produce the technique of scanning QCBED or "SQCBED", could be very powerful for analyzing the spatial variations of both atomic structure and chemical bonding as a function of position in many other materials, including alloys.

Here, it is proposed that scanning CBED be applied to SSSSs that are heat treated in situ so that scanning CBED data sets are collected for the same region of interest as it is undergoing an artificial ageing heat treatment in a TEM hot stage. This is illustrated in figure 21. In the far-left column of figure 21, figure 20 has been rearranged so that artificial ageing time runs from top to bottom. The colour codes of the different bonding types already discussed for figure 20, are shown above this sequence on the far left. The column second from the left shows a schematic of the experimental setup in the TEM. Heating coils in contact with the specimen cup heat the solid solution to the required ageing temperature. As ageing progresses in the manner previously described in the discussion of figure 20, scanning CBED data sets are collected. Circles in each of the images in the left-most column of figure 21, show the probe positions from which each CBED pattern is collected.

After the heat treatment and scanning CBED experiment have been run to completion, the CBED patterns can be pattern matched in the usual way of QCBED for very accurate and precise measurements of the relevant structure factors. The incident beam orientation can be set up from the outset to give CBED patterns that are conducive to measuring different components of the bonding electron distribution. These can be colour-coded with red, green and blue such that each structure factor map acts as a colour channel for the final image that shows the different bonding modes and their distribution within the region of interest. In other words, the Fourier sum is computed from each colour channel as shown.

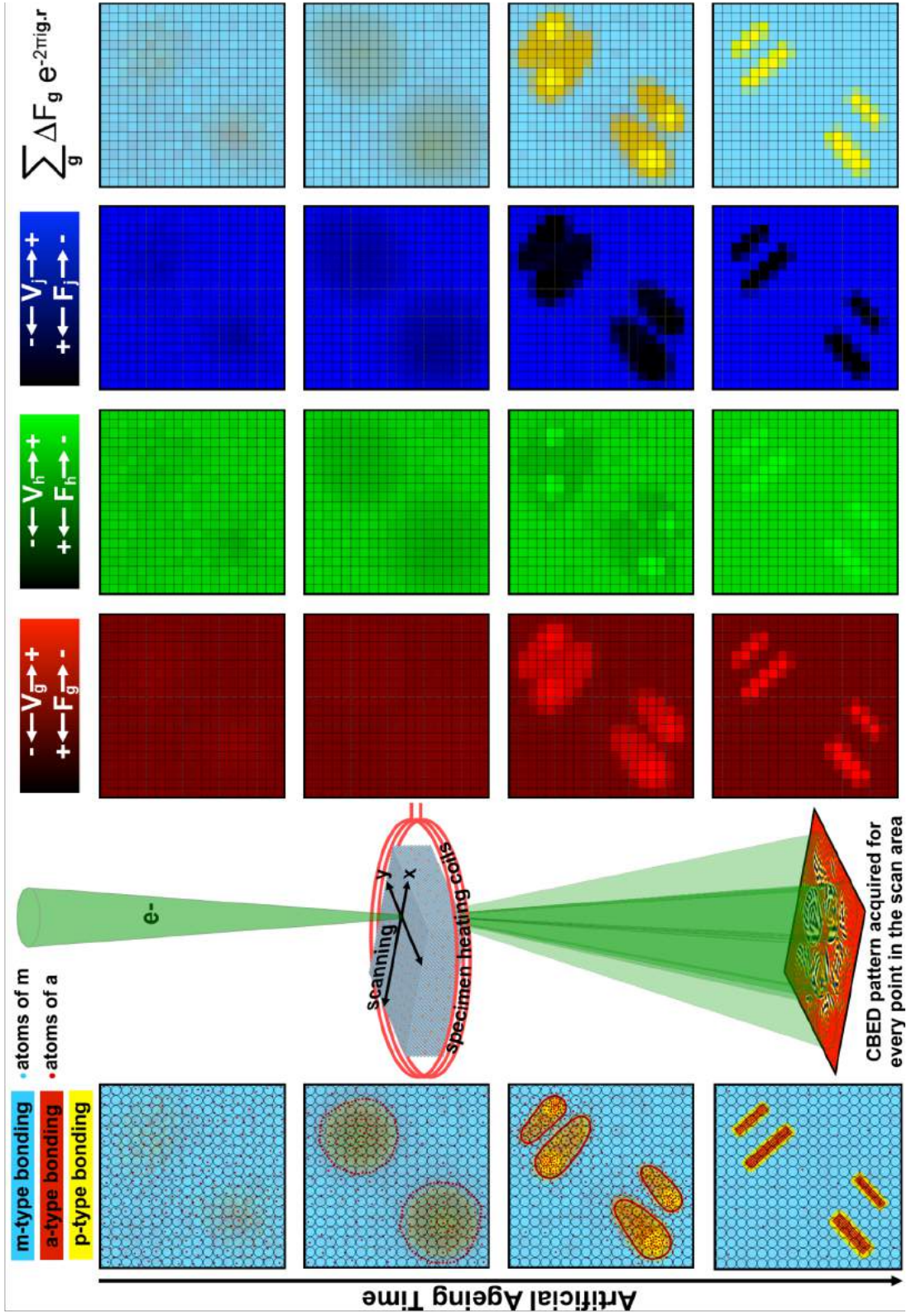

Figure 21: A schematic illustration of how a new technique that combines scanning CBED and QCBED, i.e. "scanning QCBED" or SQCBED, could be key to validating the electron density domain theory (EDDT) of secondary phase nucleation and precipitate growth. The region illustrated in figure 20 is examined here in situ (left column of the present figure) whilst performing a heat treatment that results in the solid solution evolving to form precipitates. Hot stages for transmission electron microscopes are not uncommon and this allows the heat treatment to be carried out whilst simultaneously collecting CBED patterns for each position (circled in the images in the left-most column) in a grid spanning the region of interest. CBED patterns can be collected for an orientation of the incident electron beam which maximises the sensitivity of QCBED to particular structure factors that are most sensitive to the different types of bonding structures that may be present in the specimen. In the present illustration, there are three structure factors being measured from each pattern, $V_g$, $V_h$ and $V_j$. These are structure factors of the electrostatic potential and they are directly related to structure factors of the electron distribution, $F_g$, $F_h$ and $F_j$ respectively, via the Mott formula (194) (note that as the structure factor of the potential increases, the electron distribution structure factor decreases). Each structure factor characterises a different component of the electron distribution. Maps that associate a different colour with each structure factor can be generated from the array of probe positions on the specimen as shown here in RGB format (red for $V_g$ and $F_g$, green for $V_h$ and $F_h$, and blue for $V_j$ and $F_j$). The combination of these components can then be used to map the bonding type distribution as a function of annealing time and position within the specimen (far right column).

In this way, the as yet never attempted technique of scanning QCBED is predicted to be the way ahead in expanding the horizons of crystallography in nano-structured crystalline composites such as aluminium alloys.

# Summary

This chapter has taken a hierarchic approach in describing the crystallography of pure aluminium, before bringing the same considerations to bear in the same order on the discussion of aluminium alloys.

The discussion began with a description of the basic form of the structure of pure aluminium and how it can be described by different unit cells.

This was followed by a review of as many measurements of the lattice parameter of the fcc cell of pure aluminium as could be found, spanning the full range of temperatures where aluminium is a solid. In combination with an analogous review of linear thermal expansion coefficients across the same temperature range, the related parametrizations of $a$(T) and $\alpha$(T) (the lattice parameter as a function of temperature and the linear thermal expansion coefficient as a function of temperature respectively) were developed as canonical and continuous descriptors of $a$(T) and $\alpha$(T) from 0K to 933K (the latter temperature being the melting point of aluminium). These parametrisations are intended as a reference for readers who need to obtain accurate values of these physical properties at any temperature where aluminium is a solid. They are supported by, and serve to summarise, a large number of experimental measurements spanning about a century of work.

In the analysis of linear thermal expansion coefficients at high temperatures approaching the melting point of aluminium, a new approach, named here as the "vacancy triangle", was developed and demonstrated, that allows the vacancy concentration in pure aluminium to be determined with reasonable confidence at temperatures near the melting point of 933K. In summary, a "vacancy triangle" is drawn to span the spread of points on a graph of $\alpha$ against T where each $\alpha$ is determined either from length dilation of a bulk sample (affected by vacancies), or by diffraction (relatively unaffected by vacancies – an approximation in itself). The differences between these two types of measurements as a function of temperature result in a triangular spread of points in a graph of α versus T, and the area spanned by this triangle leads to a measurement of $C_{vac}$, the vacancy concentration at the temperature that bounds the vertical edge of the triangle in the graph. Using this approach, $C_{vac}$ (927K) ≈ (1.1±0.3) x $10^{-3}$ was obtained which agrees within the bounds of uncertainty with previous measurements in the literature.

The third aspect of the crystallography of pure aluminium considered was the Debye-Waller factor, $B$, which quantifies the magnitude of the thermal vibration of the atoms in a crystal. Again, a review of as many published measurements as could be found was undertaken and a parameterization of $B$(T), valid over the range from 0K to 933K, was presented. This is intended as a resource for the reader seeking reliable values of $B$ at any temperature where aluminium is a solid. In the process of this review, the shortcomings of other published models and parametrisations of $B$(T) were discussed and it was made clear that the present parametrization should be used in preference to that by Gao and Peng (159, 160) which diverges strongly from all of the experimental measurements at higher temperatures.

The discussion of the crystallography of pure aluminium culminated in a review of the bonding electron distribution determinations spanning the last 90 years. Determinations of bonding depend very sensitively on an accurate knowledge of the structure of the lattice, lattice parameters at the temperature of the experiments, and the Debye-Waller factor of aluminiun at the relevant temperature, as pre-requisites. This defines the hierarchic approach applied in this chapter, which culminated in determinations of interatomic bonding – the ultimate resolution attainable in crystallography.

In the review of bonding in pure aluminium, it was shown that the technique of quantitative convergent-beam electron diffraction (QCBED) gives the highest resolution and that this technique was pivotal in determining unequivocally that the bonds in pure aluminium are entirely tetrahedral in nature. Before the QCBED study of (175), there was considerable disagreement in the literature as to the location and nature of bonds in elemental aluminium.

The hierarchic approach developed in considering the crystallography of pure aluminium in the first four sections of this chapter, was applied in discussing the crystallography of aluminium alloys in the last section. Research in the domain of aluminium alloys is vast and after highlighting a selection of significant papers in each alloy class and summarizing the array of techniques and theoretical approaches used in these works, the discussion turned to specific examples to illustrate the relevance of the concepts discussed in the treatment of pure aluminium in their application to aluminium alloys.

It was shown that describing the structure of the supporting aluminium matrix with an alternate unit cell can greatly simplify the structural modelling of secondary precipitate phases in relation to the surrounding aluminium matrix crystal structure. This was followed by a brief summary of work on determining lattice parameters and Debye-Waller factors in alloy solid solutions and a discussion of the correlation between Debye-Waller factors and the macroscopic property of elastic (Young's) modulus.

The correlation between the thermal vibration amplitude of atoms and Young's modulus is highly intuitive and rational but very little research has been done in relating these properties. It is suggested here that future work should be directed into the exploration of this nano-macro relationship of physical properties, especially for intermetallic phases and solid solutions. The outcome of such studies would allow mechanical properties of alloys to be used to estimate Debye-Waller factors of the constituent atoms. This would supply missing information about Debye-Waller factors for atoms in different bonding environments, removing a significant obstacle to future planned interatomic bonding studies in alloy systems.

The chapter closed with the unexplored domain of interatomic bonding measurement in aluminium alloys, introducing a new electron density-based theory for explaining the driving force behind nucleation and growth of precipitate phases from solid solutions (applied in this context to aluminium alloys but applicable to all alloys in general). This is the electron density domain theory (EDDT). In summary, it states that solid state precipitation reactions are driven by different types of bonding electron distribution that segregate into domains and strengthen within these domains via the migration of solute atoms into favoured electronic environments and out of less favourable environments. This process is accelerated by increased thermal vibration (elevated temperature) and is a self-enforcing process. I.e. as a domain strengthens in one type of bonding electron distribution, the region surrounding it weakens in that bonding type due to the associated and accelerated migration of atoms to their favoured environments.

In closing, an experimental approach was described for testing the EDDT. The method would combine QCBED with scanning transmission electron microscopy whilst an alloy solid solution is heat treated in situ in an electron microscope. This leaves a vision for the future of crystallographic research in aluminium alloys that would provide fundamental insight into the origins of properties in these extremely important materials.

# Acknowledgements

The author is grateful for funding from the Australian Research Council (FT110100427). The author is indebted to Prof. J. Etheridge, Prof. B.C. Muddle, A/Prof. L. Bourgeois, A/Prof. M. Weyland, Mr. T. Liu, Dr. Y. Guo, Mr. D. Peng at Monash University and Ms J. Shih, formerly at Monash University. The author is grateful to the staff at the Monash Centre for Electron Microscopy (MCEM) and the Department of Materials Science and Engineering at Monash University. Original work included here was mainly carried out on a JEOL 2011 TEM as well as an FEI Titan$^3$ 80-300 TEM with probe and imaging aberration correctors at MCEM, in combination with a liquid helium specimen stage. The author thanks Prof. J.-M. Zuo from the University of Illinois at Urbana-Champaign (UIUC) for long term collaborations and sharing his RefineCB Bloch-wave QCBED code. The author thanks Mr. Y.-T. Shao, also from UIUC. Finally, this work would never have been feasible without the immeasurable help and support of Mrs. K. Nakashima.

# Appendices

### Appendix A
Table of experimental and modelled lattice parameters for the fcc unit cell of pure aluminium from the literature.

### Appendix B
Table of experimental and modelled linear thermal expansion coefficients for pure aluminium from the literature.

### Appendix C
Table of experimentally and theoretically determined Debye-Waller factors for pure aluminium from the literature.

### Appendix D
Table of experimentally and theoretically determined deformation electron densities ($\Delta\rho$) at the tetrahedral, octahedral and bridge centres in the fcc unit cell of pure aluminium. These are followed by the actual structure factor determinations and associated errors (in the cases of experimental measurements) for the bonding-sensitive structure factors, $F_{111}$, $F_{200}$ and $F_{220}$.

# Appendix A

Table of experimental and modelled lattice parameters for the fcc unit cell of pure aluminium from the literature.

| kX to Å conversion factor, from: W.L. Bragg, E. Armstrong Wood, J. Am. Chem. Soc. **69** (1947), 2919 (1.00202) superseded by E.R. Cohen, J.W.M. DuMond, Rev. Mod. Phys. **37** (1965), 537 (1.002080±0.000006) | Unknown or "room" temp is assigned 25°C by default (See A.S. Cooper 1962) | | | Unknown error is assigned 0.01% or 0.001% depending on sig figs | | | | | |
|---|---|---|---|---|---|---|---|---|---|
| Source | T (°C) | T (°K) | *a* (kX) | error (kX) | *a* (Å) | error (Å) | Date | Purity of Al (%) | Notes |
| **EXPERIMENTAL MEASUREMENTS** | | | | | | | | | |
| F.C. Blake, Phys. Rev. **26** (1925), 60. | 25 | 298.15 | 4.04380 | 0.00020 | 4.05221 | 0.00020 | 1925 | 99.97 | From A. Ieviņš, M. Straumanis, Z. Phys. Chem. B **33** (1936), 265. |
| F.K. von Göler, G. Sachs, Metallwirtschaft, Berlin **8** (1929), 671. | 25 | 298.15 | 4.04020 | 0.00040 | 4.04860 | 0.00040 | 1929 | | From A.S. Cooper, Acta Cryst. **15** (1962), 578. |
| M.L.V. Gayler, G.D. Preston, J. Inst. Metals **41** (1929), 193. | 25 | 298.15 | 4.04120 | 0.00040 | 4.04961 | 0.00040 | 1929 | | From A. Ieviņš, M. Straumanis, Z. Phys. Chem. B **33** (1936), 265. |
| G. Wassermann, Z. Metallk. **22** (1930), 158. | 25 | 298.15 | 4.04040 | 0.00040 | 4.04880 | 0.00040 | 1930 | | From A.S. Cooper, Acta Cryst. **15** (1962), 578. |
| Z. Nishiyama, Sci. Rep. Tohoku Imp. Univ. **21** (1932), 364. | 25 | 298.15 | 4.04100 | 0.00040 | 4.04941 | 0.00040 | 1932 | | From A. Phillips, R.M. Brick, J. Franklin Inst. **215** (1933), 557. |
| E.A. Owen, J. Iball, Phil. Mag. **13** (1932), 1020. | 25 | 298.15 | 4.04060 | 0.00030 | 4.04900 | 0.00030 | 1932 | | From A.S. Cooper, Acta Cryst. **15** (1962), 578. |
| E.A. Owen, E.L. Yates, Phil. Mag. **15** (1933), 472. | 18 | 291.15 | 4.04060 | 0.00030 | 4.04900 | 0.00030 | 1933 | 99.6 | From A. Ieviņš, M. Straumanis, Z. Phys. Chem. B **33** (1936), 265. |
| W. Stenzel, J. Weerts, Metallwirtschaft, Berlin **12** (1933), 353. | 20 | 293.15 | 4.04110 | 0.00040 | 4.04951 | 0.00040 | 1933 | 99.9 | From A. Ieviņš, M. Straumanis, Z. Phys. Chem. B **33** (1936), 265. |
| A. Phillips, R.M. Brick, J. Franklin Inst. **215** (1933), 557. | 24 | 297.15 | 4.04180 | 0.00010 | 4.05021 | 0.00010 | 1933 | 99.97 | From A. Ieviņš, M. Straumanis, Z. Phys. Chem. B **33** (1936), 265. |
| M.C. Neuburger, Z. Kristallogr. **86** (1933), 395. | 20 | 293.15 | 4.04020 | 0.00040 | 4.04860 | 0.00040 | 1933 | | From S.S. Lu, Y.L. Chang, Proc. Phys. Soc. **53** (1941), 517. |
| E.R. Jette, F. Foote, J. Chem. Phys. **3** (1935), 605. | 25 | 298.15 | 4.04139 | 0.00008 | 4.04980 | 0.00008 | 1935 | 99.791 | From A. Ieviņš, M. Straumanis, Z. Phys. Chem. B **33** (1936), 265. & A.S. Cooper, Acta Cryst. **15** (1962), 578. & A.J.C. Wilson, Proc. Phys. Soc. **53** (1941), 235. & A. Smakula, J. Kalnajs, Phys. Rev. **99** (1955), 1737. & B.F. Figgins, G.O. Jones, D.P. Riley, Phil. Mag. **1** (1956), 747. |
| A. Ieviņš, M. Straumanis, Z. Phys. Chem. B **33** (1936), 265. | 23.1 | 296.25 | 4.04112 | 0.00003 | 4.04953 | 0.00003 | 1936 | 99.9986 | Film number 539 |
| | 23.1 | 296.25 | 4.04117 | 0.00003 | 4.04958 | 0.00003 | 1936 | 99.9986 | Film number 528 |
| | 23 | 296.15 | 4.04129 | 0.00003 | 4.04970 | 0.00003 | 1936 | 99.9986 | Film number 538 |
| | 23.1 | 296.25 | 4.04128 | 0.00003 | 4.04969 | 0.00003 | 1936 | 99.9986 | Film number 536 |
| | 23.1 | 296.27 | 4.04137 | 0.00003 | 4.04978 | 0.00003 | 1936 | 99.9986 | Film number 530 |
| | 21.6 | 294.75 | 4.04125 | 0.00003 | 4.04966 | 0.00003 | 1936 | 99.9986 | Film number 467 |
| | 23.1 | 296.25 | 4.04123 | 0.00003 | 4.04964 | 0.00003 | 1936 | 99.9986 | Film number 529 |
| | 23.1 | 296.25 | 4.04127 | 0.00003 | 4.04968 | 0.00003 | 1936 | 99.9986 | Film number 542 |
| | 23.1 | 296.25 | 4.04122 | 0.00003 | 4.04963 | 0.00003 | 1936 | 99.9986 | Film number 534 |
| | 22.5 | 295.65 | 4.04116 | 0.00002 | 4.04957 | 0.00002 | 1936 | 99.9986 | Film number 472 |
| | 22.5 | 295.65 | 4.04119 | 0.00002 | 4.04960 | 0.00002 | 1936 | 99.9986 | Film number 475 |
| | 23 | 296.15 | 4.04122 | 0.00002 | 4.04963 | 0.00002 | 1936 | 99.9986 | Film number 494 |
| | 46.8 | 319.95 | 4.04352 | 0.00002 | 4.05193 | 0.00002 | 1936 | 99.9986 | Film number 484 |
| | 47 | 320.15 | 4.04348 | 0.00002 | 4.05189 | 0.00002 | 1936 | 99.9986 | Film number 492 |
| A. Ieviņš, M. Straumanis, Z. Phys. Chem. B **34** (1936), 402. | 25.6 | 298.75 | 4.04149 | 0.00001 | 4.04990 | 0.00001 | 1936 | 99.9986 | Film number 596 |
| | 25.6 | 298.75 | 4.04153 | 0.00001 | 4.04994 | 0.00001 | 1936 | 99.9986 | Film number 597 |
| | 44.3 | 317.45 | 4.04327 | 0.00001 | 4.05168 | 0.00001 | 1936 | 99.9986 | Film number 598 |
| | 25.85 | 299.00 | 4.04154 | 0.00001 | 4.04995 | 0.00001 | 1936 | 99.9986 | Film number 599a |
| | 25.85 | 299.00 | 4.04151 | 0.00001 | 4.04992 | 0.00001 | 1936 | 99.9986 | Film number 599b |
| E.A. Owen, E.L. Yates, Phil. Mag. **21** (1936), 809. | 18 | 291.15 | 4.04060 | 0.00020 | 4.04900 | 0.00020 | 1936 | 99.992 | From S.S. Lu, Y.L. Chang, Proc. Phys. Soc. **53** (1941), 517. |
| C.S. Taylor, L.A. Willey, D.W. Smith, J.D. Edwards, Metals and Alloys **9** (1938), 189. | 26.6 | 299.75 | 4.04150 | 0.00010 | 4.04991 | 0.00010 | 1938 | 99.996 | |
| | 27.6 | 300.75 | 4.04150 | 0.00010 | 4.04991 | 0.00010 | 1938 | 99.996 | |
| | 25 | 298.15 | 4.04130 | 0.00010 | 4.04971 | 0.00010 | 1938 | 99.996 | Inferred from first two measurements |
| H. van Bergen, Ann. d. Phys. **39** (1941), 553. | 25 | 298.15 | 4.04117 | 0.00004 | 4.04958 | 0.00004 | 1941 | | From B.F. Figgins, G.O. Jones, D.P. Riley, Phil. Mag. **1** (1956), 747. |
| H. van Bergen, Ann. d. Phys. **39** (1941), 553. | 20 | 293.15 | 4.04091 | 0.00006 | 4.04932 | 0.00006 | 1941 | | From A.S. Cooper, Acta Cryst. **15** (1962), 578. & A. Smakula, J. Kalnajs, Phys. Rev. **99** (1955), 1737. |
| A.J.C. Wilson, Proc. Phys. Soc. **53** (1941), 235. | 0 | 273.15 | 4.03910 | 0.00010 | 4.04750 | 0.00010 | 1941 | 99.992 | 4 measurements. |
| | 25 | 298.15 | 4.04130 | 0.00010 | 4.04971 | 0.00010 | 1941 | 99.992 | 5 measurements. Also reported in A.S. Cooper, Acta Cryst. **15** (1962), 578. |
| | 100 | 373.15 | 4.04860 | 0.00010 | 4.05702 | 0.00010 | 1941 | 99.992 | 1 measurement. |
| | 150 | 423.15 | 4.05380 | 0.00010 | 4.06223 | 0.00010 | 1941 | 99.992 | 1 measurement. |
| | 200 | 473.15 | 4.05920 | 0.00010 | 4.06764 | 0.00010 | 1941 | 99.992 | 2 measurements. |
| | 300 | 573.15 | 4.07000 | 0.00010 | 4.07847 | 0.00010 | 1941 | 99.992 | 2 measurements. |
| | 400 | 673.15 | 4.08200 | 0.00010 | 4.09049 | 0.00010 | 1941 | 99.992 | 2 measurements. |
| | 500 | 773.15 | 4.09470 | 0.00010 | 4.10322 | 0.00010 | 1941 | 99.992 | 1 measurement. |
| | 600 | 873.15 | 4.10870 | 0.00010 | 4.11725 | 0.00010 | 1941 | 99.992 | 7 measurements. |
| | 650 | 923.15 | 4.11620 | 0.00010 | 4.12476 | 0.00010 | 1941 | 99.992 | 4 measurements. |

| Source | T (°C) | T (°K) | a (kX) | error (kX) | a (Å) | error (Å) | Date | Purity of Al (%) | Notes |
|---|---|---|---|---|---|---|---|---|---|
| kX to Å conversion factor, from: W.L. Bragg, E. Armstrong Wood, J. Am. Chem. Soc. **69** (1947), 2919 (1.00202) superseded by E.R. Cohen, J.W.M. DuMond, Rev. Mod. Phys. **37** (1965), 537 (1.002080±0.000006) | Unknown or "room" temp is assigned 25°C by default (See A.S. Cooper 1962) | | | Unknown error is assigned 0.01% or 0.001% depending on sig figs | | | | | |
| S.S. Lu, Y.L. Chang, Proc. Phys. Soc. **53** (1941), 517. | 20 | 293.15 | 4.04110 | 0.00010 | 4.04951 | 0.00010 | 1941 | | Aluminium Francaise. |
| A.J.C. Wilson, Proc. Phys. Soc. **54** (1942), 487. | 0 | 273.15 | 4.03920 | 0.00010 | 4.04760 | 0.00010 | 1942 | 99.992 | New measurements with annealed Al filings. |
| | 50 | 323.15 | 4.04370 | 0.00010 | 4.05211 | 0.00010 | 1942 | 99.992 | New measurements with annealed Al filings. |
| | 100 | 373.15 | 4.04850 | 0.00010 | 4.05692 | 0.00010 | 1942 | 99.992 | New measurements with annealed Al filings. |
| | 150 | 423.15 | 4.05340 | 0.00010 | 4.06183 | 0.00010 | 1942 | 99.992 | New measurements with annealed Al filings. |
| | 200 | 473.15 | 4.05860 | 0.00010 | 4.06704 | 0.00010 | 1942 | 99.992 | New measurements with annealed Al filings. |
| | 300 | 573.15 | 4.06970 | 0.00010 | 4.07816 | 0.00010 | 1942 | 99.992 | New measurements with annealed Al filings. |
| | 400 | 673.15 | 4.08160 | 0.00010 | 4.09009 | 0.00010 | 1942 | 99.992 | New measurements with annealed Al filings. |
| W. Hume-Rothery, D.J. Strawbridge, J. Sci. Instrum. **24** (1947), 89. | 25 | 298.15 | 4.04140 | 0.00060 | 4.04981 | 0.00060 | 1947 | | |
| | 25 | 298.15 | 4.04140 | 0.00010 | 4.04981 | 0.00010 | 1947 | | |
| | -33.2 | 239.95 | 4.03630 | 0.00040 | 4.04470 | 0.00040 | 1947 | | |
| | -47.6 | 225.55 | 4.03500 | 0.00040 | 4.04339 | 0.00040 | 1947 | | |
| | -97.5 | 175.65 | 4.03060 | 0.00050 | 4.03898 | 0.00050 | 1947 | | |
| H.J. Axon, W. Hume-Rothery, Proc. Roy. Soc. **193** (1948), 1. | 25 | 298.15 | 4.04134 | 0.00004 | 4.04950 | 0.00004 | 1948 | | From A.S. Cooper, Acta Cryst. **15** (1962), 578. |
| E.C. Ellwood, J.M. Silcock, J. Inst. Met. **74** (1948), 457 | 18 | 291.15 | 4.04090 | 0.00020 | 4.04906 | 0.00020 | 1948 | | From A.S. Cooper, Acta Cryst. **15** (1962), 578. |
| E.A. Owen, Y.H. Liu, D.P. Morris, Phil. Mag. **39** (1948), 831. | 18 | 291.15 | 4.04060 | 0.00040 | 4.04876 | 0.00040 | 1948 | | From A.S. Cooper, Acta Cryst. **15** (1962), 578. |
| H.K. Hardy, T.J. Heal, J. Inst. Metals **74** (1948), 721. | 25 | 298.15 | 4.04137 | 0.00004 | 4.04978 | 0.00004 | 1948 | | |
| W. Hume-Rothery, T.H. Boultbee, Phil. Mag. **40** (1949), 71. | 25 | 298.15 | 4.04142 | 0.00004 | 4.04958 | 0.00004 | 1949 | | From A.S. Cooper, Acta Cryst. **15** (1962), 578. |
| M.E. Straumanis, J. Appl. Phys. **20** (1949), 726. | 25.6 | 298.75 | 4.04149 | 0.00003 | 4.04990 | 0.00003 | 1949 | 99.998 | |
| | 25.6 | 298.75 | 4.04153 | 0.00003 | 4.04994 | 0.00003 | 1949 | 99.998 | |
| | 25.85 | 299.00 | 4.04154 | 0.00003 | 4.04995 | 0.00003 | 1949 | 99.998 | |
| | 25.85 | 299.00 | 4.04151 | 0.00003 | 4.04992 | 0.00003 | 1949 | 99.998 | |
| | 44.3 | 317.45 | 4.04327 | 0.00003 | 4.05168 | 0.00003 | 1949 | 99.998 | |
| A. Kochanovska, Physica **15** (1949), 191. | 22 | 295.15 | 4.04100 | 0.00020 | 4.04916 | 0.00020 | 1949 | | |
| | 64 | 337.15 | 4.04490 | 0.00020 | 4.05307 | 0.00020 | 1949 | | |
| | 107 | 380.15 | 4.04880 | 0.00020 | 4.05698 | 0.00020 | 1949 | | |
| | 171 | 444.15 | 4.05450 | 0.00020 | 4.06269 | 0.00020 | 1949 | | |
| | 197 | 470.15 | 4.05670 | 0.00020 | 4.06489 | 0.00020 | 1949 | | |
| | 239 | 512.15 | 4.06080 | 0.00020 | 4.06900 | 0.00020 | 1949 | | |
| | 287 | 560.15 | 4.06540 | 0.00020 | 4.07361 | 0.00020 | 1949 | | |
| | 22 | 295.15 | 4.04100 | 0.00020 | 4.04916 | 0.00020 | 1949 | | |
| | 55 | 328.15 | 4.04430 | 0.00020 | 4.05247 | 0.00020 | 1949 | | |
| | 79 | 352.15 | 4.04600 | 0.00020 | 4.05417 | 0.00020 | 1949 | | |
| | 102 | 375.15 | 4.04840 | 0.00020 | 4.05658 | 0.00020 | 1949 | | |
| | 144 | 417.15 | 4.05240 | 0.00020 | 4.06059 | 0.00020 | 1949 | | |
| | 196 | 469.15 | 4.05710 | 0.00020 | 4.06530 | 0.00020 | 1949 | | |
| | 200 | 473.15 | 4.05740 | 0.00020 | 4.06560 | 0.00020 | 1949 | | |
| | 236 | 509.15 | 4.06080 | 0.00020 | 4.06900 | 0.00020 | 1949 | | |
| | 273 | 546.15 | 4.06490 | 0.00020 | 4.07311 | 0.00020 | 1949 | | |
| | 280 | 553.15 | 4.06570 | 0.00020 | 4.07391 | 0.00020 | 1949 | | |
| | 22 | 295.15 | 4.04120 | 0.00020 | 4.04936 | 0.00020 | 1949 | | |
| | 68 | 341.15 | 4.04550 | 0.00020 | 4.05367 | 0.00020 | 1949 | | |
| | 84 | 357.15 | 4.04720 | 0.00020 | 4.05538 | 0.00020 | 1949 | | |
| | 117 | 390.15 | 4.04980 | 0.00020 | 4.05798 | 0.00020 | 1949 | | |
| | 119 | 392.15 | 4.05010 | 0.00020 | 4.05828 | 0.00020 | 1949 | | |
| | 144 | 417.15 | 4.05250 | 0.00020 | 4.06069 | 0.00020 | 1949 | | |
| | 217 | 490.15 | 4.05920 | 0.00020 | 4.06740 | 0.00020 | 1949 | | |
| | 257 | 530.15 | 4.06180 | 0.00020 | 4.07000 | 0.00020 | 1949 | | |
| J.E. Dorn, P. Pietrokowsky, T.E. Tietz, J. Metals **2** (1950), 933. | 42 | 315.13 | 4.04129 | 0.00020 | 4.04970 | 0.00020 | 1950 | | Read in from graph in J. Bandopadhyay, K.P Gupta, Cryogenics **18** (1978), 54 |
| D.M. Poole, H.J. Axon, J. Inst. Met. **80** (1952), 599. | 25 | 298.15 | 4.04121 | 0.00004 | 4.04937 | 0.00004 | 1952 | | From A.S. Cooper, Acta Cryst. **15** (1962), 578. |
| R.B. Hill, H.J. Axon, Research **6** (1953), 23S. | 25 | 298.15 | 4.04118 | 0.00004 | 4.04934 | 0.00004 | 1953 | | From A.S. Cooper, Acta Cryst. **15** (1962), 578. |
| H.E. Swanson, E. Tatge, Nat. Bur. Stand. Cir. **539** (vol 1) (1953), 11. | 25 | 298.15 | 4.04050 | 0.00040 | 4.04866 | 0.00040 | 1953 | | From A.S. Cooper, Acta Cryst. **15** (1962), 578. |
| A. Smakula, J. Kalnajs, Phys. Rev. **99** (1955), 1737. | 25 | 298.15 | 4.04119 | 0.00002 | 4.04960 | 0.00002 | 1955 | 99.99+ | Also in A. Smakula, J. Kalnajs, V. Sils, Phys. Rev. **99** (1955), 1747. Quoted by W.B. Pearson, Lattice Spacings and Structures of Metals and Alloys, 1st Ed. (Pergamon Press, 1967). |
| B.F. Figgins, G.O. Jones, D.P. Riley, Phil. Mag. **1** (1956), 747. | -252.75 | 20.40 | 4.02349 | 0.00002 | 4.03186 | 0.00002 | 1956 | 99.99 | 5 measurements. |
| | -240.85 | 32.30 | 4.02354 | 0.00002 | 4.03191 | 0.00002 | 1956 | 99.99 | 2 measurements. |
| | -228.75 | 44.40 | 4.02364 | 0.00002 | 4.03201 | 0.00002 | 1956 | 99.99 | 3 measurements. |
| | -218.05 | 55.10 | 4.02382 | 0.00002 | 4.03219 | 0.00002 | 1956 | 99.99 | 2 measurements. |
| | -207.15 | 66.00 | 4.02402 | 0.00002 | 4.03239 | 0.00002 | 1956 | 99.99 | 2 measurements. |
| | -198.15 | 75.00 | 4.02434 | 0.00002 | 4.03271 | 0.00002 | 1956 | 99.99 | 2 measurements. |
| | -187.45 | 85.70 | 4.02477 | 0.00002 | 4.03314 | 0.00002 | 1956 | 99.99 | 7 measurements. |
| | -166.95 | 106.20 | 4.02575 | 0.00002 | 4.03412 | 0.00002 | 1956 | 99.99 | 2 measurements. |
| | -157.95 | 115.20 | 4.02625 | 0.00002 | 4.03462 | 0.00002 | 1956 | 99.99 | 2 measurements. |
| | -148.15 | 125.00 | 4.02690 | 0.00002 | 4.03528 | 0.00002 | 1956 | 99.99 | 2 measurements. |
| | 25.55 | 298.70 | 4.04127 | 0.00002 | 4.04968 | 0.00002 | 1956 | 99.99 | 3 measurements. |
| | 25.01 | 298.16 | 4.04122 | 0.00002 | 4.04963 | 0.00002 | 1956 | 99.99 | Interpolation. |
| | -0.15 | 273.00 | 4.03896 | 0.00002 | 4.04736 | 0.00002 | 1956 | 99.99 | Interpolation. |

| Source | T (°C) | T (°K) | a (kX) | error (kX) | a (Å) | error (Å) | Date | Purity of Al (%) | Notes |
|---|---|---|---|---|---|---|---|---|---|
| kX to Å conversion factor, from: W.L. Bragg, E. Armstrong Wood, J. Am. Chem. Soc. **69** (1947), 2919 (1.00202) superseded by E.R. Cohen, J.W.M. DuMond, Rev. Mod. Phys. **37** (1965), 537 (1.002080±0.000006) | Unknown or "room" temp is assigned 25°C by default (See A.S. Cooper 1962) | | | Unknown error is assigned 0.01% or 0.001% depending on sig figs | | | | | |
| V.V. Zubenko, M.M. Umansky, Kristallografija **1** (1956), 436 | 9.85 | 283.00 | 4.04023 | 0.00040 | 4.04864 | 0.00040 | 1956 | | Read from graph in X.-G. Lu, M. Selleby, B. Sundman, CALPHAD **29** (2005), 68. (converted from molar volume) |
| | 294.85 | 568.00 | 4.06938 | 0.00040 | 4.07785 | 0.00040 | 1956 | | Read from graph in X.-G. Lu, M. Selleby, B. Sundman, CALPHAD **29** (2005), 68. (converted from molar volume) |
| W.B. Pearson, Lattice Spacings and Structures of Metals and Alloys, (Pergamon Press, 1958). | 20 | 293.15 | 4.04071 | 0.00004 | 4.04911 | 0.00004 | 1958 | | From R.O. Simmons and R.W. Balluffi, Phys. Rev. **117** (1960), 52. |
| R. Feder, A.S. Nowick, Phys. Rev. **109** (1958), 1959. | 25 | 298.15 | 4.04107 | 0.00004 | 4.04948 | 0.00004 | 1958 | 99.997 | |
| M.E. Straumanis, J. Appl. Phys. **30** (1959), 1965. | 25 | 298.15 | 4.04113 | 0.00004 | 4.04954 | 0.00004 | 1959 | | Quoted by W.B. Pearson, Lattice Spacings and Structures of Metals and Alloys, 1st Ed. (Pergamon Press, 1967). |
| S. Nenno, J,W, Kauffman, J. Phys. Soc. Jpn. **15** (1960), 220. | 0 | 273.15 | 4.03886 | 0.00004 | 4.04726 | 0.00004 | 1960 | 99.996 | Extrapolated. |
| | 25 | 298.15 | 4.04121 | 0.00004 | 4.04962 | 0.00004 | 1960 | 99.996 | |
| | 96 | 369.15 | 4.04782 | 0.00004 | 4.05624 | 0.00004 | 1960 | 99.996 | |
| | 208 | 481.15 | 4.05958 | 0.00004 | 4.06802 | 0.00004 | 1960 | 99.996 | |
| | 298 | 571.15 | 4.06942 | 0.00004 | 4.07788 | 0.00004 | 1960 | 99.996 | |
| | 383 | 656.15 | 4.07917 | 0.00004 | 4.08765 | 0.00004 | 1960 | 99.996 | |
| | 409 | 682.15 | 4.08246 | 0.00004 | 4.09095 | 0.00004 | 1960 | 99.996 | |
| | 479 | 752.15 | 4.09140 | 0.00004 | 4.09991 | 0.00004 | 1960 | 99.996 | |
| | 564 | 837.15 | 4.10249 | 0.00004 | 4.11102 | 0.00004 | 1960 | 99.996 | |
| | 575 | 848.15 | 4.10420 | 0.00004 | 4.11274 | 0.00004 | 1960 | 99.996 | |
| | 605 | 878.15 | 4.10834 | 0.00004 | 4.11689 | 0.00004 | 1960 | 99.996 | |
| | 612 | 885.15 | 4.10942 | 0.00004 | 4.11797 | 0.00004 | 1960 | 99.996 | |
| | 615 | 888.15 | 4.10974 | 0.00004 | 4.11829 | 0.00004 | 1960 | 99.996 | |
| | 626 | 899.15 | 4.11124 | 0.00004 | 4.11979 | 0.00004 | 1960 | 99.996 | |
| | 630 | 903.15 | 4.11184 | 0.00004 | 4.12039 | 0.00004 | 1960 | 99.996 | |
| | 635 | 908.15 | 4.11253 | 0.00004 | 4.12108 | 0.00004 | 1960 | 99.996 | |
| | 636 | 909.15 | 4.11237 | 0.00004 | 4.12092 | 0.00004 | 1960 | 99.996 | |
| | 641 | 914.15 | 4.11331 | 0.00004 | 4.12187 | 0.00004 | 1960 | 99.996 | |
| | 648 | 921.15 | 4.11455 | 0.00004 | 4.12311 | 0.00004 | 1960 | 99.996 | |
| | 649 | 922.15 | 4.11473 | 0.00004 | 4.12329 | 0.00004 | 1960 | 99.996 | |
| | 651 | 924.15 | 4.11487 | 0.00004 | 4.12343 | 0.00004 | 1960 | 99.996 | |
| M.E. Straumanis, C.H. Cheng, J. Inst. Metals **88** (1960), 287. | 10 | 283.15 | 4.04003 | 0.00004 | 4.04843 | 0.00004 | 1960 | 99.99+ | Perpendicular to Al wire axis |
| | 20 | 293.15 | 4.04096 | 0.00004 | 4.04937 | 0.00004 | 1960 | 99.99+ | |
| | 30 | 303.15 | 4.04193 | 0.00004 | 4.05034 | 0.00004 | 1960 | 99.99+ | |
| | 40 | 313.15 | 4.04272 | 0.00004 | 4.05113 | 0.00004 | 1960 | 99.99+ | |
| | 50 | 323.15 | 4.04381 | 0.00004 | 4.05222 | 0.00004 | 1960 | 99.99+ | |
| | 60 | 333.15 | 4.04462 | 0.00004 | 4.05303 | 0.00004 | 1960 | 99.99+ | |
| | 10 | 283.15 | 4.03983 | 0.00003 | 4.04823 | 0.00003 | 1960 | 99.99+ | Along the axis of the wire |
| | 20 | 293.15 | 4.04075 | 0.00003 | 4.04915 | 0.00003 | 1960 | 99.99+ | |
| | 30 | 303.15 | 4.04174 | 0.00003 | 4.05015 | 0.00003 | 1960 | 99.99+ | |
| | 40 | 313.15 | 4.04264 | 0.00003 | 4.05105 | 0.00003 | 1960 | 99.99+ | |
| | 50 | 323.15 | 4.04358 | 0.00003 | 4.05199 | 0.00003 | 1960 | 99.99+ | |
| | 60 | 333.15 | 4.04439 | 0.00003 | 4.05280 | 0.00003 | 1960 | 99.99+ | |
| M.E. Straumanis, T. Ejima, J. Chem. Phys. **32** (1960), 629. | 25 | 298.15 | 4.04117 | 0.00002 | 4.04958 | 0.00002 | 1960 | 99.9998 | |
| H.M. Otte, J. Appl. Phys. **32** (1961), 1536. | 24 | 297.15 | 4.04085 | 0.00005 | 4.04925 | 0.00005 | 1961 | 99.99 | |
| A.S. Cooper, Acta Cryst. **15** (1962), 578. | 24.8 | 297.95 | 4.04143 | 0.00002 | 4.04959 | 0.00002 | 1962 | 99.9997 | Quoted by W.B. Pearson, Lattice Spacings and Structures of Metals and Alloys, 1st Ed. (Pergamon Press, 1967). |
| M. Simerska, Czech. J. Phys. **12** (1962), 54. | 381 | 654.15 | 4.07950 | 0.00030 | 4.08774 | 0.00030 | 1962 | 99.99 | Read off the graph in the paper |
| | 414 | 687.15 | 4.08400 | 0.00030 | 4.09225 | 0.00030 | 1962 | 99.99 | Read off the graph in the paper |
| | 454 | 727.15 | 4.08850 | 0.00030 | 4.09676 | 0.00030 | 1962 | 99.99 | Read off the graph in the paper |
| | 498 | 771.15 | 4.09470 | 0.00030 | 4.10297 | 0.00030 | 1962 | 99.99 | Read off the graph in the paper |
| | 536 | 809.15 | 4.09950 | 0.00030 | 4.10778 | 0.00030 | 1962 | 99.99 | Read off the graph in the paper |
| | 572 | 845.15 | 4.10380 | 0.00030 | 4.11209 | 0.00030 | 1962 | 99.99 | Read off the graph in the paper |
| | 618 | 891.15 | 4.11100 | 0.00030 | 4.11930 | 0.00030 | 1962 | 99.99 | Read off the graph in the paper |
| H.M. Otte, W.G. Montague, D.O. Welch, J. Appl. Phys. **34** (1963), 3149. | 19.3 | 292.45 | 4.04040 | 0.00000 | 4.04880 | 0.00000 | 1963 | 99.99 | Quoted by Pearson's Handbook of Crystallographic Data for Intermetallic Phases **Vol 1**, 2nd Ed., P. Villar, L.D. Calvert (Eds), ASM International (1991), p648. Also originally in W.B. Pearson, Lattice Spacings and |
| | 24.2 | 297.35 | 4.04086 | 0.00000 | 4.04926 | 0.00000 | 1963 | 99.99 | |
| | 29.3 | 302.45 | 4.04136 | 0.00000 | 4.04977 | 0.00000 | 1963 | 99.99 | |
| | 39.2 | 312.35 | 4.04229 | 0.00000 | 4.05069 | 0.00000 | 1963 | 99.99 | |
| B.W. Delf, Brit. J. Appl. Phys. **14** (1963), 345. | 20 | 293.15 | 4.04056 | 0.00004 | 4.04896 | 0.00004 | 1963 | | From D. King, A.J. Cornish, J. Burke, J. Appl. Phys. **37** (1966), 4717. |
| D.N. Batchelder, R.O. Simmons, J. Appl. Phys. **36** (1965), 2864. | 25.3 | 298.45 | 4.04151 | 0.00007 | 4.04992 | 0.00007 | 1965 | 99.995 | |
| | 25.5 | 298.65 | 4.04147 | 0.00007 | 4.04988 | 0.00007 | 1965 | 99.995 | |
| | 25.5 | 298.65 | 4.04151 | 0.00007 | 4.04992 | 0.00007 | 1965 | 99.995 | |
| | 25.5 | 298.65 | 4.04146 | 0.00007 | 4.04987 | 0.00007 | 1965 | 99.995 | |

| Source | T (°C) | T (°K) | a (kX) | error (kX) | a (Å) | error (Å) | Date | Purity of Al (%) | Notes |
|---|---|---|---|---|---|---|---|---|---|
| kX to Å conversion factor, from: W.L. Bragg, E. Armstrong Wood, J. Am. Chem. Soc. **69** (1947), 2919 (1.00202) superseded by E.R. Cohen, J.W.M. DuMond, Rev. Mod. Phys. **37** (1965), 537 (1.002080±0.000006) | Unknown or "room" temp is assigned 25°C by default (See A.S. Cooper 1962) | | | Unknown error is assigned 0.01% or 0.001% depending on sig figs | | | | | |
| A.J. Cornish, J. Burke, J. Sci. Instrum. **42** (1965), 212. | 20 | 293.15 | 4.04043 | 0.00004 | 4.04883 | 0.00004 | 1965 | 99.995 | |
| | 134.9 | 408.05 | 4.05159 | 0.00004 | 4.06002 | 0.00004 | 1965 | 99.995 | |
| | 199.1 | 472.25 | 4.05834 | 0.00004 | 4.06678 | 0.00004 | 1965 | 99.995 | |
| | 275.5 | 548.65 | 4.06648 | 0.00004 | 4.07494 | 0.00004 | 1965 | 99.995 | |
| | 303.5 | 576.65 | 4.06977 | 0.00004 | 4.07824 | 0.00004 | 1965 | 99.995 | |
| | 376.5 | 649.65 | 4.07820 | 0.00004 | 4.08668 | 0.00004 | 1965 | 99.995 | |
| | 433.8 | 706.95 | 4.08534 | 0.00004 | 4.09384 | 0.00004 | 1965 | 99.995 | |
| | 468.6 | 741.75 | 4.08987 | 0.00004 | 4.09838 | 0.00004 | 1965 | 99.995 | |
| | 474.9 | 748.05 | 4.09051 | 0.00004 | 4.09902 | 0.00004 | 1965 | 99.995 | |
| | 514.4 | 787.55 | 4.09566 | 0.00004 | 4.10418 | 0.00004 | 1965 | 99.995 | |
| | 524.6 | 797.75 | 4.09716 | 0.00004 | 4.10568 | 0.00004 | 1965 | 99.995 | |
| | 575 | 848.15 | 4.10404 | 0.00004 | 4.11258 | 0.00004 | 1965 | 99.995 | |
| | 618.9 | 892.05 | 4.11029 | 0.00004 | 4.11884 | 0.00004 | 1965 | 99.995 | |
| | 634.6 | 907.75 | 4.11263 | 0.00004 | 4.12118 | 0.00004 | 1965 | 99.995 | |
| | 640.9 | 914.05 | 4.11364 | 0.00004 | 4.12220 | 0.00004 | 1965 | 99.995 | |
| | 653.6 | 926.75 | 4.11562 | 0.00004 | 4.12418 | 0.00004 | 1965 | 99.995 | |
| D. King, A.J. Cornish, J. Burke, J. Appl. Phys. **37** (1966), 4717. | 20 | 293.15 | 4.04050 | 0.00010 | 4.04890 | 0.00010 | 1966 | 99.995 | Measured. |
| | 250 | 523.15 | 4.06375 | 0.00010 | 4.07220 | 0.00010 | 1966 | 99.995 | Interpolations from fit to measurements not reported directly. |
| | 300 | 573.15 | 4.06934 | 0.00010 | 4.07780 | 0.00010 | 1966 | 99.995 | Interpolations from fit to measurements not reported directly. |
| | 350 | 623.15 | 4.07492 | 0.00010 | 4.08340 | 0.00010 | 1966 | 99.995 | Interpolations from fit to measurements not reported directly. |
| | 400 | 673.15 | 4.08101 | 0.00010 | 4.08950 | 0.00010 | 1966 | 99.995 | Interpolations from fit to measurements not reported directly. |
| | 450 | 723.15 | 4.08730 | 0.00010 | 4.09580 | 0.00010 | 1966 | 99.995 | Interpolations from fit to measurements not reported directly. |
| | 500 | 773.15 | 4.09369 | 0.00010 | 4.10220 | 0.00010 | 1966 | 99.995 | Interpolations from fit to measurements not reported directly. |
| | 550 | 823.15 | 4.10037 | 0.00010 | 4.10890 | 0.00010 | 1966 | 99.995 | Interpolations from fit to measurements not reported directly. |
| | 600 | 873.15 | 4.10746 | 0.00010 | 4.11600 | 0.00010 | 1966 | 99.995 | Interpolations from fit to measurements not reported directly. |
| | 650 | 923.15 | 4.11504 | 0.00010 | 4.12360 | 0.00010 | 1966 | 99.995 | Interpolations from fit to measurements not reported directly. |
| H. Kiendl, W. Witt, Phys. Lett. **22** (1966), 33. | 23 | 296.15 | 4.04109 | 0.00010 | 4.04950 | 0.00010 | 1966 | | |
| W. Witt, Z. Naturforsch. **22** (1967), 92. | 23 | 296.15 | 4.04109 | 0.00010 | 4.04950 | 0.00010 | 1967 | 99.999 | From electron diffraction. Quoted by Pearson's Handbook of Crystallographic Data for Intermetallic Phases **Vol 1**, 2nd Ed., P. Villar, L.D. Calvert (Eds), ASM International (1991), p648. |
| M.-Y. Adam-Vigneron, F. Jordi, H. Roulet, Comptes Rendus Heb. d. Sci. Acad. Sci. B **269** (1969), 912. | 25 | 298.15 | 4.04191 | 0.00009 | 4.05032 | 0.00009 | 1969 | 99.5 | Electron diffraction |
| | 25 | 298.15 | 4.04132 | 0.00009 | 4.04973 | 0.00009 | 1969 | 99.999 | Electron diffraction |
| | 26 | 299.15 | 4.04200 | 0.00009 | 4.05041 | 0.00009 | 1969 | 99.5 | X-ray diffraction |
| C.L. Woodard, M.E. Straumanis, J. Appl. Cryst. **4** (1971), 201. | -233.15 | 40.00 | 4.02362 | 0.00004 | 4.03199 | 0.00004 | 1971 | 99+ | |
| | -213.15 | 60.00 | 4.02392 | 0.00004 | 4.03229 | 0.00004 | 1971 | 99+ | |
| | -193.15 | 80.00 | 4.02454 | 0.00004 | 4.03291 | 0.00004 | 1971 | 99+ | |
| | -173.15 | 100.00 | 4.02543 | 0.00004 | 4.03380 | 0.00004 | 1971 | 99+ | |
| M.E. Straumanis, C.L. Woodard, Acta Cryst. **A27** (1971), 549. | -273.15 | 0.00 | 4.02348 | 0.00004 | 4.03185 | 0.00004 | 1971 | 99+ | Extrapolated. |
| | -147.65 | 125.50 | 4.02688 | 0.00004 | 4.03526 | 0.00004 | 1971 | 99+ | Read off the graph in the paper |
| B. von Guerard, H. Peisl, R, Zitzmann, Appl. Phys. **3** (1974), 37. | 21 | 294.15 | 4.04268 | 0.00003 | 4.05109 | 0.00003 | 1974 | 99.999 | Fitted value from other measurements - taken as $a_0$. |
| | 21.83 | 294.98 | 4.04280 | 0.00003 | 4.05121 | 0.00003 | 1974 | 99.999 | Measurements made during the heating up of the sample. |
| | 40.85 | 314.00 | 4.04462 | 0.00003 | 4.05303 | 0.00003 | 1974 | 99.999 | |
| | 66.66 | 339.81 | 4.04717 | 0.00003 | 4.05559 | 0.00003 | 1974 | 99.999 | |
| | 104.92 | 378.07 | 4.05081 | 0.00003 | 4.05923 | 0.00003 | 1974 | 99.999 | |
| | 156.75 | 429.90 | 4.05598 | 0.00003 | 4.06442 | 0.00003 | 1974 | 99.999 | |
| | 213.27 | 486.42 | 4.06217 | 0.00003 | 4.07062 | 0.00003 | 1974 | 99.999 | |
| | 281.06 | 554.21 | 4.06880 | 0.00003 | 4.07726 | 0.00003 | 1974 | 99.999 | |
| | 336.93 | 610.08 | 4.07632 | 0.00003 | 4.08480 | 0.00003 | 1974 | 99.999 | |
| | 391.59 | 664.74 | 4.08303 | 0.00003 | 4.09152 | 0.00003 | 1974 | 99.999 | |
| | 438.73 | 711.88 | 4.08885 | 0.00003 | 4.09735 | 0.00003 | 1974 | 99.999 | |
| | 494.07 | 767.22 | 4.09609 | 0.00003 | 4.10460 | 0.00003 | 1974 | 99.999 | |
| | 552.91 | 826.06 | 4.10397 | 0.00003 | 4.11250 | 0.00003 | 1974 | 99.999 | |
| | 553.2 | 826.35 | 4.10405 | 0.00003 | 4.11258 | 0.00003 | 1974 | 99.999 | |
| | 581.36 | 854.51 | 4.10797 | 0.00003 | 4.11652 | 0.00003 | 1974 | 99.999 | |
| | 605.95 | 879.10 | 4.11137 | 0.00003 | 4.11992 | 0.00003 | 1974 | 99.999 | |
| | 633.57 | 906.72 | 4.11541 | 0.00003 | 4.12397 | 0.00003 | 1974 | 99.999 | |
| | 637.46 | 910.61 | 4.11577 | 0.00003 | 4.12433 | 0.00003 | 1974 | 99.999 | |
| | 622.74 | 895.89 | 4.11371 | 0.00003 | 4.12227 | 0.00003 | 1974 | 99.999 | Measurements made during the cooling down of the sample. |
| | 574.21 | 847.36 | 4.10680 | 0.00003 | 4.11534 | 0.00003 | 1974 | 99.999 | |
| | 519.54 | 792.69 | 4.09908 | 0.00003 | 4.10760 | 0.00003 | 1974 | 99.999 | |
| | 467.69 | 740.84 | 4.09237 | 0.00003 | 4.10088 | 0.00003 | 1974 | 99.999 | |
| | 467.02 | 740.17 | 4.09224 | 0.00003 | 4.10076 | 0.00003 | 1974 | 99.999 | |
| | 413.68 | 686.83 | 4.08545 | 0.00003 | 4.09395 | 0.00003 | 1974 | 99.999 | |
| | 361.5 | 634.65 | 4.07927 | 0.00003 | 4.08775 | 0.00003 | 1974 | 99.999 | |
| | 303.95 | 577.10 | 4.07235 | 0.00003 | 4.08083 | 0.00003 | 1974 | 99.999 | |
| | 226.61 | 499.76 | 4.06334 | 0.00003 | 4.07179 | 0.00003 | 1974 | 99.999 | |
| | 158.23 | 431.38 | 4.05586 | 0.00003 | 4.06430 | 0.00003 | 1974 | 99.999 | |
| | 104.1 | 377.25 | 4.05052 | 0.00003 | 4.05895 | 0.00003 | 1974 | 99.999 | |
| | 68.95 | 342.10 | 4.04717 | 0.00003 | 4.05559 | 0.00003 | 1974 | 99.999 | |
| | 44.77 | 317.92 | 4.04482 | 0.00003 | 4.05324 | 0.00003 | 1974 | 99.999 | |
| | 21.71 | 294.86 | 4.04248 | 0.00003 | 4.05089 | 0.00003 | 1974 | 99.999 | |

| | Unknown or "room" temp is assigned 25°C by default (See A.S. Cooper 1962) | | | Unknown error is assigned 0.01% or 0.001% depending on sig figs | | | | |
|---|---|---|---|---|---|---|---|---|---|
| kX to Å conversion factor, from: W.L. Bragg, E. Armstrong Wood, J. Am. Chem. Soc. **69** (1947), 2919 (1.00202) superseded by E.R. Cohen, J.W.M. DuMond, Rev. Mod. Phys. **37** (1965), 537 (1.002080±0.000006) | | | | | | | | | |
| Source | T (°C) | T (°K) | a (kX) | error (kX) | a (Å) | error (Å) | Date | Purity of Al (%) | Notes |
| R. Roberge, J. Less-Common Metals **40** (1975), 161. | -269 | 4.62 | 4.02377 | 0.00060 | 4.03214 | 0.00060 | 1975 | 99+ | Read off the graph in the paper |
| | -234 | 39.60 | 4.02375 | 0.00040 | 4.03212 | 0.00040 | 1975 | 99+ | Read off the graph in the paper |
| | -193 | 79.80 | 4.02471 | 0.00060 | 4.03308 | 0.00060 | 1975 | 99+ | Read off the graph in the paper |
| | -154 | 119.55 | 4.02680 | 0.00060 | 4.03518 | 0.00060 | 1975 | 99+ | Read off the graph in the paper |
| G.K. Bichile, R.G. Kulkarni, J. Appl. Cryst. **10** (1977), 441. | 28 | 301.15 | 4.04149 | 0.00020 | 4.04990 | 0.00020 | 1977 | | |
| | 95 | 368.15 | 4.04778 | 0.00020 | 4.05620 | 0.00020 | 1977 | | |
| | 125 | 398.15 | 4.05057 | 0.00020 | 4.05900 | 0.00020 | 1977 | | |
| | 190 | 463.15 | 4.05676 | 0.00020 | 4.06520 | 0.00020 | 1977 | | |
| | 230 | 503.15 | 4.06125 | 0.00020 | 4.06970 | 0.00020 | 1977 | | |
| J. Bandopadhyay, K.P Gupta, Cryogenics **18** (1978), 54 | -273 | 0.00 | 4.02313 | 0.00020 | 4.03150 | 0.00020 | 1978 | 99.9 | Extrapolated. |
| | -180 | 93.14 | 4.02478 | 0.00020 | 4.03316 | 0.00020 | 1978 | 99.9 | Read off the graph in the paper |
| | -167 | 105.94 | 4.02535 | 0.00020 | 4.03373 | 0.00020 | 1978 | 99.9 | Read off the graph in the paper |
| | -140 | 133.28 | 4.02666 | 0.00020 | 4.03504 | 0.00020 | 1978 | 99.9 | Read off the graph in the paper |
| | -126 | 146.92 | 4.02752 | 0.00020 | 4.03589 | 0.00020 | 1978 | 99.9 | Read off the graph in the paper |
| | -80 | 193.19 | 4.03079 | 0.00020 | 4.03917 | 0.00020 | 1978 | 99.9 | Read off the graph in the paper |
| | 50 | 322.93 | 4.04176 | 0.00020 | 4.05017 | 0.00020 | 1978 | 99.9 | Read off the graph in the paper |
| S.K. Seshadri, D.B. Downie, Met. Sci. **13** (1979), 696. | 27 | 300.00 | 4.04149 | 0.00040 | 4.04989 | 0.00040 | 1979 | | Read from graph in X.-G. Lu, M. Selleby, B. Sundman, CALPHAD **29** (2005), 68. (converted from molar volume) |
| | 65 | 337.89 | 4.04526 | 0.00040 | 4.05367 | 0.00040 | 1979 | | Read from graph in X.-G. Lu, M. Selleby, B. Sundman, CALPHAD **29** (2005), 68. (converted from molar volume) |
| | 140 | 412.97 | 4.05273 | 0.00040 | 4.06116 | 0.00040 | 1979 | | Read from graph in X.-G. Lu, M. Selleby, B. Sundman, CALPHAD **29** (2005), 68. (converted from molar volume) |
| | 275 | 547.66 | 4.06693 | 0.00040 | 4.07539 | 0.00040 | 1979 | | Read from graph in X.-G. Lu, M. Selleby, B. Sundman, CALPHAD **29** (2005), 68. (converted from molar volume) |
| | 350 | 623.64 | 4.07545 | 0.00040 | 4.08393 | 0.00040 | 1979 | | Read from graph in X.-G. Lu, M. Selleby, B. Sundman, CALPHAD **29** (2005), 68. (converted from molar volume) |
| | 420 | 692.86 | 4.08238 | 0.00040 | 4.09087 | 0.00040 | 1979 | | Read from graph in X.-G. Lu, M. Selleby, B. Sundman, CALPHAD **29** (2005), 68. (converted from molar volume) |
| A.K. Giri, G.B. Mitra, J. Phys. D: Appl. Phys. **18** (1985), L75. | -273.15 | 0.00 | 4.02333 | 0.00040 | 4.03170 | 0.00040 | 1985 | | Extrapolated. |
| M. Johnsson, L. Eriksson, Z. Metallkd. **89** (1998), 478. | 49.85 | 323.00 | 4.04249 | 0.00020 | 4.05090 | 0.00020 | 1998 | 99.995 | |
| | 199.85 | 473.00 | 4.05387 | 0.00040 | 4.06230 | 0.00040 | 1998 | 99.995 | |
| | 399.85 | 673.00 | 4.07862 | 0.00090 | 4.08710 | 0.00090 | 1998 | 99.995 | |
| | 621.85 | 895.00 | 4.10576 | 0.00020 | 4.11430 | 0.00020 | 1998 | 99.995 | |
| G. Langelaan, S. Saimoto, Rev. Sci. Inst. **70** (1999), 3413. | 0 | 273.15 | 4.03920 | 0.00020 | 4.04760 | 0.00020 | 1999 | 99.9 | Multi temperature measurements were done but only shown in a plot, yielding a thermal expansion equation. This is the reference value. |
| | 31 | 304.04 | 4.04182 | 0.00020 | 4.05022 | 0.00020 | 1999 | 99.9 | |
| | 31 | 304.04 | 4.04213 | 0.00020 | 4.05054 | 0.00020 | 1999 | 99.9 | |
| | 34 | 307.06 | 4.04195 | 0.00020 | 4.05036 | 0.00020 | 1999 | 99.9 | |
| | 42 | 314.67 | 4.04290 | 0.00020 | 4.05131 | 0.00020 | 1999 | 99.9 | |
| | 50 | 323.28 | 4.04398 | 0.00020 | 4.05239 | 0.00020 | 1999 | 99.9 | |
| | 58 | 331.54 | 4.04393 | 0.00020 | 4.05234 | 0.00020 | 1999 | 99.9 | |
| | 68 | 340.73 | 4.04530 | 0.00020 | 4.05371 | 0.00020 | 1999 | 99.9 | |
| | 76 | 349.38 | 4.04575 | 0.00020 | 4.05417 | 0.00020 | 1999 | 99.9 | |
| | 103 | 376.26 | 4.04880 | 0.00020 | 4.05722 | 0.00020 | 1999 | 99.9 | |
| | 112 | 385.20 | 4.04950 | 0.00020 | 4.05793 | 0.00020 | 1999 | 99.9 | |
| | 121 | 393.88 | 4.05031 | 0.00020 | 4.05873 | 0.00020 | 1999 | 99.9 | |
| | 126 | 398.89 | 4.05108 | 0.00020 | 4.05950 | 0.00020 | 1999 | 99.9 | |
| | 137 | 410.13 | 4.05255 | 0.00020 | 4.06098 | 0.00020 | 1999 | 99.9 | |
| | 146 | 419.23 | 4.05317 | 0.00020 | 4.06160 | 0.00020 | 1999 | 99.9 | |
| | 155 | 428.07 | 4.05335 | 0.00020 | 4.06178 | 0.00020 | 1999 | 99.9 | |
| | 164 | 437.11 | 4.05462 | 0.00020 | 4.06305 | 0.00020 | 1999 | 99.9 | |
| | 173 | 445.98 | 4.05581 | 0.00020 | 4.06425 | 0.00020 | 1999 | 99.9 | |
| | 181 | 454.46 | 4.05629 | 0.00020 | 4.06473 | 0.00020 | 1999 | 99.9 | |
| | 191 | 463.69 | 4.05743 | 0.00020 | 4.06587 | 0.00020 | 1999 | 99.9 | Read off the graph in the paper. Note, two outliers were left out. |
| | 199 | 472.15 | 4.05842 | 0.00020 | 4.06686 | 0.00020 | 1999 | 99.9 | |
| | 207 | 479.98 | 4.05969 | 0.00020 | 4.06813 | 0.00020 | 1999 | 99.9 | |
| | 216 | 488.89 | 4.06031 | 0.00020 | 4.06875 | 0.00020 | 1999 | 99.9 | |
| | 225 | 497.79 | 4.06094 | 0.00020 | 4.06938 | 0.00020 | 1999 | 99.9 | |
| | 233 | 506.31 | 4.06189 | 0.00020 | 4.07034 | 0.00020 | 1999 | 99.9 | |
| | 243 | 516.25 | 4.06311 | 0.00020 | 4.07156 | 0.00020 | 1999 | 99.9 | |
| | 251 | 524.09 | 4.06352 | 0.00020 | 4.07197 | 0.00020 | 1999 | 99.9 | |
| | 259 | 532.18 | 4.06510 | 0.00020 | 4.07356 | 0.00020 | 1999 | 99.9 | |
| | 269 | 542.42 | 4.06577 | 0.00020 | 4.07423 | 0.00020 | 1999 | 99.9 | |
| | 277 | 550.64 | 4.06712 | 0.00020 | 4.07558 | 0.00020 | 1999 | 99.9 | |
| | 286 | 559.32 | 4.06838 | 0.00020 | 4.07685 | 0.00020 | 1999 | 99.9 | |
| | 295 | 568.42 | 4.06857 | 0.00020 | 4.07703 | 0.00020 | 1999 | 99.9 | |
| | 303 | 576.49 | 4.06963 | 0.00020 | 4.07810 | 0.00020 | 1999 | 99.9 | |
| | 312 | 585.10 | 4.07093 | 0.00020 | 4.07940 | 0.00020 | 1999 | 99.9 | |
| | 321 | 594.33 | 4.07239 | 0.00020 | 4.08086 | 0.00020 | 1999 | 99.9 | |
| | 330 | 603.43 | 4.07275 | 0.00020 | 4.08123 | 0.00020 | 1999 | 99.9 | |
| | 340 | 613.50 | 4.07395 | 0.00020 | 4.08242 | 0.00020 | 1999 | 99.9 | |
| | 349 | 621.79 | 4.07489 | 0.00020 | 4.08337 | 0.00020 | 1999 | 99.9 | |
| | 367 | 640.64 | 4.07752 | 0.00020 | 4.08600 | 0.00020 | 1999 | 99.9 | |
| | 385 | 658.35 | 4.07940 | 0.00020 | 4.08789 | 0.00020 | 1999 | 99.9 | |

| Source | T (°C) | T (°K) | a (kX) | error (kX) | a (Å) | error (Å) | Date | Purity of Al (%) | Notes |
|---|---|---|---|---|---|---|---|---|---|
| | 17 | 290.15 | 4.04212 | 0.00040 | 4.05053 | 0.00040 | 2001 | | |
| | 23.5 | 296.65 | 4.04283 | 0.00040 | 4.05124 | 0.00040 | 2001 | | |
| | 35.4 | 308.55 | 4.04434 | 0.00040 | 4.05276 | 0.00040 | 2001 | | |
| | 45.8 | 318.95 | 4.04566 | 0.00040 | 4.05407 | 0.00040 | 2001 | | |
| | 60.4 | 333.55 | 4.04586 | 0.00040 | 4.05428 | 0.00040 | 2001 | | |
| | 72 | 345.15 | 4.04890 | 0.00040 | 4.05732 | 0.00040 | 2001 | | |
| | 104.8 | 377.95 | 4.05205 | 0.00040 | 4.06047 | 0.00040 | 2001 | | |
| | 150.5 | 423.65 | 4.05489 | 0.00040 | 4.06333 | 0.00040 | 2001 | | |
| S. Battaglia, F. Mango, Materials Science Forum **378** (2001), 92. | 24.5 | 297.65 | 4.03224 | 0.00040 | 4.04063 | 0.00040 | 2001 | | Outlier |
| | 3.7 | 276.85 | 4.03325 | 0.00040 | 4.04164 | 0.00040 | 2001 | | Outlier |
| | 47.8 | 320.95 | 4.03486 | 0.00040 | 4.04325 | 0.00040 | 2001 | | Outlier |
| | 67 | 340.15 | 4.03667 | 0.00040 | 4.04507 | 0.00040 | 2001 | | Outlier |
| | 103.3 | 376.45 | 4.04010 | 0.00040 | 4.04850 | 0.00040 | 2001 | | Outlier |
| | 18.4 | 291.55 | 4.03395 | 0.00040 | 4.04234 | 0.00040 | 2001 | | Outlier |
| | 31.7 | 304.85 | 4.03546 | 0.00040 | 4.04386 | 0.00040 | 2001 | | Outlier |
| | 55.3 | 328.45 | 4.03778 | 0.00040 | 4.04618 | 0.00040 | 2001 | | Outlier |
| | 80 | 353.15 | 4.04020 | 0.00040 | 4.04860 | 0.00040 | 2001 | | Outlier |
| | 26 | 299.15 | 4.03516 | 0.00040 | 4.04355 | 0.00040 | 2001 | | Outlier |
| | 45.5 | 318.65 | 4.03697 | 0.00040 | 4.04537 | 0.00040 | 2001 | | Outlier |
| | 78.3 | 351.45 | 4.04010 | 0.00040 | 4.04850 | 0.00040 | 2001 | | Outlier |
| | -262 | 10.99 | 4.02354 | 0.00000 | 4.03191 | 0.00000 | 2013 | 99.5 | Value given in the paper explicitly |
| | -253 | 19.96 | 4.02356 | 0.00020 | 4.03193 | 0.00020 | 2013 | 99.5 | Read off the graph in the paper |
| | -233 | 40.05 | 4.02360 | 0.00020 | 4.03197 | 0.00020 | 2013 | 99.5 | Read off the graph in the paper |
| | -213 | 59.99 | 4.02385 | 0.00020 | 4.03222 | 0.00020 | 2013 | 99.5 | Read off the graph in the paper |
| | -193 | 80.18 | 4.02440 | 0.00020 | 4.03277 | 0.00020 | 2013 | 99.5 | Read off the graph in the paper |
| | -173 | 99.90 | 4.02523 | 0.00020 | 4.03360 | 0.00020 | 2013 | 99.5 | Read off the graph in the paper |
| | -153 | 120.01 | 4.02627 | 0.00020 | 4.03465 | 0.00020 | 2013 | 99.5 | Read off the graph in the paper |
| J. Potter, J.E. Parker, A.R. Lennie, S.P. Thompson, C.C. Tang, J. Appl. Cryst. **46** (2013), 826. | -133 | 139.88 | 4.02752 | 0.00020 | 4.03589 | 0.00020 | 2013 | 99.5 | Read off the graph in the paper |
| | -113 | 160.14 | 4.02893 | 0.00020 | 4.03731 | 0.00020 | 2013 | 99.5 | Read off the graph in the paper |
| | -93 | 180.01 | 4.03048 | 0.00020 | 4.03886 | 0.00020 | 2013 | 99.5 | Read off the graph in the paper |
| | -73 | 199.91 | 4.03216 | 0.00020 | 4.04054 | 0.00020 | 2013 | 99.5 | Read off the graph in the paper |
| | -53 | 219.89 | 4.03389 | 0.00020 | 4.04228 | 0.00020 | 2013 | 99.5 | Read off the graph in the paper |
| | -33 | 240.08 | 4.03569 | 0.00020 | 4.04409 | 0.00020 | 2013 | 99.5 | Read off the graph in the paper |
| | -13 | 260.14 | 4.03747 | 0.00020 | 4.04587 | 0.00020 | 2013 | 99.5 | Read off the graph in the paper |
| | 7 | 280.20 | 4.03921 | 0.00020 | 4.04761 | 0.00020 | 2013 | 99.5 | Read off the graph in the paper |
| | 27 | 299.78 | 4.04106 | 0.00020 | 4.04947 | 0.00020 | 2013 | 99.5 | Read off the graph in the paper |

## MODELLING and THEORY

| Source | T (°C) | T (°K) | a (kX) | a (Å) | Date | Notes |
|---|---|---|---|---|---|---|
| | -223 | 50 | 4.02436 | 4.03273 | 2000 | |
| | -173 | 100 | 4.02599 | 4.03436 | 2000 | |
| | -123 | 150 | 4.02895 | 4.03733 | 2000 | |
| | -73 | 200 | 4.03271 | 4.04110 | 2000 | |
| | -23 | 250 | 4.03695 | 4.04535 | 2000 | |
| | 27 | 300 | 4.04151 | 4.04992 | 2000 | |
| | 77 | 350 | 4.04632 | 4.05474 | 2000 | |
| | 127 | 400 | 4.05133 | 4.05976 | 2000 | |
| | 177 | 450 | 4.05652 | 4.06496 | 2000 | |
| | 227 | 500 | 4.06189 | 4.07034 | 2000 | Real Crystal Model - including crystal imperfections (vacancies). |
| | 277 | 550 | 4.06743 | 4.07589 | 2000 | |
| | 327 | 600 | 4.07315 | 4.08162 | 2000 | |
| | 377 | 650 | 4.07906 | 4.08754 | 2000 | |
| | 427 | 700 | 4.08518 | 4.09368 | 2000 | |
| | 477 | 750 | 4.09159 | 4.10010 | 2000 | |
| | 527 | 800 | 4.09832 | 4.10684 | 2000 | |
| | 577 | 850 | 4.10545 | 4.11399 | 2000 | |
| | 627 | 900 | 4.11308 | 4.12164 | 2000 | |
| | 660 | 933 | 4.11844 | 4.12701 | 2000 | |
| K. Wang, R.R. Reeber, Phil. Mag. A **80** (2000), 1629. | -223 | 50 | 4.02436 | 4.03273 | 2000 | |
| | -173 | 100 | 4.02599 | 4.03436 | 2000 | |
| | -123 | 150 | 4.02895 | 4.03733 | 2000 | |
| | -73 | 200 | 4.03271 | 4.04110 | 2000 | |
| | -23 | 250 | 4.03695 | 4.04535 | 2000 | |
| | 27 | 300 | 4.04151 | 4.04992 | 2000 | |
| | 77 | 350 | 4.04632 | 4.05474 | 2000 | |
| | 127 | 400 | 4.05133 | 4.05976 | 2000 | |
| | 177 | 450 | 4.05652 | 4.06496 | 2000 | |
| | 227 | 500 | 4.06189 | 4.07034 | 2000 | Perfect Crystal Model. |
| | 277 | 550 | 4.06742 | 4.07588 | 2000 | |
| | 327 | 600 | 4.07312 | 4.08159 | 2000 | |
| | 377 | 650 | 4.07898 | 4.08746 | 2000 | |
| | 427 | 700 | 4.08500 | 4.09350 | 2000 | |
| | 477 | 750 | 4.09121 | 4.09972 | 2000 | |
| | 527 | 800 | 4.09760 | 4.10612 | 2000 | |
| | 577 | 850 | 4.10417 | 4.11271 | 2000 | |
| | 627 | 900 | 4.11095 | 4.11950 | 2000 | |
| | 660 | 933 | 4.11554 | 4.12410 | 2000 | |

Notes on header columns:
- kX to Å conversion factor, from: W.L. Bragg, E. Armstrong Wood, J. Am. Chem. Soc. **69** (1947), 2919 (1.00202) superseded by E.R. Cohen, J.W.M. DuMond, Rev. Mod. Phys. **37** (1965), 537 (1.002080±0.000006)
- Unknown or "room" temp is assigned 25°C by default (See A.S. Cooper 1962)
- Unknown error is assigned 0.01% or 0.001% depending on sig figs

## Appendix B

Table of experimental and modelled linear thermal expansion coefficients for pure aluminium from the literature.

| Source | T (°C) | T (°K) | $\Delta L/L_0$ | $\Delta a/a_0$ | $\alpha \times 10^6$ | Date | Purity of Al (%) | Notes |
|---|---|---|---|---|---|---|---|---|
| | | | EXPERIMENTAL MEASUREMENTS | | | | | |
| F. Henning, Ann. d. Phys. **327** (1907), 631. | -191 | 82.15 | -0.003799 | | | 1907 | | $\Delta L/L_0$ relative to +16°C. Corrected to $L$(T). |
| | 16 | 289.15 | 0 | | 21.40 | 1907 | | $\Delta L/L_0$ relative to +16°C. Corrected to $L$(T). |
| | 250 | 523.15 | 0.00572 | | 26.90 | 1907 | | $\Delta L/L_0$ relative to +16°C. Corrected to $L$(T). |
| | 375 | 648.15 | 0.009429 | | 29.44 | 1907 | | $\Delta L/L_0$ relative to +16°C. Corrected to $L$(T). |
| | 500 | 773.15 | 0.013149 | | 30.82 | 1907 | | $\Delta L/L_0$ relative to +16°C. Corrected to $L$(T). |
| | 625 | 898.15 | 0.017239 | | | 1907 | | $\Delta L/L_0$ relative to +16°C. Corrected to $L$(T). |
| P. Hidnert, Sci. Pap. Bur. Stand. Wash. **19** (1925), 497. | 0 | 273.15 | | | 22.60 | 1925 | 99.95 | Obtained from A.J.C. Wilson, Proc. Phys. Soc. **54** (1942), 487 and A. Ieviņš, M. Straumanis, Z. Phys. Chem. B **33** |
| | 100 | 373.15 | | | 24.50 | 1925 | 99.95 | Obtained from A.J.C. Wilson, Proc. Phys. Soc. **54** (1942), 487 and A. Ieviņš, M. Straumanis, Z. Phys. Chem. B **33** |
| | 200 | 473.15 | | | 26.40 | 1925 | 99.95 | Obtained from A.J.C. Wilson, Proc. Phys. Soc. **54** (1942), 487 and A. Ieviņš, M. Straumanis, Z. Phys. Chem. B **33** |
| | 300 | 573.15 | | | 28.30 | 1925 | 99.95 | Obtained from A.J.C. Wilson, Proc. Phys. Soc. **54** (1942), 487 and A. Ieviņš, M. Straumanis, Z. Phys. Chem. B **33** |
| | 400 | 673.15 | | | 30.30 | 1925 | 99.95 | Obtained from A.J.C. Wilson, Proc. Phys. Soc. **54** (1942), 487 and A. Ieviņš, M. Straumanis, Z. Phys. Chem. B **33** |
| | 500 | 773.15 | | | 32.20 | 1925 | 99.95 | Obtained from A.J.C. Wilson, Proc. Phys. Soc. **54** (1942), 487 and A. Ieviņš, M. Straumanis, Z. Phys. Chem. B **33** |
| | 600 | 873.15 | | | 34.10 | 1925 | 99.95 | Obtained from A.J.C. Wilson, Proc. Phys. Soc. **54** (1942), 487 and A. Ieviņš, M. Straumanis, Z. Phys. Chem. B **33** |
| R.M. Buffington, W.M. Latimer, J. Am. Chem. Soc. **48** (1926), 2305. | | 85.00 | | | 8.81 | 1926 | | |
| | | 90.00 | | | 10.20 | 1926 | | |
| | | 100.00 | | | 11.87 | 1926 | | |
| | | 110.00 | | | 13.23 | 1926 | | |
| | | 130.00 | | | 15.39 | 1926 | | |
| | | 150.00 | | | 16.93 | 1926 | | |
| | | 170.00 | | | 18.15 | 1926 | | |
| | | 190.00 | | | 19.17 | 1926 | | |
| | | 210.00 | | | 20.00 | 1926 | | |
| | | 240.00 | | | 21.07 | 1926 | | |
| | | 270.00 | | | 21.99 | 1926 | | |
| | | 297.00 | | | 22.75 | 1926 | | |
| | | 315.00 | | | 23.19 | 1926 | | |
| Staatliches Materialprüfungsamt, Z. Metallkde. **20** (1928), 14. | 25 | 298.15 | | | 23.00 | 1928 | 99.66 | Obtained from A. Ieviņš, M. Straumanis, Z. Phys. Chem. B **33** (1936), 265. |
| H. Ebert, Z. Physik **47** (1928), 719. | 0 | 273.15 | | | 22.50 | 1928 | 99 | Obtained from A. Ieviņš, M. Straumanis, Z. Phys. Chem. B **33** (1936), 265. |
| | 100 | 373.15 | | | 23.40 | 1928 | 99 | Obtained from A. Ieviņš, M. Straumanis, Z. Phys. Chem. B **33** (1936), 265. |
| F.L. Uffelmann, Phil. Mag. **10** (1930), 633. | 100 | 373.15 | | | 23.50 | 1930 | | |
| | 140 | 413.15 | | | 24.30 | 1930 | | |
| | 180 | 453.15 | | | 25.00 | 1930 | | |
| | 220 | 493.15 | | | 25.70 | 1930 | | |
| | 260 | 533.15 | | | 26.50 | 1930 | | |
| | 300 | 573.15 | | | 27.40 | 1930 | | |
| | 340 | 613.15 | | | 28.30 | 1930 | | |
| | 380 | 653.15 | | | 28.80 | 1930 | | |
| | 420 | 693.15 | | | 29.60 | 1930 | | |
| | 460 | 733.15 | | | 30.10 | 1930 | | |
| | 500 | 773.15 | | | 31.10 | 1930 | | |
| | 530 | 803.15 | | | 32.30 | 1930 | | |
| G. Shinoda, Mem. Sci. Kyoto Univ. A **16** (1933), 193. | 60 | 333.15 | | | 22.90 | 1933 | 99.8 | Obtained from A. Ieviņš, M. Straumanis, Z. Phys. Chem. B **33** (1936), 265. |
| F. Bollenrath, Z. Metallkunde **26** (1934), 62. | 0 | 273.15 | | | 22.80 | 1934 | 99.87 | Obtained from A.J.C. Wilson, Proc. Phys. Soc. **54** (1942), 487. |
| | 100 | 373.15 | | | 23.60 | 1934 | 99.87 | Obtained from A.J.C. Wilson, Proc. Phys. Soc. **54** (1942), 487. |
| | 200 | 473.15 | | | 25.00 | 1934 | 99.87 | Obtained from A.J.C. Wilson, Proc. Phys. Soc. **54** (1942), 487. |
| | 300 | 573.15 | | | 26.60 | 1934 | 99.87 | Obtained from A.J.C. Wilson, Proc. Phys. Soc. **54** (1942), 487. |
| | 400 | 673.15 | | | 28.90 | 1934 | 99.87 | Obtained from A.J.C. Wilson, Proc. Phys. Soc. **54** (1942), 487. |
| | 500 | 773.15 | | | 31.50 | 1934 | 99.87 | Obtained from A.J.C. Wilson, Proc. Phys. Soc. **54** (1942), 487. |
| A. Ieviņš, M. Straumanis, Z. Phys. Chem. B **33** (1936), 265. | 24.5 | 297.65 | | | 23.13 | 1936 | 99.9986 | |
| A. Ieviņš, M. Straumanis, Z. Phys. Chem. B **34** (1936), 402. | 25 | 298.15 | | | 23.29 | 1936 | 99.9986 | |

| Source | T (°C) | T (°K) | $\Delta L/L_0$ | $\Delta a/a_0$ | $\alpha \times 10^6$ | Date | Purity of Al (%) | Notes |
|---|---|---|---|---|---|---|---|---|
| C.S. Taylor, L.A. Willey, D.W. Smith, J.D. Edwards, Metals and Alloys **9** (1938), 189. | 0 | 273.15 | | | 23.20 | 1938 | 99.996 | Obtained from A.J.C. Wilson, Proc. Phys. Soc. **54** (1942), 487. |
| | 100 | 373.15 | | | 24.30 | 1938 | 99.996 | Obtained from A.J.C. Wilson, Proc. Phys. Soc. **54** (1942), 487. |
| | 200 | 473.15 | | | 25.90 | 1938 | 99.996 | Obtained from A.J.C. Wilson, Proc. Phys. Soc. **54** (1942), 487. |
| | 300 | 573.15 | | | 27.90 | 1938 | 99.996 | Obtained from A.J.C. Wilson, Proc. Phys. Soc. **54** (1942), 487. |
| | 400 | 673.15 | | | 30.40 | 1938 | 99.996 | Obtained from A.J.C. Wilson, Proc. Phys. Soc. **54** (1942), 487. |
| | 500 | 773.15 | | | 33.40 | 1938 | 99.996 | Obtained from A.J.C. Wilson, Proc. Phys. Soc. **54** (1942), 487. |
| M.E. Straumanis, A. Ieviņš, Z. Anorg. Chem. **238** (1938), 175. | 24 | 297.15 | | | 23.31 | 1938 | 99.9986 | Obtained from H.M. Otte, W.G. Montague, D.O. Welch, J. Appl. Phys. **34** (1963), 3149. |
| H. Esser, H. Eusterbrock, Arch. Eisenhuttenw. **14** (1940-41), 341. | 0 | 273.15 | | | 23.40 | 1940 | 99.87 | Obtained from A.J.C. Wilson, Proc. Phys. Soc. **54** (1942), 487. |
| | 100 | 373.15 | | | 24.50 | 1940 | 99.87 | Obtained from A.J.C. Wilson, Proc. Phys. Soc. **54** (1942), 487. |
| | 200 | 473.15 | | | 25.80 | 1940 | 99.87 | Obtained from A.J.C. Wilson, Proc. Phys. Soc. **54** (1942), 487. |
| | 300 | 573.15 | | | 27.40 | 1940 | 99.87 | Obtained from A.J.C. Wilson, Proc. Phys. Soc. **54** (1942), 487. |
| | 400 | 673.15 | | | 29.20 | 1940 | 99.87 | Obtained from A.J.C. Wilson, Proc. Phys. Soc. **54** (1942), 487. |
| | 500 | 773.15 | | | 31.30 | 1940 | 99.87 | Obtained from A.J.C. Wilson, Proc. Phys. Soc. **54** (1942), 487. |
| | 600 | 873.15 | | | 33.70 | 1940 | 99.87 | Obtained from A.J.C. Wilson, Proc. Phys. Soc. **54** (1942), 487. |
| F.C. Nix, D. MacNair, Phys. Rev. **60** (1941), 597. | | 94.48 | | | 10.75 | 1941 | 99.997 | Read directly from graph ($\Delta L/L_0$ measurements are too noisy) $T_0$=273.15K |
| | | 94.42 | | | 11.41 | 1941 | 99.997 | Read directly from graph ($\Delta L/L_0$ measurements are too noisy) $T_0$=273.15K |
| | | 105.52 | | | 12.27 | 1941 | 99.997 | Read directly from graph ($\Delta L/L_0$ measurements are too noisy) $T_0$=273.15K |
| | | 114.47 | | | 13.53 | 1941 | 99.997 | Read directly from graph ($\Delta L/L_0$ measurements are too noisy) $T_0$=273.15K |
| | | 124.81 | | | 14.75 | 1941 | 99.997 | Read directly from graph ($\Delta L/L_0$ measurements are too noisy) $T_0$=273.15K |
| | | 135.60 | | | 15.58 | 1941 | 99.997 | Read directly from graph ($\Delta L/L_0$ measurements are too noisy) $T_0$=273.15K |
| | | 143.83 | | | 16.51 | 1941 | 99.997 | Read directly from graph ($\Delta L/L_0$ measurements are too noisy) $T_0$=273.15K |
| | | 156.62 | | | 17.48 | 1941 | 99.997 | Read directly from graph ($\Delta L/L_0$ measurements are too noisy) $T_0$=273.15K |
| | | 166.45 | | | 18.10 | 1941 | 99.997 | Read directly from graph ($\Delta L/L_0$ measurements are too noisy) $T_0$=273.15K |
| | | 175.91 | | | 18.74 | 1941 | 99.997 | Read directly from graph ($\Delta L/L_0$ measurements are too noisy) $T_0$=273.15K |
| | | 186.74 | | | 19.31 | 1941 | 99.997 | Read directly from graph ($\Delta L/L_0$ measurements are too noisy) $T_0$=273.15K |
| | | 195.15 | | | 19.83 | 1941 | 99.997 | Read directly from graph ($\Delta L/L_0$ measurements are too noisy) $T_0$=273.15K |
| | | 216.22 | | | 20.80 | 1941 | 99.997 | Read directly from graph ($\Delta L/L_0$ measurements are too noisy) $T_0$=273.15K |
| | | 236.39 | | | 21.83 | 1941 | 99.997 | Read directly from graph ($\Delta L/L_0$ measurements are too noisy) $T_0$=273.15K |
| | | 243.65 | | | 22.09 | 1941 | 99.997 | Read directly from graph ($\Delta L/L_0$ measurements are too noisy) $T_0$=273.15K |
| | | 253.60 | | | 22.39 | 1941 | 99.997 | Read directly from graph ($\Delta L/L_0$ measurements are too noisy) $T_0$=273.15K |
| | | 273.41 | | | 23.00 | 1941 | 99.997 | Read directly from graph ($\Delta L/L_0$ measurements are too noisy) $T_0$=273.15K |
| | | 285.20 | | | 23.31 | 1941 | 99.997 | Read directly from graph ($\Delta L/L_0$ measurements are too noisy) $T_0$=273.15K |
| | | 295.18 | | | 23.53 | 1941 | 99.997 | Read directly from graph ($\Delta L/L_0$ measurements are too noisy) $T_0$=273.15K |
| | | 312.72 | | | 23.87 | 1941 | 99.997 | Read directly from graph ($\Delta L/L_0$ measurements are too noisy) $T_0$=273.15K |
| | | 354.24 | | | 24.69 | 1941 | 99.997 | Read directly from graph ($\Delta L/L_0$ measurements are too noisy) $T_0$=273.15K |
| | | 373.89 | | | 25.17 | 1941 | 99.997 | Read directly from graph ($\Delta L/L_0$ measurements are too noisy) $T_0$=273.15K |
| | | 389.39 | | | 25.26 | 1941 | 99.997 | Read directly from graph ($\Delta L/L_0$ measurements are too noisy) $T_0$=273.15K |
| | | 413.54 | | | 25.99 | 1941 | 99.997 | Read directly from graph ($\Delta L/L_0$ measurements are too noisy) $T_0$=273.15K |
| | | 433.71 | | | 26.23 | 1941 | 99.997 | Read directly from graph ($\Delta L/L_0$ measurements are too noisy) $T_0$=273.15K |
| | | 471.51 | | | 26.61 | 1941 | 99.997 | Read directly from graph ($\Delta L/L_0$ measurements are too noisy) $T_0$=273.15K |
| | | 511.79 | | | 26.84 | 1941 | 99.997 | Read directly from graph ($\Delta L/L_0$ measurements are too noisy) $T_0$=273.15K |
| | | 550.98 | | | 27.46 | 1941 | 99.997 | Read directly from graph ($\Delta L/L_0$ measurements are too noisy) $T_0$=273.15K |
| | | 571.03 | | | 27.68 | 1941 | 99.997 | Read directly from graph ($\Delta L/L_0$ measurements are too noisy) $T_0$=273.15K |
| | | 598.34 | | | 28.09 | 1941 | 99.997 | Read directly from graph ($\Delta L/L_0$ measurements are too noisy) $T_0$=273.15K |
| | | 611.15 | | | 28.27 | 1941 | 99.997 | Read directly from graph ($\Delta L/L_0$ measurements are too noisy) $T_0$=273.15K |
| | | 649.38 | | | 28.81 | 1941 | 99.997 | Read directly from graph ($\Delta L/L_0$ measurements are too noisy) $T_0$=273.15K |

| Source | T (°C) | T (°K) | $\Delta L/L_0$ | $\Delta a/a_0$ | $\alpha \times 10^6$ | Date | Purity of Al (%) | Notes |
|---|---|---|---|---|---|---|---|---|
| A.J.C. Wilson, Proc. Phys. Soc. **53** (1941), 235. | 0 | 273.15 | | | 22.00 | 1941 | 99.992 | |
| | 100 | 373.15 | | | 25.40 | 1941 | 99.992 | |
| | 200 | 473.15 | | | 26.50 | 1941 | 99.992 | |
| | 300 | 573.15 | | | 27.80 | 1941 | 99.992 | |
| | 400 | 673.15 | | | 29.90 | 1941 | 99.992 | |
| | 500 | 773.15 | | | 32.50 | 1941 | 99.992 | |
| | 600 | 873.15 | | | 35.50 | 1941 | 99.992 | |
| | 650 | 923.15 | | | 37.20 | 1941 | 99.992 | |
| A.J.C. Wilson, Proc. Phys. Soc. **54** (1942), 487. | 0 | 273.15 | | | 22.80 | 1942 | 99.992 | |
| | 100 | 373.15 | | | 24.00 | 1942 | 99.992 | |
| | 200 | 473.15 | | | 26.00 | 1942 | 99.992 | |
| | 300 | 573.15 | | | 28.30 | 1942 | 99.992 | |
| | 400 | 673.15 | | | 30.30 | 1942 | 99.992 | |
| | 500 | 773.15 | | | 32.70 | 1942 | 99.992 | |
| | 600 | 873.15 | | | 35.00 | 1942 | 99.992 | |
| M.E. Straumanis, J. Appl. Phys. **20** (1949), 726. | 25 | 298.15 | | | 23.13 | 1949 | 99.998 | |
| J.L. Snoek, Phil. Mag. **41** (1950), 1188. | 20 | 293.15 | | | 23.29 | 1950 | | Obtained from A. Smakula, V. Sils, Phys. Rev. **99** (1955), 1744. |
| D. Bijl, H. Pullan, Physica **21** (1955), 285. | | 40.00 | | | 1.50 | 1955 | 99.994 | |
| | | 50.00 | | | 3.00 | 1955 | 99.994 | |
| | | 60.00 | | | 4.90 | 1955 | 99.994 | |
| | | 70.00 | | | 7.00 | 1955 | 99.994 | |
| | | 80.00 | | | 9.40 | 1955 | 99.994 | |
| | | 90.00 | | | 11.20 | 1955 | 99.994 | |
| | | 110.00 | | | 13.90 | 1955 | 99.994 | |
| | | 130.00 | | | 15.60 | 1955 | 99.994 | |
| | | 150.00 | | | 17.10 | 1955 | 99.994 | |
| | | 170.00 | | | 18.40 | 1955 | 99.994 | |
| | | 190.00 | | | 19.50 | 1955 | 99.994 | |
| | | 210.00 | | | 20.60 | 1955 | 99.994 | |
| | | 230.00 | | | 21.40 | 1955 | 99.994 | |
| | | 250.00 | | | 22.20 | 1955 | 99.994 | |
| | | 270.00 | | | 22.40 | 1955 | 99.994 | |
| M.E. Straumanis, C.H. Cheng, J. Inst. Metals **88** (1960), 287. | 35 | 308.15 | | | 22.80 | 1960 | 99.99+ | |
| S. Nenno, J,W, Kauffman, J. Phys. Soc. Jpn. **15** (1960), 220. | 25 | 298.15 | | | 23.12 | 1960 | 99.996 | |
| | 96 | 369.15 | | | 24.30 | 1960 | 99.996 | |
| | 208 | 481.15 | | | 26.26 | 1960 | 99.996 | |
| | 298 | 571.15 | | | 27.93 | 1960 | 99.996 | |
| | 383 | 656.15 | | | 29.59 | 1960 | 99.996 | |
| | 409 | 682.15 | | | 30.11 | 1960 | 99.996 | |
| | 479 | 752.15 | | | 31.54 | 1960 | 99.996 | |
| | 564 | 837.15 | | | 33.36 | 1960 | 99.996 | |
| | 575 | 848.15 | | | 33.60 | 1960 | 99.996 | Values obtained from fitted polynomial for smoothness. $\Delta a/a_0$ where $a_0$ is at 0°C - results corrected to $a(T)$ |
| | 605 | 878.15 | | | 34.26 | 1960 | 99.996 | |
| | 612 | 885.15 | | | 34.41 | 1960 | 99.996 | |
| | 615 | 888.15 | | | 34.48 | 1960 | 99.996 | |
| | 626 | 899.15 | | | 34.72 | 1960 | 99.996 | |
| | 630 | 903.15 | | | 34.81 | 1960 | 99.996 | |
| | 635 | 908.15 | | | 34.93 | 1960 | 99.996 | |
| | 636 | 909.15 | | | 34.95 | 1960 | 99.996 | |
| | 641 | 914.15 | | | 35.06 | 1960 | 99.996 | |
| | 648 | 921.15 | | | 35.22 | 1960 | 99.996 | |
| | 649 | 922.15 | | | 35.24 | 1960 | 99.996 | |
| | 651 | 924.15 | | | 35.29 | 1960 | 99.996 | |
| R.O. Simmons, R.W. Balluffi, Phys. Rev. **117** (1960), 52. | 225 | 498.15 | 0.00506 | | | 1960 | 99.995 | $\Delta L/L_0$ where $L_0$ is at 20°C - results corrected to $L(T)$ |
| | 250 | 523.15 | 0.00575 | | 27.45 | 1960 | 99.995 | $\Delta L/L_0$ where $L_0$ is at 20°C - results corrected to $L(T)$ |
| | 275 | 548.15 | 0.00644 | | 27.43 | 1960 | 99.995 | $\Delta L/L_0$ where $L_0$ is at 20°C - results corrected to $L(T)$ |
| | 300 | 573.15 | 0.00713 | | 27.80 | 1960 | 99.995 | $\Delta L/L_0$ where $L_0$ is at 20°C - results corrected to $L(T)$ |
| | 325 | 598.15 | 0.00784 | | 28.38 | 1960 | 99.995 | $\Delta L/L_0$ where $L_0$ is at 20°C - results corrected to $L(T)$ |
| | 350 | 623.15 | 0.00856 | | 28.95 | 1960 | 99.995 | $\Delta L/L_0$ where $L_0$ is at 20°C - results corrected to $L(T)$ |
| | 375 | 648.15 | 0.0093 | | 29.52 | 1960 | 99.995 | $\Delta L/L_0$ where $L_0$ is at 20°C - results corrected to $L(T)$ |
| | 400 | 673.15 | 0.01005 | | 30.10 | 1960 | 99.995 | $\Delta L/L_0$ where $L_0$ is at 20°C - results corrected to $L(T)$ |
| | 425 | 698.15 | 0.01082 | | 30.86 | 1960 | 99.995 | $\Delta L/L_0$ where $L_0$ is at 20°C - results corrected to $L(T)$ |
| | 450 | 723.15 | 0.01161 | | 31.23 | 1960 | 99.995 | $\Delta L/L_0$ where $L_0$ is at 20°C - results corrected to $L(T)$ |
| | 475 | 748.15 | 0.0124 | | 32.00 | 1960 | 99.995 | $\Delta L/L_0$ where $L_0$ is at 20°C - results corrected to $L(T)$ |
| | 500 | 773.15 | 0.01323 | | 32.96 | 1960 | 99.995 | $\Delta L/L_0$ where $L_0$ is at 20°C - results corrected to $L(T)$ |
| | 525 | 798.15 | 0.01407 | | 33.52 | 1960 | 99.995 | $\Delta L/L_0$ where $L_0$ is at 20°C - results corrected to $L(T)$ |
| | 550 | 823.15 | 0.01493 | | 34.68 | 1960 | 99.995 | $\Delta L/L_0$ where $L_0$ is at 20°C - results corrected to $L(T)$ |
| | 575 | 848.15 | 0.01583 | | 36.03 | 1960 | 99.995 | $\Delta L/L_0$ where $L_0$ is at 20°C - results corrected to $L(T)$ |
| | 600 | 873.15 | 0.01676 | | 37.18 | 1960 | 99.995 | $\Delta L/L_0$ where $L_0$ is at 20°C - results corrected to $L(T)$ |
| | 625 | 898.15 | 0.01772 | | 38.52 | 1960 | 99.995 | $\Delta L/L_0$ where $L_0$ is at 20°C - results corrected to $L(T)$ |
| | 650 | 923.15 | 0.01872 | | | 1960 | 99.995 | |
| E. Huzan, C.P. Abbiss, G.O. Jones, Phil. Mag. **62** (1961), 277. | | 15.00 | | | 0.12 | 1961 | 99.99 | Volume expansion coefficients given (divided by 3 to give linear coefficient) |
| | | 20.00 | | | 0.33 | 1961 | 99.99 | Volume expansion coefficients given (divided by 3 to give linear coefficient) |
| | | 25.00 | | | 0.62 | 1961 | 99.99 | Volume expansion coefficients given (divided by 3 to give linear coefficient) |
| | | 30.00 | | | 1.10 | 1961 | 99.99 | Volume expansion coefficients given (divided by 3 to give linear coefficient) |
| | | 35.00 | | | 1.62 | 1961 | 99.99 | Volume expansion coefficients given (divided by 3 to give linear coefficient) |
| | | 40.00 | | | 2.25 | 1961 | 99.99 | Volume expansion coefficients given (divided by 3 to give linear coefficient) |
| | | 50.00 | | | 3.87 | 1961 | 99.99 | Volume expansion coefficients given (divided by 3 to give linear coefficient) |
| | | 60.00 | | | 5.65 | 1961 | 99.99 | Volume expansion coefficients given (divided by 3 to give linear coefficient) |
| | | 70.00 | | | 7.43 | 1961 | 99.99 | Volume expansion coefficients given (divided by 3 to give linear coefficient) |
| | | 80.00 | | | 9.13 | 1961 | 99.99 | Volume expansion coefficients given (divided by 3 to give linear coefficient) |
| | | 90.00 | | | 10.80 | 1961 | 99.99 | Volume expansion coefficients given (divided by 3 to give linear coefficient) |
| | | 100.00 | | | 12.43 | 1961 | 99.99 | Volume expansion coefficients given (divided by 3 to give linear coefficient) |

| Source | T (°C) | T (°K) | $\Delta L/L_0$ | $\Delta a/a_0$ | $\alpha \times 10^6$ | Date | Purity of Al (%) | Notes |
|---|---|---|---|---|---|---|---|---|
| H.M. Otte, W.G. Montague, D.O. Welch, J. Appl. Phys. **34** (1963), 3149. | 24 | 297.15 | | | 23.40 | 1963 | 99.99 | |
| D.B. Fraser, A.C. Hollis Hallett, Can. J. Phys. **43** (1965), 193. | | 25.00 | | | 0.62 | 1965 | | Smoothed values from two data collection runs |
| | | 30.00 | | | 1.02 | 1965 | | |
| | | 35.00 | | | 1.57 | 1965 | | |
| | | 40.00 | | | 2.18 | 1965 | | |
| | | 45.00 | | | 2.83 | 1965 | | |
| | | 50.00 | | | 3.60 | 1965 | | |
| | | 60.00 | | | 5.35 | 1965 | | |
| | | 70.00 | | | 7.32 | 1965 | | |
| | | 80.00 | | | 9.20 | 1965 | | |
| A.J. Cornish, J. Burke, J. Sci. Instrum. **42** (1965), 212. | 20 | 293.15 | | 0 | | 1965 | 99.995 | |
| | 134.9 | 408.05 | | 0.0028 | 24.95 | 1965 | 99.995 | $\Delta a/a_0$ where $a_0$ is at 20°C - results corrected to $a$(T) - temp intervals equalised |
| | 199.1 | 472.25 | | 0.0044 | 26.11 | 1965 | 99.995 | $\Delta a/a_0$ where $a_0$ is at 20°C - results corrected to $a$(T) - temp intervals equalised |
| | 275.5 | 548.65 | | 0.0065 | 27.51 | 1965 | 99.995 | $\Delta a/a_0$ where $a_0$ is at 20°C - results corrected to $a$(T) - temp intervals equalised |
| | 303.5 | 576.65 | | 0.0073 | 28.58 | 1965 | 99.995 | $\Delta a/a_0$ where $a_0$ is at 20°C - results corrected to $a$(T) - temp intervals equalised |
| | 376.5 | 649.65 | | 0.0094 | 29.48 | 1965 | 99.995 | $\Delta a/a_0$ where $a_0$ is at 20°C - results corrected to $a$(T) - temp intervals equalised |
| | 433.8 | 706.95 | | 0.0111 | 30.67 | 1965 | 99.995 | $\Delta a/a_0$ where $a_0$ is at 20°C - results corrected to $a$(T) - temp intervals equalised |
| | 468.6 | 741.75 | | 0.0122 | 31.32 | 1965 | 99.995 | $\Delta a/a_0$ where $a_0$ is at 20°C - results corrected to $a$(T) - temp intervals equalised |
| | 474.9 | 748.05 | | 0.0124 | 31.26 | 1965 | 99.995 | $\Delta a/a_0$ where $a_0$ is at 20°C - results corrected to $a$(T) - temp intervals equalised |
| | 514.4 | 787.55 | | 0.0137 | 32.70 | 1965 | 99.995 | $\Delta a/a_0$ where $a_0$ is at 20°C - results corrected to $a$(T) - temp intervals equalised |
| | 524.6 | 797.75 | | 0.0140 | 32.90 | 1965 | 99.995 | $\Delta a/a_0$ where $a_0$ is at 20°C - results corrected to $a$(T) - temp intervals equalised |
| | 575 | 848.15 | | 0.0157 | 33.98 | 1965 | 99.995 | $\Delta a/a_0$ where $a_0$ is at 20°C - results corrected to $a$(T) - temp intervals equalised |
| | 618.9 | 892.05 | | 0.0173 | 36.05 | 1965 | 99.995 | $\Delta a/a_0$ where $a_0$ is at 20°C - results corrected to $a$(T) - temp intervals equalised |
| | 634.6 | 907.75 | | 0.0179 | 37.64 | 1965 | 99.995 | $\Delta a/a_0$ where $a_0$ is at 20°C - results corrected to $a$(T) - temp intervals equalised |
| | 640.9 | 914.05 | | 0.0181 | 38.44 | 1965 | 99.995 | $\Delta a/a_0$ where $a_0$ is at 20°C - results corrected to $a$(T) - temp intervals equalised |
| | 653.6 | 926.75 | | 0.0186 | | 1965 | 99.995 | |
| D. King, A.J. Cornish, J. Burke, J. Appl. Phys. **37** (1966), 4717. | 250 | 523.15 | | 0.0058 | | 1966 | 99.995 | |
| | 300 | 573.15 | | 0.0071 | 27.61 | 1966 | 99.995 | $\Delta a/a_0$ where $a_0$ is at 20°C - results corrected to $a$(T) |
| | 350 | 623.15 | | 0.0085 | 28.75 | 1966 | 99.995 | $\Delta a/a_0$ where $a_0$ is at 20°C - results corrected to $a$(T) |
| | 400 | 673.15 | | 0.0100 | 30.19 | 1966 | 99.995 | $\Delta a/a_0$ where $a_0$ is at 20°C - results corrected to $a$(T) |
| | 450 | 723.15 | | 0.0116 | 31.04 | 1966 | 99.995 | $\Delta a/a_0$ where $a_0$ is at 20°C - results corrected to $a$(T) |
| | 500 | 773.15 | | 0.0132 | 32.07 | 1966 | 99.995 | $\Delta a/a_0$ where $a_0$ is at 20°C - results corrected to $a$(T) |
| | 550 | 823.15 | | 0.0148 | 33.60 | 1966 | 99.995 | $\Delta a/a_0$ where $a_0$ is at 20°C - results corrected to $a$(T) |
| | 600 | 873.15 | | 0.0166 | 35.51 | 1966 | 99.995 | $\Delta a/a_0$ where $a_0$ is at 20°C - results corrected to $a$(T) |
| | 650 | 923.15 | | 0.0184 | | 1966 | 99.995 | |
| G. Bianchi, D. Mallejac, C. Janot, G. Champier, Comptes Rendus Heb. d. Sci. Acad. Sci. B **263** (1966), 1404. | 485 | 758.15 | | 0.0127 | | 1966 | 99.998 | |
| | 514.8 | 787.95 | | 0.0137 | 32.67 | 1966 | 99.998 | $\Delta a/a_0$ where $a_0$ is at 20°C - results corrected to $a$(T) |
| | 538 | 811.15 | | 0.0145 | 33.00 | 1966 | 99.998 | $\Delta a/a_0$ where $a_0$ is at 20°C - results corrected to $a$(T) |
| | 563 | 836.15 | | 0.0153 | 33.73 | 1966 | 99.998 | $\Delta a/a_0$ where $a_0$ is at 20°C - results corrected to $a$(T) |
| | 574.5 | 847.65 | | 0.0157 | 34.48 | 1966 | 99.998 | $\Delta a/a_0$ where $a_0$ is at 20°C - results corrected to $a$(T) |
| | 592.1 | 865.25 | | 0.0163 | 35.57 | 1966 | 99.998 | $\Delta a/a_0$ where $a_0$ is at 20°C - results corrected to $a$(T) |
| | 601.2 | 874.35 | | 0.0166 | 35.84 | 1966 | 99.998 | $\Delta a/a_0$ where $a_0$ is at 20°C - results corrected to $a$(T) |
| | 611 | 884.15 | | 0.0170 | 34.98 | 1966 | 99.998 | $\Delta a/a_0$ where $a_0$ is at 20°C - results corrected to $a$(T) |
| | 621.6 | 894.75 | | 0.0174 | 35.89 | 1966 | 99.998 | $\Delta a/a_0$ where $a_0$ is at 20°C - results corrected to $a$(T) |
| | 630.8 | 903.95 | | 0.0177 | 36.82 | 1966 | 99.998 | $\Delta a/a_0$ where $a_0$ is at 20°C - results corrected to $a$(T) |
| | 643.6 | 916.75 | | 0.0182 | 36.23 | 1966 | 99.998 | $\Delta a/a_0$ where $a_0$ is at 20°C - results corrected to $a$(T) |
| | 652.9 | 926.05 | | 0.0185 | | 1966 | 99.998 | |
| P.D. Pathak, N.G. Vasavada, J. Phys. C **3** (1970), L44. | | 300.00 | | | 23.40 | 1970 | 99.999 | |
| | | 400.00 | | | 25.20 | 1970 | 99.999 | |
| | | 500.00 | | | 26.80 | 1970 | 99.999 | |
| | | 600.00 | | | 28.50 | 1970 | 99.999 | |
| | | 700.00 | | | 30.60 | 1970 | 99.999 | |
| | | 800.00 | | | 33.30 | 1970 | 99.999 | |
| | | 900.00 | | | 37.80 | 1970 | 99.999 | |
| F.G. Awad, D. Gugan, Cryogenics **11** (1971), 414. | | 10.00 | | | 0.03 | 1971 | 99.999 | |
| | | 15.00 | | | 0.13 | 1971 | 99.999 | |
| | | 20.00 | | | 0.26 | 1971 | 99.999 | |
| | | 25.00 | | | 0.47 | 1971 | 99.999 | |
| | | 30.00 | | | 0.87 | 1971 | 99.999 | |
| | | 35.00 | | | 1.44 | 1971 | 99.999 | |
| | | 40.00 | | | 2.15 | 1971 | 99.999 | |
| | | 45.00 | | | 2.99 | 1971 | 99.999 | |
| | | 50.00 | | | 3.90 | 1971 | 99.999 | |
| | | 60.00 | | | 5.68 | 1971 | 99.999 | |
| | | 65.00 | | | 6.56 | 1971 | 99.999 | |
| | | 70.00 | | | 7.46 | 1971 | 99.999 | |
| | | 75.00 | | | 8.33 | 1971 | 99.999 | |

| Source | T (°C) | T (°K) | $\Delta L/L_0$ | $\Delta a/a_0$ | $\alpha \times 10^6$ | Date | Purity of Al (%) | Notes |
|---|---|---|---|---|---|---|---|---|
| M.E. Straumanis, C.L. Woodard, Acta Cryst. A**27** (1971), 549. | | 38.00 | | | 2.45 | 1971 | 99+ | Read from graph of **α** |
| | | 49.00 | | | 4.00 | 1971 | 99+ | Read from graph of **α** |
| | | 59.00 | | | 6.05 | 1971 | 99+ | Read from graph of **α** |
| | | 70.00 | | | 7.75 | 1971 | 99+ | Read from graph of **α** |
| | | 80.00 | | | 9.45 | 1971 | 99+ | Read from graph of **α** |
| | | 110.00 | | | 13.90 | 1971 | 99+ | Read from graph of **α** |
| | | 120.00 | | | 15.10 | 1971 | 99+ | Read from graph of **α** |
| | | 140.00 | | | 16.80 | 1971 | 99+ | Read from graph of **α** |
| | | 160.00 | | | 17.40 | 1971 | 99+ | Read from graph of **α** |
| J.G. Collins, G.K. White, C.A. Swenson, J. Low Temp. Phys. **10** (1973), 69. | | 1.00 | | | 0.00 | 1973 | 99.97+ | |
| | | 2.00 | | | 0.00 | 1973 | 99.97+ | |
| | | 3.00 | | | 0.00 | 1973 | 99.97+ | |
| | | 4.00 | | | 0.01 | 1973 | 99.97+ | |
| | | 5.00 | | | 0.01 | 1973 | 99.97+ | |
| | | 6.00 | | | 0.01 | 1973 | 99.97+ | |
| | | 7.00 | | | 0.02 | 1973 | 99.97+ | |
| | | 8.00 | | | 0.02 | 1973 | 99.97+ | |
| | | 9.00 | | | 0.03 | 1973 | 99.97+ | |
| | | 10.00 | | | 0.04 | 1973 | 99.97+ | |
| | | 12.00 | | | 0.06 | 1973 | 99.97+ | |
| | | 14.00 | | | 0.09 | 1973 | 99.97+ | |
| | | 16.00 | | | 0.12 | 1973 | 99.97+ | |
| | | 18.00 | | | 0.17 | 1973 | 99.97+ | |
| | | 20.00 | | | 0.23 | 1973 | 99.97+ | |
| | | 22.00 | | | 0.31 | 1973 | 99.97+ | |
| | | 24.00 | | | 0.41 | 1973 | 99.97+ | |
| | | 26.00 | | | 0.53 | 1973 | 99.97+ | |
| | | 28.00 | | | 0.67 | 1973 | 99.97+ | |
| | | 30.00 | | | 0.84 | 1973 | 99.97+ | |
| | | 32.00 | | | 1.04 | 1973 | 99.97+ | |
| | | 34.00 | | | 1.25 | 1973 | 99.97+ | |
| B. von Guerard, H. Peisl, R, Zitzmann, Appl. Phys. **3** (1974), 37. | 21.83 | 294.98 | | 0 | | 1974 | 99.999 | |
| | 40.85 | 314.00 | | 0.00045 | 23.64 | 1974 | 99.999 | |
| | 66.66 | 339.81 | | 0.00106 | 23.81 | 1974 | 99.999 | |
| | 104.92 | 378.07 | | 0.00198 | 24.42 | 1974 | 99.999 | |
| | 156.75 | 429.90 | | 0.00327 | 25.46 | 1974 | 99.999 | |
| | 213.27 | 486.42 | | 0.00475 | 26.84 | 1974 | 99.999 | |
| | 281.06 | 554.21 | | 0.00663 | 28.36 | 1974 | 99.999 | Measurements during heating up, Δa/a₀ from smoothed fit of polynomial to expt measurements and where $a_0$ taken at 21°C - Results corrected to a(T). Temperature intervals made more uniform where necessary |
| | 336.93 | 610.08 | | 0.00827 | 29.62 | 1974 | 99.999 | |
| | 391.59 | 664.74 | | 0.00993 | 30.48 | 1974 | 99.999 | |
| | 438.73 | 711.88 | | 0.0114 | 31.32 | 1974 | 99.999 | |
| | 494.07 | 767.22 | | 0.01318 | 32.31 | 1974 | 99.999 | |
| | 552.91 | 826.06 | | 0.01514 | 33.39 | 1974 | 99.999 | |
| | 553.2 | 826.35 | | 0.01515 | 33.95 | 1974 | 99.999 | |
| | 581.36 | 854.51 | | 0.01612 | 34.16 | 1974 | 99.999 | |
| | 605.95 | 879.10 | | 0.01698 | 35.00 | 1974 | 99.999 | |
| | 633.57 | 906.72 | | 0.01798 | 35.47 | 1974 | 99.999 | |
| | 637.46 | 910.61 | | 0.01812 | 35.70 | 1974 | 99.999 | |
| | 622.74 | 895.89 | | 0.01758 | 35.34 | 1974 | 99.999 | |
| | 574.21 | 847.36 | | 0.01587 | 34.00 | 1974 | 99.999 | |
| | 519.54 | 792.69 | | 0.01402 | 32.76 | 1974 | 99.999 | |
| | 467.69 | 740.84 | | 0.01233 | 31.83 | 1974 | 99.999 | |
| | 467.02 | 740.17 | | 0.01231 | 31.82 | 1974 | 99.999 | Measurements during cooling down, Δa/a₀ from smoothed fit of polynomial to expt measurements and where $a_0$ taken at 21°C - Results corrected to a(T). Temperature intervals made more uniform where necessary |
| | 413.68 | 686.83 | | 0.01061 | 30.94 | 1974 | 99.999 | |
| | 361.5 | 634.65 | | 0.00901 | 30.01 | 1974 | 99.999 | |
| | 303.95 | 577.10 | | 0.00729 | 28.83 | 1974 | 99.999 | |
| | 226.61 | 499.76 | | 0.00511 | 27.12 | 1974 | 99.999 | |
| | 158.23 | 431.38 | | 0.00331 | 25.55 | 1974 | 99.999 | |
| | 104.1 | 377.25 | | 0.00196 | 24.37 | 1974 | 99.999 | |
| | 68.95 | 342.10 | | 0.00112 | 23.91 | 1974 | 99.999 | |
| | 44.77 | 317.92 | | 0.00054 | 23.69 | 1974 | 99.999 | |
| | 21.71 | 294.86 | | 0 | | 1974 | 99.999 | |

| Source | T (°C) | T (°K) | $\Delta L/L_0$ | $\Delta a/a_0$ | $\alpha \times 10^6$ | Date | Purity of Al (%) | Notes |
|---|---|---|---|---|---|---|---|---|
| F.R. Kroeger, C.A. Swenson, J. Appl. Phys. **48** (1977), 853. | | 2.00 | | | 0.00 | 1977 | 99.97+ | |
| | | 4.00 | | | 0.01 | 1977 | 99.97+ | |
| | | 6.00 | | | 0.01 | 1977 | 99.97+ | |
| | | 8.00 | | | 0.02 | 1977 | 99.97+ | |
| | | 10.00 | | | 0.04 | 1977 | 99.97+ | |
| | | 12.00 | | | 0.06 | 1977 | 99.97+ | |
| | | 14.00 | | | 0.09 | 1977 | 99.97+ | |
| | | 16.00 | | | 0.12 | 1977 | 99.97+ | |
| | | 18.00 | | | 0.17 | 1977 | 99.97+ | |
| | | 20.00 | | | 0.23 | 1977 | 99.97+ | |
| | | 25.00 | | | 0.46 | 1977 | 99.97+ | |
| | | 30.00 | | | 0.84 | 1977 | 99.97+ | |
| | | 35.00 | | | 1.36 | 1977 | 99.97+ | |
| | | 40.00 | | | 2.03 | 1977 | 99.97+ | |
| | | 45.00 | | | 2.81 | 1977 | 99.97+ | |
| | | 50.00 | | | 3.66 | 1977 | 99.97+ | |
| | | 60.00 | | | 5.50 | 1977 | 99.97+ | |
| | | 70.00 | | | 7.35 | 1977 | 99.97+ | |
| | | 80.00 | | | 9.11 | 1977 | 99.97+ | |
| | | 90.00 | | | 10.71 | 1977 | 99.97+ | |
| | | 100.00 | | | 12.16 | 1977 | 99.97+ | |
| | | 120.00 | | | 14.55 | 1977 | 99.97+ | |
| | | 140.00 | | | 16.42 | 1977 | 99.97+ | |
| | | 160.00 | | | 17.88 | 1977 | 99.97+ | |
| | | 180.00 | | | 19.05 | 1977 | 99.97+ | |
| | | 200.00 | | | 20.01 | 1977 | 99.97+ | |
| | | 220.00 | | | 20.80 | 1977 | 99.97+ | |
| | | 240.00 | | | 21.48 | 1977 | 99.97+ | |
| | | 260.00 | | | 22.05 | 1977 | 99.97+ | |
| | | 273.15 | | | 22.40 | 1977 | 99.97+ | |
| | | 280.00 | | | 22.56 | 1977 | 99.97+ | |
| | | 293.15 | | | 22.87 | 1977 | 99.97+ | |
| | | 300.00 | | | 23.02 | 1977 | 99.97+ | |
| | | 320.00 | | | 23.47 | 1977 | 99.97+ | |
| S. Battaglia, F. Mango, Materials Science Forum **378** (2001), 92. | 25 | 298.15 | | | 23.44 | 2001 | | Only 1 result given here as others appear to be associated with significant systematic errors. |

## MODELLING and THEORY

| Source | T (°C) | T (°K) | $\Delta L/L_0$ | $\Delta a/a_0$ | $\alpha \times 10^6$ | Date | Purity of Al (%) | Notes |
|---|---|---|---|---|---|---|---|---|
| K. Wang, R.R. Reeber, Phil. Mag. A **80** (2000), 1629. | -223 | 50 | | | 3.97 | 2000 | | |
| | -173 | 100 | | | 11.9 | 2000 | | |
| | -123 | 150 | | | 17.04 | 2000 | | |
| | -73 | 200 | | | 20.03 | 2000 | | |
| | -23 | 250 | | | 21.91 | 2000 | | |
| | 27 | 300 | | | 23.26 | 2000 | | |
| | 77 | 350 | | | 24.28 | 2000 | | |
| | 127 | 400 | | | 25.19 | 2000 | | |
| | 177 | 450 | | | 26.03 | 2000 | | Real Crystal Model - including crystal imperfections (vacancies). |
| | 227 | 500 | | | 26.84 | 2000 | | |
| | 277 | 550 | | | 27.65 | 2000 | | |
| | 327 | 600 | | | 28.51 | 2000 | | |
| | 377 | 650 | | | 29.49 | 2000 | | |
| | 427 | 700 | | | 30.64 | 2000 | | |
| | 477 | 750 | | | 32.04 | 2000 | | |
| | 527 | 800 | | | 33.77 | 2000 | | |
| | 577 | 850 | | | 35.89 | 2000 | | |
| | 627 | 900 | | | 38.46 | 2000 | | |
| | 660 | 933 | | | 40.43 | 2000 | | |
| | -223 | 50 | | | 3.97 | 2000 | | |
| | -173 | 100 | | | 11.9 | 2000 | | |
| | -123 | 150 | | | 17.04 | 2000 | | |
| | -73 | 200 | | | 20.03 | 2000 | | |
| | -23 | 250 | | | 21.91 | 2000 | | |
| | 27 | 300 | | | 23.26 | 2000 | | |
| | 77 | 350 | | | 24.28 | 2000 | | |
| | 127 | 400 | | | 25.19 | 2000 | | |
| | 177 | 450 | | | 26.03 | 2000 | | |
| | 227 | 500 | | | 26.82 | 2000 | | Perfect Crystal Model. |
| | 277 | 550 | | | 27.6 | 2000 | | |
| | 327 | 600 | | | 28.37 | 2000 | | |
| | 377 | 650 | | | 29.15 | 2000 | | |
| | 427 | 700 | | | 29.94 | 2000 | | |
| | 477 | 750 | | | 30.77 | 2000 | | |
| | 527 | 800 | | | 31.63 | 2000 | | |
| | 577 | 850 | | | 32.53 | 2000 | | |
| | 627 | 900 | | | 33.47 | 2000 | | |
| | 660 | 933 | | | 34.13 | 2000 | | |

## Appendix C

Table of experimentally and theoretically determined Debye-Waller factors for pure aluminium from the literature.

| Source | T (K) | Error in T (where not given, ±1K assumed) | B (Å$^2$) | Error in B (where not given, ±0.01 is assumed) | Notes |
|---|---|---|---|---|---|
| | | | EXPERIMENTS | | |
| R.W. James, G.W. Brindley, R.G. Wood, *Proc. Roy. Soc. Lond. A* **125** (1929), 401. | 290 | 1 | 0.77 | 0.01 | Single crystal X-ray diffraction |
| | 86 | 1 | 0.32 | 0.01 | Single crystal X-ray diffraction |
| G.W. Brindley, *Philos. Mag.* **21** (1936), 778 | 293 | 1 | 0.74 | 0.01 | Taken from N.N. Sirota, *Acta Cryst. A* **25** (1969), 223. |
| | 293 | 1 | 0.85 | 0.01 | Taken from N.N. Sirota, *Acta Cryst. A* **25** (1969), 223. |
| E.A. Owen, R.W. Williams, *Proc. Roy. Soc. Lond. A* **188** (1947), 509. | 293 | 1 | 0.84 | 0.01 | Taken from R.E. Dingle, E.H. Medlin, Acta Cryst. A**28** (1972), 22. Powder X-ray diffraction |
| | 426 | 1 | 1.28 | 0.01 | |
| | 543 | 1 | 1.70 | 0.01 | |
| | 617 | 1 | 1.87 | 0.01 | |
| | 688 | 1 | 2.12 | 0.01 | Read off graph in J. Prakash, M.P. Hemkar, *J. Phys. Soc. Jpn* **34** (1973), 1583. |
| | 757 | 1 | 2.73 | 0.01 | |
| | 823 | 1 | 3.10 | 0.01 | |
| | 909 | 1 | 3.65 | 0.01 | |
| N.V. Ageev, D.L. Ageeva, *Izv. Akad. Nauk SSSR, Otd. Khim. Nauk* **1** (1948), 17. | 293 | 1 | 0.77 | 0.01 | Scaled to R.W. James, G.W. Brindley, R.G. Wood, Proc. Roy. Soc. Lond. A **125** (1929), 401. Taken from N.N. Sirota, Acta Cryst. A **25** (1969), 223. |
| H. Bensch, H. Witte, E. Wölfel, *Z. Phys. Chem.* **4** (1955), 65. | 293 | 1 | 0.78 | 0.01 | Single crystal X-ray diffraction |
| | 293 | 1 | 0.82 | 0.01 | Single crystal X-ray diffraction |
| C.B. Walker, *Phys. Rev.* **103** (1956), 547. | 4 | 1 | 0.27 | 0.01 | C(T), scaled to zero point B(T) using 4K data point. Taken from R.E. DeWames, T. Wolfram, G.W. Lehman, Phys. Rev. **131** (1963), 528. |
| | 20 | 1 | 0.28 | 0.01 | C(T), scaled to zero point B(T) using 4K data point. Taken from R.E. DeWames, T. Wolfram, G.W. Lehman, Phys. Rev. **131** (1963), 528. |
| | 80 | 1 | 0.35 | 0.01 | C(T), scaled to zero point B(T) using 4K data point. Taken from R.E. DeWames, T. Wolfram, G.W. Lehman, Phys. Rev. **131** (1963), 528. NB Scaling done against Gao and Peng's fit to Gilat and Nicklow, claimed by E. Rantavuori, V.-P. Tanninen, *Physica Scr.* **15** (1977), 273 to be too high. |
| | 300 | 1 | 0.89 | 0.01 | C(T), scaled to zero point B(T) using 4K data point. Taken from R.E. DeWames, T. Wolfram, G.W. Lehman, Phys. Rev. **131** (1963), 528. |
| | 400 | 1 | 1.17 | 0.01 | C(T), scaled to zero point B(T) using 4K data point. Taken from R.E. DeWames, T. Wolfram, G.W. Lehman, Phys. Rev. **131** (1963), 528. |
| D.R. Chipman, *J. Appl. Phys.* **31** (1960), 2012. | 293 | 1 | 0.87 | 0.01 | Taken from R.E. Dingle, E.H. Medlin, Acta Cryst. A**28** (1972), 22. Powder X-ray diffraction |
| | 293 | 1 | 0.79 | 0.01 | Taken from R.E. Dingle, E.H. Medlin, Acta Cryst. A**28** (1972), 22. Powder X-ray diffraction |
| | 68 | 1 | 0.30 | 0.01 | |
| | 300 | 1 | 0.90 | 0.01 | |
| | 403 | 1 | 1.29 | 0.01 | |
| | 444 | 1 | 1.48 | 0.01 | |
| | 508 | 1 | 1.69 | 0.01 | Read off graph in J. Prakash, M.P. Hemkar, *J. Phys. Soc. Jpn* **34** (1973), 1583. |
| | 641 | 1 | 2.31 | 0.01 | |
| | 711 | 1 | 2.65 | 0.01 | |
| | 753 | 1 | 2.97 | 0.01 | |
| | 863 | 1 | 3.68 | 0.01 | |
| | 877 | 1 | 3.68 | 0.01 | |
| B.W. Batterman, D.R. Chipman, J.J. DeMarco, *Phys. Rev.* **122** (1961), 68. | 293 | 1 | 0.85 | 0.01 | Taken from A.G. Fox, M.A. Tabbernor, R.M. Fisher, *J. Phys. Chem. Solids* **51** (1990), 1323. |
| P.A. Flinn, G.M. McManus, *Phys. Rev.* **132** (1963), 2458. | 293 | 1 | 0.79 | 0.01 | Taken from R.E. Dingle, E.H. Medlin, Acta Cryst. A**28** (1972), 22. Single crystal X-ray diffraction |
| | 95 | 1 | 0.32 | 0.01 | |
| | 195 | 1 | 0.57 | 0.01 | Read off graph in H.L. Kharoo, O.P. Gupta, M.P. Hemkar, *Z. Naturforsch.* **32a** (1977), 570. |
| | 310 | 1 | 0.85 | 0.01 | |
| | 395 | 1 | 1.11 | 0.01 | |
| N. Mothersole, E.A. Owen, *Brit. J. Appl. Phys.* **16** (1965), 1113. | 293 | 1 | 0.84 | 0.01 | Taken from R.E. Dingle, E.H. Medlin, Acta Cryst. A**28** (1972), 22. Powder X-ray diffraction |
| R. Stedman, G. Nilsson, *Inelastic Scattering of Neutrons in Solids and Liquids* **Vol. 1** (IAEA: Vienna, 1965) and G. Gilat, R. Nicklow, Phys. Rev. **143** (1966), 487 | 80 | 1 | 0.32 | 0.01 | Taken from R.C.G. Killean, *J. Phys. F: Metal Phys.* **4** (1974), 1908. Claimed too high in experimental X-ray diffraction analysis by E. Rantavuori, V.-P. Tanninen, *Physica Scr.* **15** (1977), 273 |
| | 300 | 1 | 0.89 | 0.01 | Taken from R.C.G. Killean, *J. Phys. F: Metal Phys.* **4** (1974), 1908. |
| R.M. Nicklow, R.A. Young, *Phys. Rev.* **152** (1966), 591. | 80 | 1 | 0.33 | 0.01 | |
| | 130 | 1 | 0.42 | 0.01 | |
| | 180 | 1 | 0.53 | 0.01 | Obtained from intensity data in their paper |
| | 230 | 1 | 0.65 | 0.01 | |
| | 280 | 1 | 0.77 | 0.01 | |
| | 330 | 1 | 0.90 | 0.01 | |
| J.J. De Marco, *Philos. Mag.* **15** (1967), 483. | 293 | 1 | 0.88 | 0.01 | Single crystal X-ray diffraction |

| Source | T (K) | Error in T (where not given, ±1K assumed) | $B$ (Å$^2$) | Error in $B$ (where not given, ±0.01 is assumed) | Notes |
|---|---|---|---|---|---|
| | 80 | 1 | 0.32 | 0.01 | From PDS measurements by G. Gilat, R. Nicklow, Phys. Rev. **143** (1966), 487. |
| | 300 | 1 | 0.89 | 0.01 | From PDS measurements by G. Gilat, R. Nicklow, Phys. Rev. **143** (1966), 487. |
| | 294 | 1 | 0.90 | 0.01 | |
| | 375 | 1 | 1.21 | 0.01 | |
| | 475 | 1 | 1.57 | 0.01 | |
| D.L. McDonald, Acta Cryst. **23** (1967), 185. | 486 | 1 | 1.58 | 0.01 | |
| | 552 | 1 | 1.87 | 0.01 | |
| | 590 | 1 | 2.04 | 0.01 | Read in off graph - corrected for TDS |
| | 657 | 1 | 2.40 | 0.01 | |
| | 700 | 1 | 2.58 | 0.01 | |
| | 733 | 1 | 2.73 | 0.01 | |
| | 811 | 1 | 3.19 | 0.01 | |
| | 830 | 1 | 3.29 | 0.01 | |
| | 861 | 1 | 3.58 | 0.01 | |
| E.H. Medlin, R.E. Dingle, D.W. Field, Nature **224** (1969), 581. | 293 | 1 | 0.816 | 0.003 | Single crystal X-ray diffraction |
| O. Inkinen, A. Pesonen, T. Paakkari, Ann. Acad. Sci. Fenn. A **VI** Physica (1970), 344. | 293 | 1 | 0.89 | 0.01 | Taken from A.G. Fox, M.A. Tabbernor, R.M. Fisher, J. Phys. Chem. Solids **51** (1990), 1323. |
| | 293 | 1 | 0.85 | 0.01 | |
| R.E. Dingle, E.H. Medlin, Acta Cryst. A **28** (1972), 22. | 370 | 2 | 1.09 | 0.01 | |
| | 477 | 3 | 1.45 | 0.01 | |
| | 559 | 9 | 1.84 | 0.01 | |
| | 400 | 1 | 1.25 | 0.01 | |
| | 400 | 1 | 1.30 | 0.01 | |
| | 500 | 1 | 1.61 | 0.01 | |
| | 500 | 1 | 1.67 | 0.01 | |
| G. Albanese, C. Ghezzi, Phys. Rev. B **8** (1973), 1315. | 599 | 1 | 2.16 | 0.01 | Read in off graph from R.C. Shukla, C.A. Plint, Phys. Rev. B **40** (1989), 10337. |
| | 599 | 1 | 2.16 | 0.01 | |
| | 700 | 1 | 2.72 | 0.01 | |
| | 700 | 1 | 2.65 | 0.01 | |
| | 799 | 1 | 3.32 | 0.01 | |
| | 799 | 1 | 3.26 | 0.01 | |
| E. Rantavuori, V.-P. Tanninen, Physica Scr. **15** (1977), 273 | 80 | 1 | 0.22 | 0.01 | Take this as the most reliable measured value of $B$ (80K) |
| | 323 | 1 | 1.19 | 0.01 | |
| | 383 | 1 | 1.30 | 0.01 | |
| | 439 | 1 | 1.48 | 0.01 | |
| | 505 | 1 | 1.68 | 0.01 | |
| | 625 | 1 | 2.08 | 0.01 | |
| | 668 | 1 | 2.34 | 0.01 | Read in off graph - measured from (222) |
| | 679 | 1 | 2.36 | 0.01 | |
| | 733 | 1 | 2.75 | 0.01 | |
| | 790 | 1 | 3.00 | 0.01 | |
| | 848 | 1 | 3.30 | 0.01 | |
| C.J. Martin, D.A. O'Connor, Acta Cryst. A **34** (1978), 500. | 898 | 1 | 3.47 | 0.01 | |
| | 290 | 1 | 0.90 | 0.01 | |
| | 345 | 1 | 1.08 | 0.01 | |
| | 400 | 1 | 1.27 | 0.01 | |
| | 453 | 1 | 1.46 | 0.01 | |
| | 507 | 1 | 1.65 | 0.01 | |
| | 566 | 1 | 1.90 | 0.01 | Read in off graph - measured from (333) |
| | 622 | 1 | 2.12 | 0.01 | |
| | 680 | 1 | 2.41 | 0.01 | |
| | 734 | 1 | 2.63 | 0.01 | |
| | 794 | 1 | 2.99 | 0.01 | |
| | 848 | 1 | 3.35 | 0.01 | |
| | 909 | 1 | 3.77 | 0.01 | |
| A.G. Fox, R.M. Fisher, Aust. J. Phys. **41** (1988), 461. | 293 | 1 | 0.85 | 0.01 | Probably obtained from R.E. Dingle, E.H. Medlin, Acta Cryst. A**28** (1972), 22. |
| | 0 | 0 | 0.27 | | |
| | 1 | 0 | 0.27 | | |
| | 2 | 0 | 0.27 | | |
| | 3 | 0 | 0.27 | | |
| | 4 | 0 | 0.27 | | |
| | 5 | 0 | 0.27 | | |
| H.X. Gao, L.M. Peng, Acta Cryst. A **55** (1999), 926 in conjunction with L.-M. Peng, G. Ren, S.L. Dudarev, M.J. Whelan, Acta Cryst. A **52** (1996), 456. Also published in L.-M. Peng, S.L. Dudarev, M.J. Whelan, High-Energy Elecytron Diffraction and Microscopy (Oxford University Press, 2004), 454. | 10 | 0 | 0.27 | | |
| | 15 | 0 | 0.27 | | |
| | 20 | 0 | 0.27 | | Parameterised fit (T<80K) to phonon density of states measurements by G. Gilat, R. Nicklow, Phys. Rev. **143** (1966), 487. |
| | 25 | 0 | 0.28 | | |
| | 30 | 0 | 0.28 | | |
| | 35 | 0 | 0.28 | | |
| | 40 | 0 | 0.29 | | |
| | 45 | 0 | 0.29 | | |
| | 50 | 0 | 0.29 | | |
| | 60 | 0 | 0.31 | | |
| | 70 | 0 | 0.32 | | |

| Source | T (K) | Error in T (where not given, ±1K assumed) | B (Å²) | Error in B (where not given, ±0.01 is assumed) | Notes |
|---|---|---|---|---|---|
| H.X. Gao, L.M. Peng, Acta Cryst. A **55** (1999), 926 in conjunction with L.-M. Peng, G. Ren, S.L. Dudarev, M.J. Whelan, Acta Cryst. A **52** (1996), 456. Also published in L.-M. Peng, S.L. Dudarev, M.J. Whelan, High-Energy Elecytron Diffraction and Microscopy (Oxford University Press, 2004), 454. | 80 | 0 | 0.33 | | Parameterised fit (T<80K) to phonon density of states measurements by G. Gilat, R. Nicklow, *Phys. Rev.* **143** (1966), 487. Note that E. Rantavuori, V.-P. Tanninen, *Physica Scr.* **15** (1977), 273 state that this value was too high to be able to fit their experimental powder X-ray data. |
| | 90 | 0 | 0.35 | | |
| | 100 | 0 | 0.37 | | |
| | 120 | 0 | 0.41 | | |
| | 140 | 0 | 0.45 | | |
| | 160 | 0 | 0.50 | | |
| | 180 | 0 | 0.54 | | |
| | 200 | 0 | 0.59 | | |
| | 220 | 0 | 0.64 | | |
| | 240 | 0 | 0.69 | | |
| | 260 | 0 | 0.73 | | |
| | 280 | 0 | 0.78 | | |
| | 300 | 0 | 0.83 | | Parameterised fit (T>80K) to phonon density of states measurements by G. Gilat, R. Nicklow, *Phys. Rev.* **143** (1966), 487. |
| | 350 | 0 | 0.96 | | |
| | 400 | 0 | 1.09 | | |
| | 450 | 0 | 1.22 | | |
| | 500 | 0 | 1.35 | | |
| | 550 | 0 | 1.48 | | |
| | 600 | 0 | 1.62 | | |
| | 650 | 0 | 1.75 | | |
| | 700 | 0 | 1.88 | | |
| | 750 | 0 | 2.01 | | |
| | 800 | 0 | 2.14 | | |
| | 850 | 0 | 2.27 | | |
| | 900 | 0 | 2.40 | | |
| | 933 | 0 | 2.49 | | |
| C. Sternemann, T. Buslaps, A. Shukla, P. Suortti, G. Döring, W. Schülke, *Phys. Rev. B* **63** (2001), 094301. | 15 | 1 | 0.15 | 0.01 | In better agreement with E. Rantavuori, V.-P. Tanninen, *Physica Scr.* **15** (1977), 273 |
| | 560 | 1 | 2.08 | 0.01 | |

## THEORETICAL DETERMINATIONS

| Source | T (K) | | B (Å²) | | Notes |
|---|---|---|---|---|---|
| R.C. Shukla, H. Hübschle, *Sol. Sta. Commun.* **72** (1989), 1135. (supercedes R.C. Shukla, C.A. Plint, *Phys. Rev. B* **40** (1989), 10337.) | 300 | | 0.8506 | | AVS QH - Ashcroft pseudopotential with Vashishta-Singwi electron gas screening functions, ε(q) [P.V.S. Rao, *J. Phys. Chem. Solids* **35** (1974), 669.]. Applies quasi-harmonic theory (QH). |
| | 450 | | 1.2759 | | |
| | 600 | | 1.7012 | | |
| | 750 | | 2.1265 | | |
| | 850 | | 2.41 | | |
| | 300 | | 0.9413 | | AVS λ²PT - Ashcroft pseudopotential with Vashishta-Singwi electron gas screening functions, ε(q) [P.V.S. Rao, J. Phys. Chem. Solids **35** (1974), 669.]. The anharmonic contributions are evaluated in the lowest order (λ²) perturbtion theory and includes the quasi harmonic components (QH) (cubic and quartic contributions which result in the T and T² terms in *B* (T)) (PT) [A.A. Maradudin, P.A. Flinn, *Phys. Rev.* **129** (1963), 2529]. |
| | 450 | | 1.4585 | | |
| | 600 | | 1.9855 | | |
| | 750 | | 2.5057 | | |
| | 850 | | 2.8388 | | |
| | 300 | | 0.9554 | | AVS RE - Ashcroft pseudopotential with Vashishta-Singwi electron gas screening functions, ε(q). All orders of anharmonicity are included via a Green's function method (RE) [R.C. Shukla, H. Hübschle, *Phys. Rev. B* **40** (1989), 1555. G.A. Heiser, R.C. Shukla, E.R. Cowley, *Phys. Rev. B* **33** (1986), 2158.] |
| | 450 | | 1.5007 | | |
| | 600 | | 2.0703 | | |
| | 750 | | 2.6399 | | |
| | 850 | | 3.0022 | | |
| | 300 | | 0.8497 | | AH QH - Ashcroft pseudopotential with Hubbard electron gas screening function, ε(q) [R.C. Shukla, C.A. Plint, *Int. J. Thermophys.* **1** (1980), 299; R.C. Shukla, C.A. Plint, D.A. Ditmars, *Int. J. Thermophys.* **6** (1985), 517.]. Applies quasi-harmonic theory (QH). |
| | 450 | | 1.2745 | | |
| | 600 | | 1.6994 | | |
| | 750 | | 2.1242 | | |
| | 850 | | 2.4075 | | |

| Source | T (K) | Error in T (where not given, ±1K assumed) | B (Å$^2$) | Error in B (where not given, ±0.01 is assumed) | Notes |
|---|---|---|---|---|---|
| R.C. Shukla, H. Hübschle, *Sol. Sta. Commun.* **72** (1989), 1135. (supercedes R.C. Shukla, C.A. Plint, *Phys. Rev. B* **40** (1989), 10337.) | 300 | | 0.9423 | | AH λ$^2$PT - Ashcroft pseudopotential with Hubbard electron gas screening function, ε(q) [R.C. Shukla, C.A. Plint, Int. J. Thermophys. **1** (1980), 299; R.C. Shukla, C.A. Plint, D.A. Ditmars, Int. J. Thermophys. **6** (1985), 517.]. The anharmonic contributions are evaluated in the lowest order (λ$^2$) perturbtion theory and includes the quasi harmonic components (QH) (cubic and quartic contributions which result in the T and T$^2$ terms in B (T)) (PT) [A.A. Maradudin, P.A. Flinn, Phys. Rev. **129** (1963), 2529]. |
| | 450 | | 1.4619 | | |
| | 600 | | 1.993 | | |
| | 750 | | 2.5178 | | |
| | 850 | | 2.8574 | | |
| | 300 | | 0.9573 | | AH RE - Ashcroft pseudopotential with Hubbard electron gas screening function, ε(q) [R.C. Shukla, C.A. Plint, Int. J. Thermophys. **1** (1980), 299; R.C. Shukla, C.A. Plint, D.A. Ditmars, Int. J. Thermophys. **6** (1985), 517.]. All orders of anharmonicity are included via a Green's function method (RE) [R.C. Shukla, H. Hübschle, Phys. Rev. B **40** (1989), 1555. G.A. Heiser, R.C. Shukla, E.R. Cowley, Phys. Rev. B **33** (1986), 2158.] |
| | 450 | | 1.5074 | | |
| | 600 | | 2.086 | | |
| | 750 | | 2.6668 | | |
| | 850 | | 3.0422 | | |
| | 300 | | 0.8671 | | HHS QH - Harrison modified point ion pseudopotential with Hubbard-Sham electron gas screening function, ε(q) [R.C. Shukla, C.A. Plint, Int. J. Thermophys. **1** (1980), 299; R.C. Shukla, C.A. Plint, D.A. Ditmars, Int. J. Thermophys. **6** (1985), 517.]. Applied quasi-harmonic theory (QH). |
| | 450 | | 1.3007 | | |
| | 600 | | 1.7343 | | |
| | 750 | | 2.1679 | | |
| | 850 | | 2.4569 | | |
| | 300 | | 0.9687 | | HHS λ$^2$PT - Harrison modified point ion pseudopotential with Hubbard-Sham electron gas screening function, ε(q) [R.C. Shukla, C.A. Plint, Int. J. Thermophys. **1** (1980), 299; R.C. Shukla, C.A. Plint, D.A. Ditmars, Int. J. Thermophys. **6** (1985), 517.]. The anharmonic contributions are evaluated in the lowest order (λ$^2$) perturbtion theory and includes the quasi harmonic components (QH) (cubic and quartic contributions which result in the T and T$^2$ terms in B (T)) (PT) [A.A. Maradudin, P.A. Flinn, Phys. Rev. **129** (1963), 2529]. |
| | 450 | | 1.5154 | | |
| | 600 | | 2.0752 | | |
| | 750 | | 2.6342 | | |
| | 850 | | 2.9976 | | |
| R.C. Shukla, H. Hübschle, *Sol. Sta. Commun.* **72** (1989), 1135. (supercedes R.C. Shukla, C.A. Plint, *Phys. Rev. B* **40** (1989), 10337.) | 300 | | 0.9875 | | HHS RE - Harrison modified point ion pseudopotential with Hubbard-Sham electron gas screening function, ε(q) [R.C. Shukla, C.A. Plint, Int. J. Thermophys. **1** (1980), 299; R.C. Shukla, C.A. Plint, D.A. Ditmars, Int. J. Thermophys. **6** (1985), 517.]. All orders of anharmonicity are included via a Green's function method (RE) [R.C. Shukla, H. Hübschle, Phys. Rev. B **40** (1989), 1555. G.A. Heiser, R.C. Shukla, E.R. Cowley, Phys. Rev. B **33** (1986), 2158.] |
| | 450 | | 1.5816 | | |
| | 600 | | 2.2199 | | |
| | 750 | | 2.8864 | | |
| | 850 | | 3.3285 | | |
| | 300 | | 0.8267 | | Morse QH - Morse potential parameters [R.C. Shukla, R.A. MacDonald, *High Temp. High Press.* **12** (1980), 291.] Applied quasi-harmonic theory (QH). |
| | 450 | | 1.27 | | |
| | 600 | | 1.7395 | | |
| | 750 | | 2.2425 | | |
| | 850 | | 2.6009 | | |
| | 300 | | 0.843 | | Morse λ$^2$PT - Morse potential parameters [R.C. Shukla, R.A. MacDonald, High Temp. High Press. **12** (1980), 291.] The anharmonic contributions are evaluated in the lowest order (λ$^2$) perturbtion theory and includes the quasi harmonic components (QH) (cubic and quartic contributions which result in the T and T$^2$ terms in B (T)) (PT) [A.A. Maradudin, P.A. Flinn, Phys. Rev. **129** (1963), 2529]. |
| | 450 | | 1.3073 | | |
| | 600 | | 1.8077 | | |
| | 750 | | 2.3521 | | |
| | 850 | | 2.7445 | | |
| | 300 | | 0.8434 | | Morse RE - Morse potential parameters [R.C. Shukla, R.A. MacDonald, High Temp. High Press. **12** (1980), 291.] All orders of anharmonicity are included via a Green's function method (RE) [R.C. Shukla, H. Hübschle, Phys. Rev. B **40** (1989), 1555. G.A. Heiser, R.C. Shukla, E.R. Cowley, Phys. Rev. B **33** (1986), 2158.] |
| | 450 | | 1.3088 | | |
| | 600 | | 1.8114 | | |
| | 750 | | 2.3595 | | |
| | 850 | | 2.7558 | | |
| R.C.G. Killean, *J. Phys. F: Metal Phys.* **4** (1974), 1908. | 80 | | 0.32 | | Nearest neighbour central force pair interactions model |
| | 295 | | 0.89 | | |
| | 300 | | 0.9 | | |
| | 375 | | 1.16 | | |
| | 475 | | 1.54 | | |
| | 485 | | 1.58 | | |
| | 552 | | 1.87 | | |
| | 588 | | 2.03 | | |
| | 655 | | 2.36 | | |
| | 700 | | 2.59 | | |
| | 730 | | 2.77 | | |
| | 810 | | 3.25 | | |
| | 830 | | 3.37 | | |
| | 860 | | 3.57 | | |

## Appendix D

Table of experimentally and theoretically determined deformation electron densities (Δρ) at the tetrahedral, octahedral and bridge centres in the fcc unit cell of pure aluminium. These are followed by the actual structure factor determinations and associated errors (in the cases of experimental measurements) for the bonding-sensitive structure factors, $F_{111}$, $F_{200}$ and $F_{220}$.

| Source | Δρ tetrahedral interstice (e⁻/Å³) | error Δρ tetrahedral | Δρ octahedral interstice (e⁻/Å³) | error Δρ octahedral | Δρ bridge bond (e⁻/Å³) | error Δρ bridge | $F_{111}$ (e⁻/atom) | error $F_{111}$ (if no error given, error taken as ±1% - the mean error of all the X-ray meas.) | $F_{002}$ (e⁻/atom) | error $F_{002}$ (if no error given, error taken as ±1% - the mean error of all the X-ray meas.) | $F_{220}$ (e⁻/atom) | error $F_{220}$ (if no error given, error taken as ±1% - the mean error of all the X-ray meas.) | Notes |
|---|---|---|---|---|---|---|---|---|---|---|---|---|---|
| **Experiments** (NB All values (where necessary) converted to T=0K using a unified value of B=0.82 Å² for room temp.) | | | | | | | | | | | | | |
| R.W. James, G.W. Brindley, R.G. Wood, *Proc. Roy. Soc. A* **125** (1929), 401. | 0.018 | 0.016 | 0.006 | 0.020 | 0.026 | 0.005 | 8.78 | 0.09 | 8.37 | 0.08 | 7.27 | 0.07 | Expt 1: X-ray diffraction (Mo). Δρ are calculated directly from the quoted structure factors using the IAM of P.A. Doyle, P.S. Turner, *Acta Cryst. A* **24** (1968), 390. |
| | 0.059 | 0.017 | 0.068 | 0.020 | -0.015 | 0.006 | 8.9 | 0.09 | 8.48 | 0.08 | 7.38 | 0.07 | Expt 2: X-ray diffraction (Cu). Δρ are calculated directly from the quoted structure factors using the IAM of P.A. Doyle, P.S. Turner, *Acta Cryst. A* **24** (1968), 390. |
| G.W. Brindley, *Philos. Mag.* **21** (1936), 778. | 0.037 | 0.016 | -0.095 | 0.020 | 0.051 | 0.005 | 8.83 | 0.09 | 8.24 | 0.08 | 7.23 | 0.07 | Expt 1: X-ray diffraction (Cu). Δρ are calculated directly from the quoted structure factors using the IAM of P.A. Doyle, P.S. Turner, *Acta Cryst. A* **24** (1968), 390. |
| | 0.073 | 0.016 | -0.044 | 0.020 | 0.029 | 0.005 | 8.86 | 0.09 | 8.28 | 0.08 | 7.3 | 0.07 | Expt 2: X-ray diffraction (Cu). Δρ are calculated directly from the quoted structure factors using the IAM of P.A. Doyle, P.S. Turner, *Acta Cryst. A* **24** (1968), 390. |
| G.W. Brindley, P. Ridley, *Proc. Phys. Soc.* **50** (1938), 96. | 0.092 | 0.017 | 0.011 | 0.020 | -0.004 | 0.006 | 8.95 | 0.09 | 8.39 | 0.08 | 7.38 | 0.07 | X-ray powder diffraction. Δρ are calculated directly from the quoted structure factors using the IAM of P.A. Doyle, P.S. Turner, Acta Cryst. A **24** (1968), 390. |
| N.V. Ageev, D.L. Ageeva, *Izv. Akad. Nauk SSSR, Otd. Khim. Nauk* **1** (1948), 17. | 0.000 | 0.016 | -0.063 | 0.020 | 0.054 | 0.005 | 8.75 | 0.09 | 8.28 | 0.08 | 7.2 | 0.07 | Expt 1: X-ray diffraction (Cu). Δρ are calculated directly from the quoted structure factors using the IAM of P.A. Doyle, P.S. Turner, Acta Cryst. A **24** (1968), 390. |
| | -0.011 | 0.016 | -0.033 | 0.020 | 0.050 | 0.005 | 8.71 | 0.09 | 8.31 | 0.08 | 7.2 | 0.07 | Expt 2: X-ray diffraction (Fe). Δρ are calculated directly from the quoted structure factors using the IAM of P.A. Doyle, P.S. Turner, Acta Cryst. A **24** (1968), 390. |
| H. Bensch, H. Witte, E. Wölfel, *Z. Phys. Chem.* **4** (1955), 65. | 0.022 | 0.016 | 0.046 | 0.020 | 0.037 | 0.005 | 8.63 | 0.09 | 8.32 | 0.08 | 7.25 | 0.07 | Single Crystal X-ray Diffraction. Δρ are calculated directly from the quoted structure factors using the IAM of P.A. Doyle, P.S. Turner, Acta Cryst. A **24** (1968), 390. |
| R.B. Roof Jr., *J. Appl. Phys.* **30** (1959), 1599. | -0.110 | 0.038 | 0.061 | 0.041 | -0.066 | 0.013 | 9.23 | 0.13 | 8.92 | 0.18 | 7.37 | 0.17 | Powder X-ray diffraction (Mo, Cu, Cr radiation). Results differ from those used in P.N.H. Nakashima, A.E. Smith, J. Etheridge, B.C. Muddle, *Science* **331** (2011), 1583 because absorption correction had not been included in the values used in that reference. Δρ are calculated directly from the quoted structure factors using the IAM of P.A. Doyle, P.S. Turner, Acta Cryst. A **24** (1968), 390. |
| B.W. Batterman, D.R. Chipman, J.J. DeMarco, *Phys. Rev.* **122** (1961), 68. | -0.044 | 0.029 | -0.054 | 0.034 | 0.059 | 0.010 | 8.7 | 0.14 | 8.32 | 0.14 | 7.16 | 0.13 | Powder X-ray diffraction (Mo radiation). Δρ are calculated directly from the quoted structure factors using the IAM of P.A. Doyle, P.S. Turner, Acta Cryst. A **24** (1968), 390. |
| J.J. DeMarco, *Philos. Mag.* **15** (1967), 483. | 0.062 | 0.014 | -0.023 | 0.015 | 0.050 | 0.005 | 8.69 | 0.04 | 8.21 | 0.07 | 7.25 | 0.06 | Single Crystal X-ray Diffraction (Mo radiation). Δρ are calculated directly from the quoted structure factors using the IAM of P.A. Doyle, P.S. Turner, Acta Cryst. A **24** (1968), 390. |
| M. Järvinen, M. Merisalo, O. Inkinen, *Phys. Rev.* **178** (1969), 1108. | -0.007 | 0.020 | -0.095 | 0.022 | 0.066 | 0.007 | 8.74 | 0.06 | 8.24 | 0.1 | 7.17 | 0.09 | Powder X-ray diffraction (Mo radiation). Δρ are calculated directly from the quoted structure factors using the IAM of P.A. Doyle, P.S. Turner, Acta Cryst. A **24** (1968), 390. |
| P.M. Raccah, V.E. Henrich, *Phys. Rev.* **184** (1969), 607. | 0.015 | 0.013 | 0.000 | 0.015 | 0.024 | 0.004 | 8.8 | 0.06 | 8.38 | 0.06 | 7.27 | 0.06 | Powder X-ray diffraction (Cu radiation). Δρ are calculated directly from the quoted structure factors using the IAM of P.A. Doyle, P.S. Turner, Acta Cryst. A **24** (1968), 390. |

| Source | Δρ tetrahedral interstice (e⁻/Å³) | error Δρ tetrahedral | Δρ octahedral interstice (e⁻/Å³) | error Δρ octahedral | Δρ bridge bond (e⁻/Å³) | error Δρ bridge | $F_{111}$ (e⁻/atom) | error $F_{111}$ (if no error given, error taken as ±1% - the mean error of all the X-ray meas.) | $F_{002}$ (e⁻/atom) | error $F_{002}$ (if no error given, error taken as ±1% - the mean error of all the X-ray meas.) | $F_{220}$ (e⁻/atom) | error $F_{220}$ (if no error given, error taken as ±1% - the mean error of all the X-ray meas.) | Notes |
|---|---|---|---|---|---|---|---|---|---|---|---|---|---|
| O. Inkinen, A. Pesonen, T. Paakkari, *Ann. Acad. Sci. Fenn. A* **6** (1970), 344. | 0.026 | 0.018 | 0.013 | 0.020 | 0.018 | 0.006 | 8.81 | 0.07 | 8.39 | 0.08 | 7.29 | 0.08 | X-ray diffraction (powder pressed in a way so as to reduce all orientaiotnal texture effects) - values taken from R.J. Temkin, V.E. Henrich, P.M. Raccah, *Sol. St. Commun.* **13** (1973), 811. Δρ are calculated directly from the quoted structure factors using the IAM of P.A. Doyle, P.S. Turner, *Acta Cryst.* A **24** (1968), 390. |
| E. Rantavuori, V.-P. Tanninen, *Phys. Scr.* **15** (1977), 273. | 0.033 | 0.009 | -0.062 | 0.010 | 0.045 | 0.003 | 8.8 | 0.04 | 8.27 | 0.04 | 7.24 | 0.04 | Powder X-ray diffraction at T=80K (Cu radiation). Δρ are calculated directly from the quoted structure factors using the IAM of P.A. Doyle, P.S. Turner, *Acta Cryst.* A **24** (1968), 390. |
| T. Takama, K. Kobayashi, S. Sato, *Trans. Jap. Inst. Metals* **23** (1982), 153. | 0.029 | 0.008 | 0.068 | 0.009 | -0.015 | 0.003 | 8.9 | 0.03 | 8.52 | 0.05 | 7.36 | 0.03 | Pendellösung white radiation X-ray diffraction (single crystal). Values taken from A.G. Fox, M.A. Tabbernor, R.M. Fisher, *J. Phys. Chem. Solids* **51** (1990), 1323. Δρ are calculated directly from the quoted structure factors using the IAM of P.A. Doyle, P.S. Turner, *Acta Cryst.* A **24** (1968), 390. |
| D. Watanabe, R. Uyeda, A. Fukuhara, *Acta Cryst. A* **24** (1968), 580. & T. Arii, R. Uyeda, O. Terasaki, D. Watanabe, *Acta Cryst. A* **29** (1973), 295. | 0.019 | 0.015 | -0.045 | 0.018 | 0.028 | 0.005 | 8.87 | 0.08 | 8.36 | 0.05 | 7.27 | 0.07 | Critical Voltage method (Electron diffraction). Cannot measure higher orders than $V_{200}$ so $F_{220}$ is taken from the average of preceeding X-ray measurements. Δρ are calculated directly from the quoted structure factors using the IAM of P.A. Doyle, P.S. Turner, *Acta Cryst.* A **24** (1968), 390. |
| A.G. Fox, R.M. Fisher, *Aust. J. Phys.* **41** (1988), 461. | 0.011 | 0.014 | -0.016 | 0.015 | 0.023 | 0.005 | 8.84 | 0.04 | 8.39 | 0.02 | 7.27 | 0.07 | Critical Voltage method (Electron diffraction). Cannot measure higher orders than $V_{200}$ so $F_{220}$ is taken from the average of preceeding X-ray measurements. Δρ are calculated directly from the quoted structure factors using the IAM of P.A. Doyle, P.S. Turner, *Acta Cryst.* A **24** (1968), 390. |
| P.N.H. Nakashima, A.E. Smith, J. Etheridge, B.C. Muddle, *Science* **331** (2011), 1583. | 0.048 | 0.006 | -0.009 | 0.006 | 0.016 | 0.002 | 8.87 | 0.01 | 8.37 | 0.01 | 7.31 | 0.03 | Quantitative Convergent-Beam Electron Diffraction (QCBED) (Bloch-wave formalism, 50, 120, 200, 300kV). Δρ are calculated directly from the quoted structure factors using the IAM of P.A. Doyle, P.S. Turner, *Acta Cryst.* A **24** (1968), 390. |
| **Theoretical Calculations** (at T=0K by default) | | | | | | | | | | | | | |
| F.J. Arlinghaus, *Phys. Rev.* **153** (1967), 743 | 0.018 | 0.000 | 0.016 | 0.000 | -0.009 | 0.000 | 8.97 | 0 | 8.51 | 0 | 7.34 | 0 | Band structure calculation based on augmented plane wave method. Δρ are calculated directly from the quoted structure factors using the IAM of P.A. Doyle, P.S. Turner, *Acta Cryst.* A **24** (1968), 390. |
| P. Ascarelli, P.M. Raccah, *Phys. Lett. A* **31** (1970), 549. | 0.062 | 0.000 | 0.026 | 0.000 | 0.016 | 0.000 | 8.8 | 0 | 8.35 | 0 | 7.32 | 0 | Developed their own pseudopotential model (early DFT). Δρ are calculated directly from the quoted structure factors using the IAM of P.A. Doyle, P.S. Turner, *Acta Cryst.* A **24** (1968), 390. |
| M. Cooper, B. Williams, *Philos. Mag.* **26** (1972), 1441. | 0.062 | 0.000 | 0.109 | 0.000 | -0.023 | 0.000 | 8.87 | 0 | 8.51 | 0 | 7.4 | 0 | Early DFT based on a set of local orbital (lo) wavefunctions. Δρ are calculated directly from the quoted structure factors using the IAM of P.A. Doyle, P.S. Turner, *Acta Cryst.* A **24** (1968), 390. |
| J.P. Walter, C.Y. Fong, M.L. Cohen, *Solid St. Commun.* **12** (1973), 303. | 0.121 | 0.000 | 0.065 | 0.000 | 0.001 | 0.000 | 8.81 | 0 | 8.33 | 0 | 7.39 | 0 | Early DFT based on an Ashcroft pseudopotential. Δρ are calculated directly from the quoted structure factors using the IAM of P.A. Doyle, P.S. Turner, *Acta Cryst.* A **24** (1968), 390. |
| F. Perrot, *Solid St. Commun.* **14** (1974), 1041. | 0.029 | 0.000 | 0.005 | 0.000 | 0.015 | 0.000 | 8.85 | 0 | 8.4 | 0 | 7.3 | 0 | DFT with APW and Kohn-Sham (2/3 Slater exchange) approach. Δρ are calculated directly from the quoted structure factors using the IAM of P.A. Doyle, P.S. Turner, *Acta Cryst.* A **24** (1968), 390. |
| | 0.059 | 0.000 | 0.049 | 0.000 | 0.000 | 0.000 | 8.85 | 0 | 8.42 | 0 | 7.35 | 0 | Hartree-Fock approach with approximations for core and valence electrons. Δρ are calculated directly from the quoted structure factors using the IAM of P.A. Doyle, P.S. Turner, *Acta Cryst.* A **24** (1968), 390. |
| R.A. Tawil, *Phys. Rev. B* **11** (1975), 4891. | -0.004 | 0.000 | -0.104 | 0.000 | 0.062 | 0.000 | 8.78 | 0 | 8.25 | 0 | 7.18 | 0 | Tight binding approximation. Δρ are calculated directly from the quoted structure factors using the IAM of P.A. Doyle, P.S. Turner, *Acta Cryst.* A **24** (1968), 390. |

| Source | Δρ tetrahedral interstice (e⁻/Å³) | error Δρ tetrahedral | Δρ octahedral interstice (e⁻/Å³) | error Δρ octahedral | Δρ bridge bond (e⁻/Å³) | error Δρ bridge | $F_{111}$ (e⁻/atom) | error $F_{111}$ (if no error given, error taken as ±1% - the mean error of all the X-ray meas.) | $F_{002}$ (e⁻/atom) | error $F_{002}$ (if no error given, error taken as ±1% - the mean error of all the X-ray meas.) | $F_{220}$ (e⁻/atom) | error $F_{220}$ (if no error given, error taken as ±1% - the mean error of all the X-ray meas.) | Notes |
|---|---|---|---|---|---|---|---|---|---|---|---|---|---|
| B. Dawson, *Studies of Atomic Charge Density by X-ray and Neutron Diffraction - A Perspective*, in *Advances in Structure Research by Diffraction Methods*, W. Hoppe, R. Mason, Eds. (Pergamon Press, Oxford, New York, Toronto, Sydney, 1975), 221. | 0.051 | 0.000 | 0.076 | 0.000 | -0.007 | 0.000 | 8.84 | 0 | 8.46 | 0 | 7.36 | 0 | Hartree-Fock approach with 4S excitation - presumably to simulate the bonding electrons. Δρ are calculated directly from the quoted structure factors using the IAM of P.A. Doyle, P.S. Turner, Acta Cryst. A **24** (1968), 390. |
| P.J. Bunyan, J.A. Nelson, *J. Phys. F: Metal Phys.* **7** (1977), 2323. | 0.004 | 0.000 | -0.004 | 0.000 | 0.016 | 0.000 | 8.86 | 0 | 8.43 | 0 | 7.28 | 0 | Band structure eigenfunctions with an improved Heine-Abarenkov potential with depletion hole. Δρ are calculated directly from the quoted structure factors using the IAM of P.A. Doyle, P.S. Turner, Acta Cryst. A **24** (1968), 390. |
| J. Hafner, *Solid St. Commun.* **27** (1978), 263. | 0.033 | 0.000 | -0.043 | 0.000 | 0.035 | 0.000 | 8.82 | 0 | 8.31 | 0 | 7.26 | 0 | OPW-pseudopotential approach. Δρ are calculated directly from the quoted structure factors using the IAM of P.A. Doyle, P.S. Turner, Acta Cryst. A **24** (1968), 390. |
| S. Chakraborty, A. Manna, A.K. Ghosh, *Phys. Stat. Sol. B* **129** (1985), 211. | -0.040 | 0.000 | -0.150 | 0.000 | 0.040 | 0.000 | 9.01 | 0 | 8.39 | 0 | 7.2 | 0 | Lowdin alpha expansion method. Δρ are calculated directly from the quoted structure factors using the IAM of P.A. Doyle, P.S. Turner, Acta Cryst. A **24** (1968), 390. |
| T.T. Rantala, *J. Phys. F: Metal Phys.* **17** (1987), 877. | 0.022 | 0.000 | 0.007 | 0.000 | 0.012 | 0.000 | 8.86 | 0 | 8.42 | 0 | 7.3 | 0 | An atom in jellium model. Δρ are calculated directly from the quoted structure factors using the IAM of P.A. Doyle, P.S. Turner, Acta Cryst. A **24** (1968), 390. |
| P.N.H. Nakashima, A.E. Smith, J. Etheridge, B.C. Muddle, *Science* **331** (2011), 1583. | 0.037 | 0.000 | -0.012 | 0.000 | 0.017 | 0.000 | 8.87 | 0 | 8.38 | 0 | 7.3 | 0 | FP-LAPW +lo +ls DFT calculation using WIEN2K [P. Blaha, K. Schwarz, P. Sorantin, S.B. Trickey, *Comput. Phys. Commun.* **59** (1990), 399.]. Δρ are calculated directly from the quoted structure factors using the IAM of P.A. Doyle, P.S. Turner, Acta Cryst. A **24** (1968), 390. |
| **Independent Atom Model** (the most widely accepted - IUCr) | | | | | | | | | | | | | |
| P.A. Doyle, P.S. Turner, *Acta Cryst. A* **24** (1968), 390. | 0.000 | 0.000 | 0.000 | 0.000 | 0.000 | 0.000 | 8.95 | 0 | 8.5 | 0 | 7.31 | 0 | Relativistic Hartree-Fock calculations of independent neutral atoms taken as the standard model for unbonded atoms. |